\documentclass[%
 reprint,
 amsmath,amssymb,
 aps
]{revtex4-2}

\usepackage{graphicx}
\usepackage{dcolumn}
\usepackage{bm}

\usepackage{dsfont}
\usepackage{bm}
\usepackage{color}
\usepackage{hyperref}%
\usepackage{esint}
\usepackage{float}
\usepackage{subcaption}
\usepackage{caption}
\graphicspath{{./figs/}}

\begin{document}

\title{
Mobile impurity in a one-dimensional gas at finite temperatures}

\author{Oleksandr Gamayun}
\email[Correspondence to: ]{oleksandr.gamayun@fuw.edu.pl}
\affiliation{
Faculty of Physics, University of Warsaw, ul. Pasteura 5, 02-093 Warsaw, Poland
}%
\affiliation{Bogolyubov Institute for Theoretical Physics, 03143 Kyiv, Ukraine}

\author{Mi{\l}osz Panfil}
\affiliation{
Faculty of Physics, University of Warsaw, ul. Pasteura 5, 02-093 Warsaw, Poland
}%

\author{Felipe Taha Sant'Ana}
\affiliation{
Faculty of Physics, University of Warsaw, ul. Pasteura 5, 02-093 Warsaw, Poland
}%

\date{\today}		
	
	\begin{abstract}
		We consider the McGuire model of a one-dimensional gas of free fermions interacting with a single impurity. We compute the static one-body function and momentum distribution of the impurity at finite temperatures. The results involve averages over Fredholm determinants that we further analyse using the effective form factors approach. With this approach, we derive the large-distance behaviour of the one-body function, which takes the form of an averaged exponential decay. This method allows us to study an experimentally important regime of small momenta of the impurity's momentum distribution. We also consider the one-body function at short distances and compute finite temperature Tan's contact.
	\end{abstract}

	\date{\today}
	\maketitle

	\section{Introduction}
	
	Mobile impurities are ubiquitous in physical systems and are usually referred as polarons \cite{landau1965,Devreese_2009,kuper_book}.
	Recent years witness increase of the interest in mobile impurities due to the advances in fabrications and manipulations of systems of cold atoms \cite{RevModPhys.80.885,Trm2014}.
	The mobile impurities appear there either in imbalanced mixtures of two different gases \cite{PhysRevLett.89.190404}, or they can be created after the application of the 
	rf pulse on the system, transferring part of atoms to a different hyperfine state 
\cite{PhysRevLett.78.586,Stenger1998}. Such systems were created and explored in number of studies \cite{PhysRevLett.102.230402,Kohstall2012,PhysRevLett.117.055301,PhysRevLett.117.055302,PhysRevLett.122.093401}.
	Especially intriguing are experiments with one-dimensional systems \cite{PhysRevLett.91.250402,Kinoshita2004,Paredes2004,PhysRevA.85.023623}, as the increased role of interactions leads to prominent effects including in particular, quantum Newton Cradle \cite{Kinoshita2006} and Bloch oscillations without lattice \cite{Meinert2017}.
	
	To theoretically explore the properties of a quantum particle propagating in a one-dimensional medium different schemes and approximations were developed. For instance, various mean-field approaches
\cite{Panochko2019,panochko2021static,hryhorchak2021polaron,Koutentakis2021,PhysRevLett.122.183001} can be used to describe properties of the ground state. Instead, perturbation theories for the weak impurity-gas coupling 
allows one to describe dynamics of the impurity \cite{Massignan2014,Burovski2014,PhysRevA.89.063627,PhysRevE.90.032132,PhysRevB.101.104503}. 
Numerical methods such as the time-dependent density-matrix renormalization-group has been successfully applied to extract breathing mode \cite{PhysRevLett.110.015302} and 
time-evolving block decimation methods can be used to describe various non-equilibrium aspects of the impurity \cite{Massel2013}, including 
the quantum flutter phenomenon \cite{PhysRevLett.112.015302}.  
Aproaches based on variational methods, in which the wave function is parameterized by a finite number of particle-hole excitations \cite{Chevy2006,PhysRevLett.98.180402,PhysRevA.79.043615}, and with more sophisticated parametrizations \cite{PhysRevA.93.043606,Grusdt2015}, made it possible to address static and dynamic properties of the impurity \cite{PhysRevA.79.043615,PhysRevA.87.033616,PhysRevX.11.041015,Mathy2012}. 
Finally, the properties of the polaron can be addressed by the Monte Carlo methods \cite{PROKOF_EV_1993,PhysRevB.77.020408,PhysRevA.95.023619,Grusdt2017}. 
For a pedagogical review of numerical approaches to impurity physics see Ref. \cite{grusdt2015new}.

In addition to numerical methods truly non-perturbative treatment of one-dimensional quantum system can be achieved for models that are solvable  
by Bethe Ansatz methods. 
An example of such model is the McGuire model \cite{McGuire1965,McGuire1966}
that describes a spin down particle interacting with the gas of spin up particles via the contact interaction. This model represents a specific sector of the fermionic Yang-Gaudin model 
\cite{Gaudin1967,PhysRevLett.19.1312}. Contrary to generic models the wave functions of the McGuire model can be written as a single determinant resembling the Slater determinant for the free Fermi
gas \cite{Edwards1990,PhysRevB.47.16186}. This allows one to find exact analytic expressions for the various physical quantities in the thermodynamic limit. In addition to the simplest correlation functions and the effective mass computed already by McGuire in Refs. \cite{McGuire1965,McGuire1966}, one can compute the large time asymptotics of the average momentum of the impurity injected in the gas with some initial velocity \cite{PhysRevLett.120.220605}, 
two-point correlation functions \cite{Gamayun2015,Gamayun2016} and the impurity's momentum distribution in the ground state (or a boosted ground state) of the whole system \cite{Recher2012,10.21468/SciPostPhys.8.4.053}.  

In this manuscript we consider impurity's momentum distribution at finite temperature in the McGuire model.
It is computed as a Fourier transform of the one-body function. 
We find that similar to \cite{10.21468/SciPostPhys.8.4.053} the answer in the thermodynamic limit can be expressed via the Fredholm determinants that additionally have to be integrated over 
an additional degree of freedom related to the impurity's momentum (the spin rapidity). 
We also explore large distance asymptotic of these Fredholm determinants by employing the effective form factors methods \cite{EffectiveXY,zhuravlev2021large,chernowitz2022dynamics}. 
This allows us to find analytically the prefactor and the correlation length (before integration of the spin rapidity). 
We also analytically compute Tan's contact as a function of coupling constant and the the temperature. 
In our derivations we never use the specific form of the thermal distribution and the same approach presented here can be used to compute the impurity's correlation function for a system in generalized Gibbs ensemble, that could appear, for instance, after the quench protocol~\cite{PhysRevA.89.033601}.

The structure of the manuscript is as follows. In Section \ref{model} we introduce the model for an impurity, recall main results form the Bethe ansatz and compute exact answers for the one-body function at finite temperature. In Section \ref{effective} we analyze large distance behavior of the one-body function with the help of effective form factors. Section \ref{sec:nk} is devoted to the impurity's momentum distribution. 
Finally, Section \ref{sec:f} contains conclusions and an outlook. In appendices we gathered more technical results. In Appendix~\ref{app:thermodynamics} on the thermodynamics of the impurity, in Appendix~\ref{app:factorD} on thermodynamic limit of the form factors and in Appendix~\ref{appC} on an analytic structure of the one-body function in the asymptotic regime.

	\section{The model and the correlation functions}
	
	\label{model}
	
	The Hamiltonian of the McGuire model \cite{McGuire1965,McGuire1966} is given by the following expression,
 	\begin{equation}\label{Hham}
		H = \frac{P_{\rm imp}^2}{2m} + \sum_{j=1}^N \frac{P_j^2}{2m} + g \sum_{j=1}^N \delta (x_j - x_{\rm imp}).
	\end{equation}
	The model describes a gas of spin up particles with momenta $P_j$ and coordinates $x_j$ interacting with a single mobile impurity (the spin down particle), with the momentum $P_{\rm imp}$ and the coordinate $X_{\rm imp}$. 
	The gas particles are assumed to be either fermions or, equivalently, the hardcore bosons. 
   The impurity-gas coupling strength $g$ in the dimensionless form is
	\begin{equation}
		\gamma = \frac{m g}{\rho_0},
	\end{equation}
	where $\rho_0 = N/L$ is the gas density.  We also set $m=1$. 
	Introducing the creation $\psi_{k, \downarrow}^+$ and annihilation operators for the impurity $\psi_{k, \downarrow}$ one can formally write an impurity's momentum distribution in an eigenstate $| \{k_j\}, \Lambda \rangle$ as 
	\begin{equation}
		n(k; \{k_j\}, \Lambda) = \langle \{k_j\}, \Lambda |\psi_{k, \downarrow}^+ \psi_{k, \downarrow}| \{k_j\}, \Lambda \rangle.
	\end{equation}
	The eigenstate $| \{k_j\}, \Lambda \rangle$ of the Hamiltonian \eqref{Hham}  is specified by a set of rapidities $\{k_j\}$ and $\Lambda$. For a system of length $L$ with periodic boundary conditions the rapidities obey the Bethe equations
	\begin{equation}
		k_j = \frac{2\pi}{L} \left( n_j - \frac{\delta(k_j)}{\pi} \right), \qquad j = 1, \dots, N+1, \label{bethe}
	\end{equation}
	where quantum numbers $n_j$ are integers and obey the Pauli principle. The phase shift is
	\begin{equation}
		\delta(k) = \frac{\pi}{2} - {\rm arctan}\left(\Lambda - \alpha k \right), \qquad \alpha = \frac{2\pi}{\gamma}.
	\end{equation}
	The rapidity $\Lambda$, called the spin rapidity, can be fixed by specifying values of other integrals of motions in this model. Traditionally, we require that the total momentum given by 
	\begin{equation}\label{PP}
		P(\{k_j\}, \Lambda) = \sum_{k=1}^{N+1} k_j,
	\end{equation}
	is fixed, i.e. $P(\{k_j\}, \Lambda) = Q$. The $\Lambda$ dependence in \eqref{PP} is implicit through $k_j$ as solutions to the Bethe equations. Therefore, for given $Q$ and the set of integers 
	one can resolve condition \eqref{PP} and thus solve the Bethe equations. Notice, however, that sometimes there are no solutions, therefore not all sets of integers are allowed for a fixed total momentum. For a detailed description of the spectrum of the McGuire model we refer to~\cite{McGuire1965}. Finally, the energy of a given state is
	\begin{equation}
		E(\{k_j\}, \Lambda) = \frac{1}{2} \sum_{k=1}^{N+1} k_j^2. \label{energy_finite}
	\end{equation}
	
	The impurity's momentum distribution is a Fourier transform of the static one-body function
		\begin{equation} \label{static_expectation}
	    \rho(x; \{k_j\}, \Lambda) \equiv \langle \{k_j\}, \Lambda| \psi_{\downarrow}^{\dagger}(x) \psi_{\downarrow}(0) |\{k_j\}, \Lambda \rangle.
	\end{equation}
	This function was computed and thoroughly analyzed in Refs. 
\cite{Recher2012,10.21468/SciPostPhys.8.4.053}.


The aim of our work is to compute and study the impurity one-body function at finite temperatures and in the thermodynamic limit,
	\begin{equation}
	    \rho_T(x) = \frac{1}{\mathcal{Z}}{\rm Tr}\left(e^{-\beta H} \psi_{\downarrow}^{\dagger}(x) \psi_{\downarrow}(0) \right), \label{T_corr_def}
	\end{equation}
	where $\mathcal{Z} = {\rm Tr} \exp(-\beta H)$ is the partition function. 
	
	To do so we need to characterize the eigenstates of the thermodynamically large system. In a finite system the Hilbert space is spanned by choices of quantum numbers $\{n_j\}$ and $Q$. Equivalently, by rapidities $\{k_j\}$ and $\Lambda$. In the thermodynamic limit, $N,L \rightarrow \infty$ such that $N/L = \rho_0$, we introduce a density function $\rho_{\rm p}(k)$ such that $L \rho_{\rm p}(k) {\rm d}k$ gives the number of rapidities in the range $[k, k + {\rm d}k]$. In fact much more convenient is to deal with the distribution $\sigma= \rho_{\rm p}/\rho_{\rm tot}$ normalized by the total density in the 
	rapidities space. In our case an impurity disturbs the underlying gas only in the subleading in the system size order such that $\rho_{\rm tot}=1/2\pi + \mathcal{O}(1/L)$. Instead, the quantum number $Q$ remains as a single parameter specifying, by fixing the momentum of the system, the impurity. The extensive part of the Gibbs free energy $\mathcal{F}$ depends only on distribution $\sigma(k)$ and is independent of $\Lambda$,
	\begin{equation}
	    \mathcal{F}[\sigma, \Lambda] = L \mathcal{F}_{\rm th}[\sigma] + \mathcal{F}_{0}[\sigma, \Lambda] + {\rm const} + \mathcal{O}(1/L),
	\end{equation}
	where ${\rm const}$ stands for intensive contributions to the free energy independent of $\Lambda$. The derivation of this relation is presented in Appendix~\ref{app:thermodynamics}. The expression for the free energy implies that the impurity affects the thermodynamics at the subleading, in the system size, level. On the other hand, the leading thermodynamics is this of a free fermions. Therefore, at the thermal equilibrium the density $\sigma(k)$ is just the usual Fermi-Dirac distribution. 
	
	Let us denote 
	\begin{equation} \label{static_th}
	    \rho(x; \sigma, \Lambda) = Z(\sigma, \Lambda)\lim_{\rm th} \rho(x; \{k_j\}, \Lambda),
	\end{equation}
	where $\{k_j\}$ is such that in the thermodynamic limit it corresponds to the distribution $\sigma(k)$ and ${Z(\sigma,\Lambda) = \partial Q/\partial \Lambda}$ the Jacobian of transformation between the quantum number $Q$ and rapidity $\Lambda$. It is introduced here for future convenience. Then, in view of the above discussion on the free energy, the thermal expectation value is 
	\begin{equation}
	    \rho_T(x) = \frac{\sum_{Q} e^{-\beta \mathcal{F}_{0}[\sigma, \Lambda]} Z(\sigma, \Lambda)^{-1}\rho(x; \sigma, \Lambda)}{\sum_{Q} e^{-\beta \mathcal{F}_{0}[\sigma, \Lambda]}}. \label{rhoT_with_Q}
	\end{equation}
	For the partition function we have then
    \begin{equation}
        \sum_{Q} e^{-\beta \mathcal{F}_{0}[\sigma, \Lambda]} =  \frac{L}{2\pi} \int {\rm d}\Lambda  e^{-\beta \mathcal{F}_{0}[\sigma , \Lambda]} Z(\sigma, \Lambda).
    \end{equation}
   The detailed account of how to perform summation over the eigenstates can be found in \cite{Gamayun2015,Gamayun2016}. All the integrals, unless explicitly specified otherwise, extend over the real line. Performing the same transformation in the numerator of the correlation function~\eqref{rhoT_with_Q} we find
    \begin{equation}\label{rhoTfull}
        \rho_T(x) = \frac{\int {\rm d}\Lambda  e^{-\beta \mathcal{F}_{0}[\sigma , \Lambda]} \rho(x; \sigma, \Lambda)}{\int {\rm d}\Lambda  e^{-\beta \mathcal{F}_{0}[\sigma, \Lambda]}  Z(\sigma, \Lambda)}. 
    \end{equation}
    Later in this section we show that the denominator is responsible for the correct normalization of the one-body function giving $\rho_T(0) =1$.
    
    Eq. \eqref{rhoTfull} is the main result of this section. It expresses normalized one-body function of the impurity as an averaged one-body function over the spin rapidity. The various contributions are weighted by the correlation energy $\mathcal{F}_0$.
    We discuss now the ingredients of this formula.
    
    The intensive contribution to the free energy, the correlation energy, according to the derivation presented in Appendix~\ref{app:thermodynamics} is 
    \begin{multline} \label{corr_energy_intro}
    	\mathcal{F}_0[\sigma, \Lambda] = - 4 \int \frac{{\rm d}k}{2\pi}\; k \sigma(k) \delta(k) \\ = - 4 \int \frac{{\rm d}k}{2\pi}  \, k\left[\frac{\pi}{2} -\arctan\left(\Lambda - \alpha k\right)  \right] \sigma(k),
    \end{multline}
    where $\sigma(k)$ is a Fermi-Dirac distribution, which we choose to parametrize as follows
    \begin{equation}
        \sigma(k) =  \frac{1}{1+e^{\beta(k^2-\mu)}}.
    \end{equation}
    An analogous expression for the correlation energy $\mathcal{F}_0$, for a lattice model, was also derived in Ref. \cite{PhysRevLett.74.972}.
    
    The expression for $\rho(x;\sigma, \Lambda)$ can be deduced from the finite $N$ results found in Ref. \cite{10.21468/SciPostPhys.8.4.053}. 
    There it was expressed in the forms of determinants and valid for any set of the momenta that specify the eigenstate. 
    Furthermore, the ensemble average can be replaced by the average in the typical state \cite{Korepin1993, 2014_PRA_Panfil} (see also appendix A in \cite{Gamayun2016}).
    In the thermodynamic limit this results in the Fredholm determinants with kernels multiplied by the Fermi distribution, namely we can present  
    	\begin{equation} \label{intro_final_rho}
        \rho(x; \sigma, \Lambda) = \det \left(1 + \sigma\hat{K} +  \sigma\hat{W}\right) - \det \left(1 + \sigma\hat{K}\right),
    \end{equation}
    where operators act on $L^2(\mathds{R})$ via the convolution, for example
    \begin{equation}
        \sigma\hat{K} u(q)  = \sigma(q) \int dq' K(q,q') u(q'). 
    \end{equation}
    The explicit form of the kernels can be found in \cite{10.21468/SciPostPhys.8.4.053},
    	\begin{align}
		\nonumber	\hat{K}(q,q') &= \frac{e_+(q)e_-(q') -  e_-(q)e_+(q') }{q-q'}, \\
		\hat{W}(q,q') &= \frac{1}{\pi}  e_-(q)e_-(q'),\\	\nonumber e_+(q) &= \frac{1}{\pi} e^{iqx/2 + i\delta(q)}, \quad e_-(q) = e^{-iqx/2} \sin \delta(q).
		\end{align}
	One can make kernels in ~\eqref{intro_final_rho} symmetric, that is $\sigma\hat{K} \to \sqrt{\sigma}\hat{K} \sqrt{\sigma}$ and similarly for $\hat{W}$, by the conjugation with diagonal matrices.
	The Fredholm determinants can be evaluated numerically by using methods from Ref.~\cite{Bornemann2009}. Finally, we can evaluate the Jacobian by computing $\rho(0; \sigma, \Lambda)$ by using an observation that at $x=0$ the kernels 
	become rank-one operators. 
	This way, we obtain 
	\begin{equation}
	    Z(\sigma, \Lambda) = {\rm Tr}\left[ \sigma \hat{W}\right] = \int \frac{dk}{\pi} \frac{\sigma (k)}{1+(\alpha k - \Lambda)^2} .
	\end{equation}
	Another justification of this formula comes directly from the formal form of the Jacobian $Z = \partial Q /\partial\Lambda$. Indeed, we can present it as $Z(\sigma,\Lambda) = -\partial_\Lambda \int dk \sigma(k) \delta(k)/\pi$, which is a derivative of $\Lambda$ dependent part of the total momentum \eqref{PP}. We also note that $\rho(x; \sigma, -\Lambda)$ is a complex function such that $\rho(x; \sigma, -\Lambda) = \rho^*(x; \sigma, \Lambda)$. Therefore the resulting one-body function $\rho_T(x)$ is a real function.
	
In the zero temperature limit, the contributions to the integrals in~\eqref{rhoTfull} localize at the minimum of the correlation energy, that is at $\Lambda = 0$. In the same time the Fermi-Dirac distribution becomes the ground state distribution and $\rho_{T=0}(x)$ is given by the expectation value in the ground state of the McGuire model.

	Finally, formula \eqref{rhoTfull} expresses the finite temperature correlation function through averaging over correlation functions in different impurity states labelled by $\Lambda$. Different contributions are weighted with the correlation energy $\mathcal{F}_0[\sigma, \Lambda]$. The derivation of this formula presented here does not rely on the specific correlation functions and therefore this structure is universal for finite temperature impurity correlation functions. A relevant example, and a generalization of the static case considered here, is the finite temperature dynamic one-body function $\rho_T(x, t)$ which then involves averaging over $\rho(x, t; \sigma, \Lambda)$ given by straightforward adaptations of~\eqref{static_expectation} and~\eqref{static_th}.

     \section{Effective form-factors and long distance asymptotic}
     \label{effective}
         
	\begin{figure}
		\begin{subfigure}{0.49\columnwidth}
			\includegraphics[width=1\columnwidth]{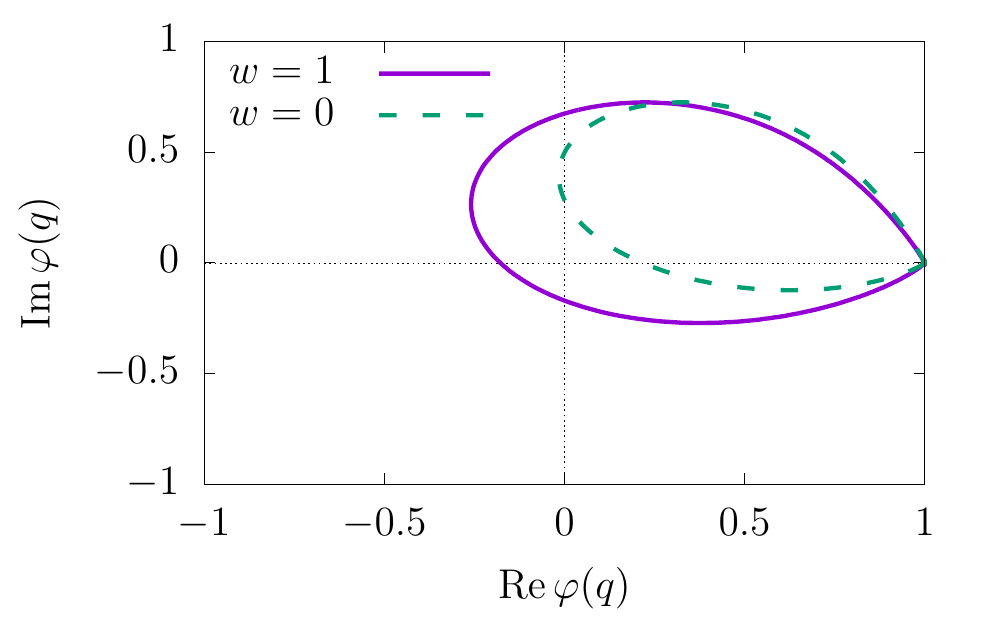}
		\end{subfigure}
		\begin{subfigure}{0.49\columnwidth}
			\includegraphics[width=1\columnwidth]{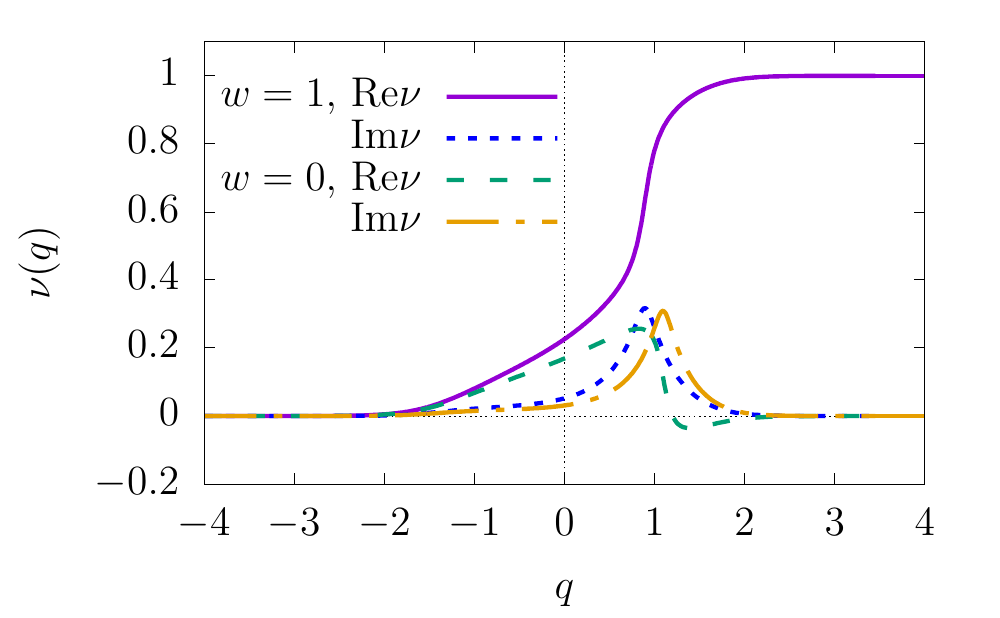}
		\end{subfigure}	
		\caption{On the left  panel we plot $\varphi(q)$ in the complex plane for $\alpha=1$, $\mu=1$, and $\beta=1$. In the $|\Lambda| > \Lambda_c$ case, $\varphi(q)$ does not encircle the origin and therefore its winding number is $w=0$. Instead, for $|\Lambda| < \Lambda_c$, $\varphi(q)$ winds once around the origin. On the right panel, we plot $\nu(q)$ as a function of $q$.}
		\label{fig:winding}
	\end{figure}

    In this section, we study the asymptotic expansion of the one-body function at large distances. The kernel $\hat{K}$ in~\eqref{intro_final_rho}
    is nothing but a generalized sine-kernel and the asymptotic of $\det(1+\sigma\hat{K})$ can be found by solving the corresponding Riemann-Hilbert problem (RHP) \cite{Kitanine_2009,Slavnov_2010}. 
    In principle, accounting for the $\hat{W}$ can be also done via RHP, however, this approach is technically involved and we prefer to employ instead recently developed heuristic methods of the effective form factors 
\cite{EffectiveXY,zhuravlev2021large,chernowitz2022dynamics}. 
In this approach an exact form factor (spectral) series that describes the correlation function at finite temperature and that in the thermodynamic limit leads to 
\eqref{intro_final_rho} is replaced with the effective one. The effective series formally corresponds to the zero temperature but the form factors depend on an effective phase shift in which information about the distribution $\sigma$ is contained. To find this phase shift we perform summation of the effective form factor series and again express the result as Fredholm determinants of the similar kind as the exact one. Asymptotically identifying the kernels we compute the desired effective phase shift. 
The advantage of the effective form factor series lies in the fact that the corresponding Fredholm determinants turn out to be an elementary functions, or integrals of elementary functions, thus, allowing us to find the desired asymptotic expression for the exact correlation function. 
    
More explicitly, the effective form factors $|\langle {\bf k}|{\bf q}\rangle|$ are
	\begin{equation}\label{effective_ff}
		|\langle {\bf k}|{\bf q}\rangle|^2 = \prod\limits_{i=1}^{N+1}
		\frac{2 e^{g(k_i)} \sin^2 \pi \nu(k_i)}{L+2\pi \nu'(k_i)}
		\prod\limits_{i=1}^N \frac{2 e^{-g(q_i)}}{L} 
		(\det D)^2,
	\end{equation}
	with
	\begin{equation} \label{D_def}
	\det D \equiv
		\begin{vmatrix}
			\frac{1}{k_1-q_1} &\dots & \frac{1}{k_{N+1}-q_1}\\
			\vdots & \ddots & \vdots \\
			\frac{1}{k_1-q_N} &\dots & \frac{1}{k_{N+1}-q_N}\\
			1& \dots  & 1\\
		\end{vmatrix}.
	\end{equation}
	Here $g(k)$ and $\nu(k)$ are arbitrary smooth functions that we are going to specify below. The momenta ${\bf k} = \{k_i\}$ and ${\bf q} = \{q_i\}$ are understood as solutions of 
	\begin{equation}
		e^{ikL} = e^{-2\pi i \nu(k)}, \qquad e^{iqL} = 1,
	\end{equation}
	respectively. They can be parametrized in terms of the quantum numbers as
	\begin{align}
	    k_j &= \frac{2\pi}{L} \left(n^{(k)}_j - \nu(k_j)\right), \quad j = 1, \dots , N+1,\\
	    q_j &= \frac{2\pi}{L} n^{(q)}_j, \quad j=1, \dots, N.
	\end{align}
	In principle we could formulate the problem for any set of $\{k_i\}$. For the impurity problem the relevant scenario is when $\{k_i\}$ takes the Fermi sea configuration at zero temperature. That is, the corresponding quantum numbers are
	\begin{equation}
	    n^{(k)}_j = -\frac{N}{2} +j, \quad j=1, \dots, N+1.
	\end{equation}

	The tau function of interest is defined as 
	\begin{equation}
		\tau_N(x) = \sum_{\bf q} |\langle {\bf k}|{\bf q}\rangle|^2  e^{-ix (P({\bf k}) - P({\bf q}))}, \label{def_tau_N}
	\end{equation}
	where the summation extends over possible values of $\{q_j\}$ or possible quantum numbers $\{n_j^{(q)}\}$ keeping in mind the Pauli principle. The total momentum $P({\bf k}) = \sum k_j$. The summation in~\eqref{def_tau_N} can be performed exactly, using a slight variation of the Cauchy–Binet formula~\cite{Shafarevich2013} and the result is
	\begin{equation}
		\tau_N(x) =  \det_N(\delta_{ij}+ A_{ij} + B_{ij}) - \det_N(\delta_{ij}+A_{ij}),
	\end{equation}
	with $A_{ij} = A(k_i, k_j)$ and $B_{ij} = B(k_i, k_j)$ where
	\begin{equation}
	\begin{aligned}
		A(q, q') =& - \frac{e(q)-e(q')}{q-q'} B(q,q'), \\
		B(q,q') =& \frac{2}{L} \exp\left({\frac{g(q)+g(q')-ix(q+q')}{2}}\right) \\
		\times &\sin \pi \nu(q)\sin \pi \nu(q') ,
	\end{aligned}
	\end{equation}
	and 
	\begin{equation}
		e(k) = \frac{2ie^{ikx-g(k)}}{e^{-2\pi i \nu(k)}-1} + \int \frac{dq}{\pi} \frac{e^{iqx-g(q)}}{k-q-i0}.
	\end{equation}
	In the thermodynamic limit, $N, L \rightarrow \infty$ with $N/L$ fixed the determinants turn into Fredholm determinants acting on $L^2([-q_F, q_F])$,
	\begin{equation}\label{tauEFF}
	    \tau(x) = \lim_{\rm th} \tau_N(x) = \det(1 + A + B) - \det(1 + A).
	\end{equation}
    where $q_F = \pi N/L$ is the Fermi momentum of the auxiliary problem. 
    
    We consider now the asymptotic expansion, $ x \rightarrow \infty$. The integral in $e(k)$ is then exponentially suppressed and
	\begin{equation}
		e(k) \approx \frac{2ie^{ikx-g(k)}}{e^{-2\pi i \nu(k)}-1},
	\end{equation}
	with corrections exponentially small in $x$. Within this approximation the kernels can be presented in the following form
	\begin{equation}
	\begin{aligned}
	    A(q,q') =& \frac{E_+(q)E_-(q') -  E_-(q)E_+(q') }{q-q'}, \\ B(q, q') =& \frac{1}{\pi}  E_-(q)E_-(q'),
	\end{aligned}
	\end{equation}
	where
		\begin{equation}
	\begin{aligned}
			E_+(q) =& \frac{1}{\pi} e^{iqx/2 + i\pi \nu(q)-g(q)/2}, \\  E_-(q) = & e^{-iqx/2+g(q)/2} \sin \pi \nu(q).
\end{aligned}
	\end{equation}
	We now compare this asymptotic structure for $\tau(x)$ with formula~\eqref{intro_final_rho} for $\rho(x; \sigma, \Lambda)$. Recall that in $\tau(x)$ functions $g(k)$ and $\nu(k)$ are arbitrary smooth functions. The two expressions match if $g(q)$ and $\nu(q)$ obey the following relations
	\begin{equation}
	\begin{aligned}
		\sqrt{\sigma(q)} &\sin \delta(q) = e^{g(q)/2} \sin \pi \nu(q), \\ \sqrt{\sigma(q)} &e^{ i\delta(q)} =  e^{i\pi \nu(q)-g(q)/2}, \label{matching_kernels}
	\end{aligned}
	\end{equation}
	which gives 
	\begin{equation} \label{nug}
	\begin{aligned}
		\nu(q) =&\frac{1}{2\pi i} \ln \varphi(q), \\
		g(q) = &\ln \varphi(q) - \ln \sigma(q) - 2 i \delta(q) ,
		\end{aligned}
	\end{equation}
	where 
	\begin{equation} \label{varphi}
		\varphi(q) \equiv 1 + (e^{2i \delta(q)}-1)\sigma(q) = 1 + \frac{2 i \sigma(q)}{\Lambda - \alpha q - i} .
	\end{equation}
    With such chosen $\nu(k)$ and $g(k)$ we have
    \begin{equation}
        \rho(x; \sigma, \Lambda) \sim \lim_{q_F \rightarrow \infty}\tau(x), \quad {\rm as\;\;} x\rightarrow \infty.
    \end{equation}
    On the other hand  $\tau_N(x)$ function has a spectral representation. We shall see that for $q_F\to \infty$, only small subset 
    of the spectral (form factor) sums matters, thus allowing us to find $\tau(x)$ exactly, and in this way to understand the asymptotic expansion of $\rho(x; \sigma, \Lambda)$.
    
    For the further understanding of the spectral series we discuss function $\nu(k)$.
    This function enters the expression for the form factor and, more importantly, the rapidities $k_j$. 
    As we shall see its analytic properties of $\nu(q)$ are determining for the form factor summation. This function, besides explicit dependence on $q$ depends on all the parameters in the problem. Through $\delta(q) \equiv \delta(q; \Lambda, \alpha)$ it depends on the coupling parameter $\alpha$ and on the impurity rapidity $\Lambda$. It also depends on the thermodynamic properties of the system through the filling function $\sigma(q)$. At the thermal equilibrium these are the temperature $T$ and the chemical potential $\mu$. The analytic properties of $\nu(q)$ are determined by analytic properties of the complex logarithm, which is defined on a Riemann surface spiraling around the origin of the complex plane. On this surface $\nu(q)$ is continuous but depends on the number of windings around the origin. More precisely, the plot of the $\varphi(q)$ (the argument of the logarithm see \eqref{nug}, \eqref{varphi}) for all available $q$ forms a loop with origin at $z=1$ in the complex plane, see Fig.~\eqref{fig:winding}.
    Whether the loop winds around the origin depends on the parameters of the system which we now inspect. The loop extends over the origin when, for some value $q_*$, $\varphi(q_*)$ has negative real part and zero imaginary part. This gives two conditions
    
    \begin{figure}
	       \includegraphics[scale=0.27]{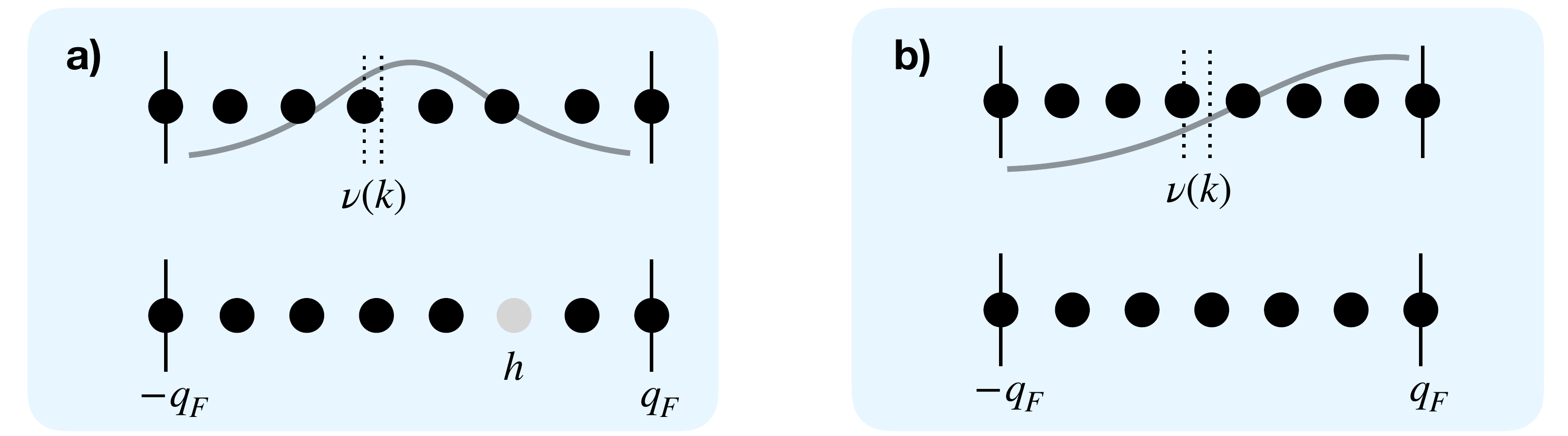}
	       \caption{The structure of the ground state and relevant excited states in the non-winding case a) and winding case b).}
	       \label{fig:excitations}
	\end{figure}
	   
    \begin{equation}
    	\tanh (\beta \epsilon(q_*)/2)<0,  \quad \alpha q_* = \Lambda.
    \end{equation} 
    At the thermal equilibrium $\epsilon(q) = q^2 - \mu$ which leads to two cases, formulated in terms of $\Lambda$,
    \begin{align}
        \quad &|\Lambda| \leq \Lambda_c: \quad {\rm winding}\, w = 1 \\
        \quad &|\Lambda| > \Lambda_c: \quad {\rm winding }\, w = 0,
    \end{align}
    where 
    \begin{equation}
        \Lambda_c \equiv \alpha \sqrt{\mu}.
    \end{equation}
    The winding number $w$ is defined as (see Fig.~\eqref{fig:winding})
    \begin{equation}
        w = \int dq \partial_q \nu(q) = \nu(\infty) - \nu(-\infty)
    \end{equation}
     Note that the imaginary part of $\nu(q)$ has always asymptote $0$ as $q\to \pm\infty$.
    
    \begin{figure*}
		\begin{subfigure}{1\columnwidth}
			\includegraphics[width=1\columnwidth]{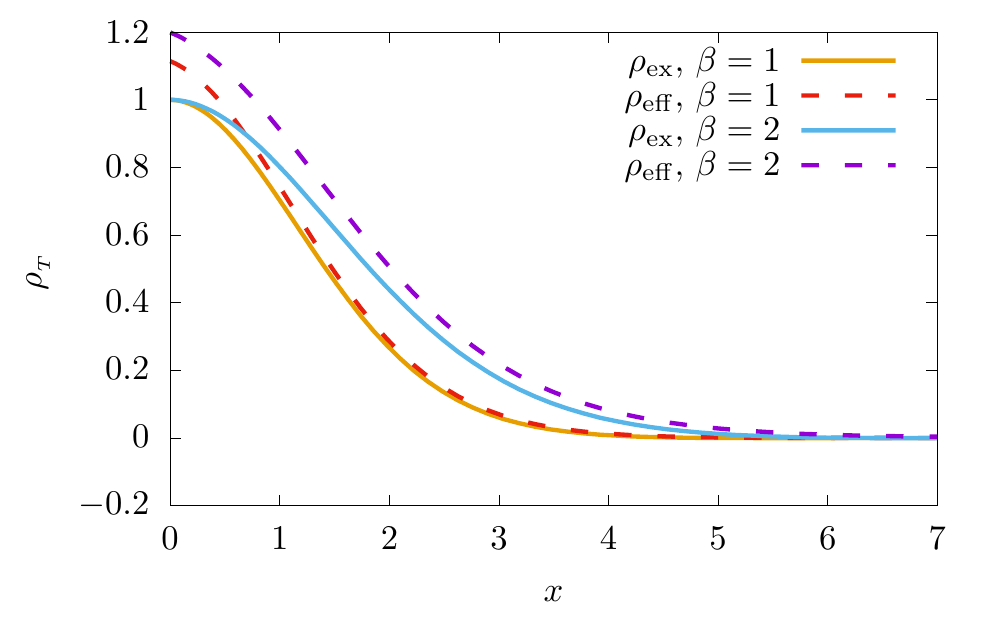}
		\end{subfigure}
		\begin{subfigure}{1\columnwidth}
			\includegraphics[width=1\columnwidth]{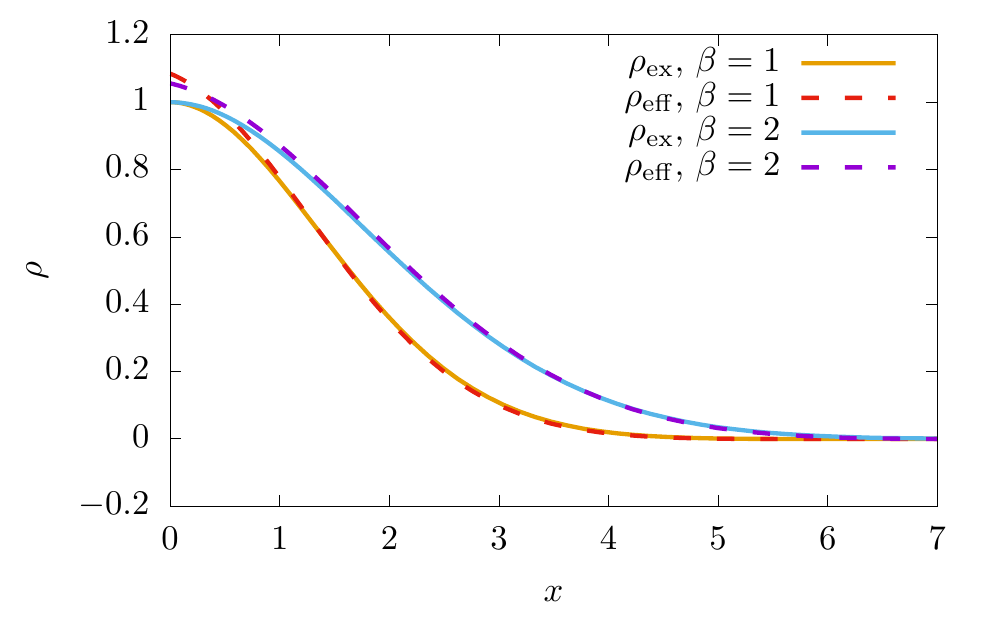}
		\end{subfigure}	
		\caption{Comparison between $\rho_T(x)$ calculated from the exact Fredholm determinants (solid lines) and the effective form factors (dashed lines) for $\alpha=0$ (left panel) and $\alpha=1$ (right panel).}
		\label{fig:rho_x}
	\end{figure*}
	
	        	\begin{figure*}
		\begin{subfigure}{\columnwidth}
			\includegraphics[width=1\columnwidth]{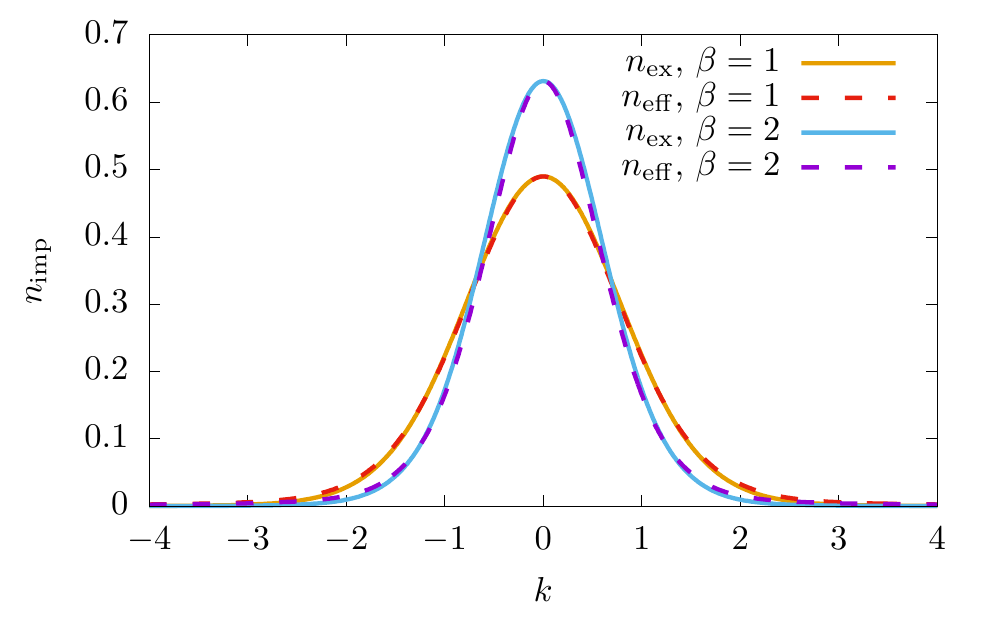}
		\end{subfigure}
		\begin{subfigure}{\columnwidth}
			\includegraphics[width=1\columnwidth]{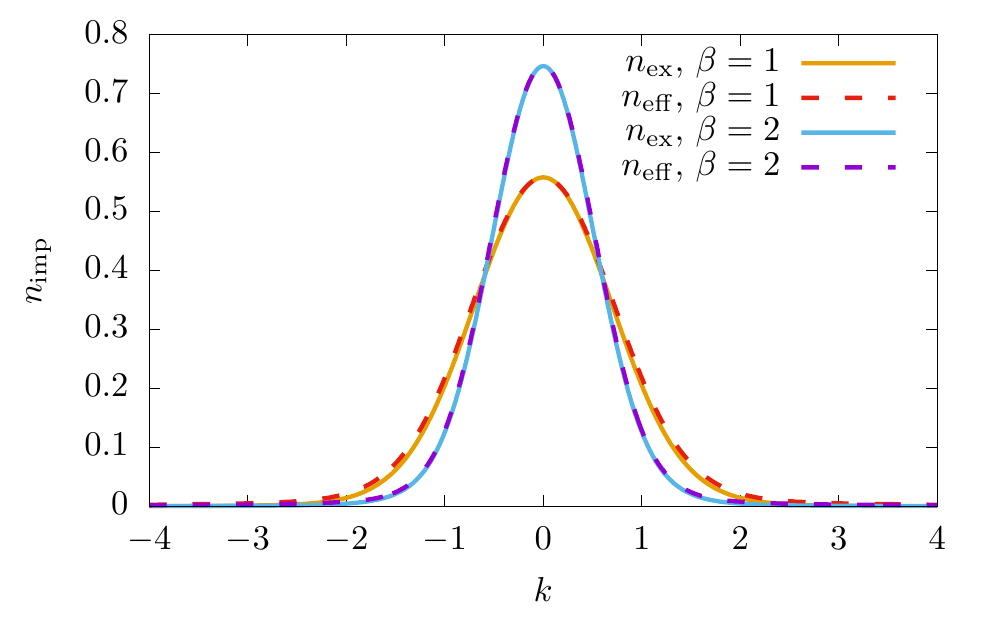}
		\end{subfigure}	
		\caption{Comparison between the exact impurity's momentum distribution $n_{\rm imp}(k)$ (solid lines) and the normalized distribution obtained from the effective form factors (dashed lines)  for $\alpha=0$ (left panel) and $\alpha=1$ (right panel).}
		\label{fig:momentum}
	\end{figure*}
    

    Function $\nu(q)$ enters the expression for the form factor through $\exp(i \pi \nu(q))$ and therefore the non-zero winding has little effect on it. On the other hand $\nu(q)$ enters directly the expression for rapidities $k_j$ and affects the structure of the relevant excitations in the spectral sum of $\tau(x)$. 
    Namely, intuitively it is clear that the largest contributions come from those $q_j$ whose quantum numbers are identical to $k_j$. In the non-winding case we cannot achieve this for all $q_j$, simply because the number of $k_j$ in the set $|\textbf{k}\rangle$ is larger by one that possible $q_j$ in $|\textbf{q}\rangle$. So there will be at least one-hole as we demonstrate in Fig.~(\ref{fig:excitations}a). In principle in the spectral sum there also different excitations, involving neccesarily particle-hole pairs. We will argue later that such excitations do not contribute in the limit $q_F \rightarrow \infty$. Therefore in the non-winding region $|\Lambda| > \Lambda_c$ we obtain the exact expression for $\tau(x)$ and asymptotic for $\rho(x; \sigma, \Lambda)$
	\begin{equation}\label{asymptNON}
		\rho(x; \sigma,  \Lambda) \approx 2J(x) A(\Lambda)\exp\left(-i x \int k \nu'(k) {\rm d}k \right),
	\end{equation}
	where
	\begin{multline}\label{A}
		\ln A(\Lambda) =\int g(k)\nu'(k)dk -\ln 2 \\+\int dq \int dp\,\nu'(p)\nu'(q)\ln|q-p| .
	\end{multline}
	and
	\begin{equation} \label{J_def}
		J(x) =  \int \frac{dq}{\pi} \frac{\sigma(q)e^{-iqx}}{(\alpha q - \Lambda)^2+1} 
		\exp\left(\fint dp \frac{ 2\nu(p)}{q-p}\right).
	\end{equation} 
	   The appearance of the integral $J(x)$ reflects the fact that we have to sum over the hole positions. Detailed derivation of this identity is given in Appendix~\ref{app:factorD1}. 
	   One can further analyze $J(x)$ asymptotically by the saddle point method, which is done in Appendix~\ref{appC}. For practical purposes, however, we leave it in the integral form.

	      If winding $w=1$ then the state $|{\bf k}\rangle$ is effectively compressed, see again Fig.~(\ref{fig:excitations}b),
    and contrary to the non-winding case, there is no extra space for a hole in the state $|{\bf q}\rangle$. The possible excitations are then only particle-hole excitations which we again can neglect in view of the $q_F \rightarrow \infty$ limit. So in fact the sum \eqref{tauEFF} reduces to one term, which we evaluate in Appendix~\ref{app:factorD2}. The asymptotic for $|\Lambda|<\Lambda_c$ 
    is 
    	\begin{equation}\label{asymptW}
		\rho(x; \sigma , \Lambda) = A(\Lambda)\exp\left(-i x \int k \nu'(k) {\rm d}k \right),
	\end{equation}
Here $A(\Lambda)$ is given by the Eq. \eqref{A}. 

We discuss now the irrelevance of other excited states in the spectral sum of $\tau(x)$ for $q_F \rightarrow \infty$. The other excited states necessarily involve particle-hole excitations. Given the Fermi sea structure of $|{\bf q}\rangle$ the rapiditiy of the particle excitation is $|k| > q_F$. Therefore in the limit $q_F \rightarrow \infty$ there is no space for such excitations and they do not contribute to the spectral sum. The formal proof of this argument, in the context of the XY spin chain can be found in \cite{EffectiveXY}. Alternatively, one can numerically check that the result of our summation (Eqs. \eqref{asymptNON}, \eqref{asymptW}) is identical to the Fredholm determinants \eqref{tauEFF} for any $x$. 
    
It might look as if the asymptotics \eqref{asymptNON} and \eqref{asymptW} are discontinuous as function of $\Lambda$.
This is apparent and connected with the fact that the function $\nu(q)$ behaves very differently for different winding numbers. In Appendix~\ref{appC}
we prove that while changing $\Lambda$ over $\Lambda_c$ one expression smoothly transforms into the other one. 
This is somewhat similar to the gap dependence of the finite temperature correlation functions during the crossover over the critical point. 
For the one-dimensional Ising model this can be seen in Ref. \cite{Sachdev1996}.

    In Fig.~\eqref{fig:rho_x} we compare the one-body function $\rho_T(x)$ computed according to~\eqref{rhoTfull} with exact Fredholm determinant expression for $\rho(x; \sigma, \Lambda)$ of Eq.~\eqref{intro_final_rho}, referred to as $\rho_{ex}$, with the results effective form-factors given by Eqs.~\eqref{asymptNON} and~\eqref{asymptW}, referred to as $\rho_{eff}$.
    We see that the deviation happens only at small $x$ and is less for smaller $\beta$ (higher temperatures). The reason for this is that the effective form factors gives only the leading contribution with the smallest decay rate. At small temperatures, both leading and subleading decay rates become small so the latter cannot be ignored anymore. At exactly zero temperature all decay rates vanish and the correlation function has a
    power law behavior with the separation distance \cite{10.21468/SciPostPhys.8.4.053}.

	\section{Momentum distribution and Tan's contact}
	\label{sec:nk}
	 The momentum distribution function of the impurity $n_{\rm imp}(k)$ can be evaluated from the one-body function through its Fourier transform, which for the real and symmetric $\rho_T(x)$ takes the following form
    \begin{equation}\label{eq:f}
    	n_{\rm imp}(k) = \int_0^{\infty} \frac{{\rm d}x}{\pi} \cos(kx)\rho_T(x).
    \end{equation}
    In Fig.~\ref{fig:momentum} we show the results for different temperatures and coupling constants. We compare the exact results with the results obtained from the effective form factors after 
    normalizing it such that for $k=0$ values of the impurity's distributions coincide.
    We see that at small momenta this two distributions are almost identical. This reflects the fact that the effective description introduced in the previous section decently approximates the exact $\rho_T(x)$ away from the origin.

\begin{figure}[h!]
	     \includegraphics[width=\columnwidth]{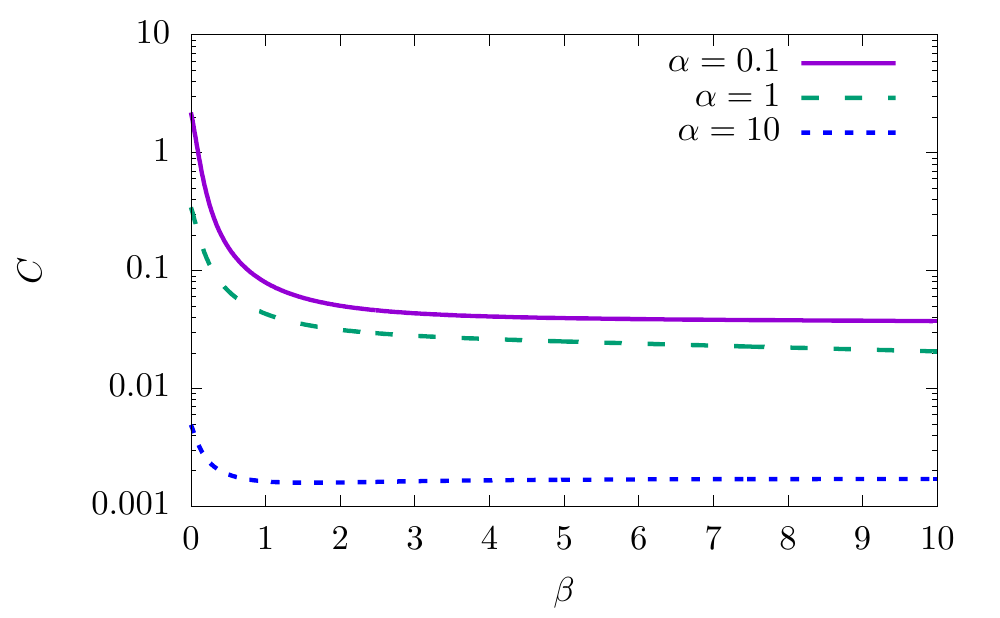}
	     \caption{Tan's contact from Eq. \eqref{ctan} as a function of  inverse temperature for different interaction strengths.}
	     \label{figTan}
	 \end{figure}

	  The short distance expansion (for  ${\mu x\ll 1}$) is however not captured by the effective form factors. For the momentum distribution 
	  this behavior is responsible for the large $k$ tails. Indeed, integrating by parts we see that the leading asymptotic expansion is governed by the 
	  odd derivatives of $\rho_T(x)$ and $x=0$. One can explicitly demonstrate that the first derivative at zero vanishes, so the leading asymptotics is giving by the third derivative, which gives rise to the famous 
	  $k^{-4}$ decay
	  \begin{equation}
	     n_{\rm imp} (k) \approx\frac{\partial_x^3\rho_T}{\pi k^4}\Big|_{x=0} \equiv \frac{C(g,\beta)}{k^4}.
	 \end{equation}
	 The constant $C(g,\beta)$ is called the Tan's contact and is  related to the thermodynamic properties of the system \cite{Tan_2008,Barth2011}. 
	 Such a constant was obtained in Ref. \cite{10.21468/SciPostPhys.8.4.053} by means of Taylor series of the kernels around $x=0$, rendering them into finite-rank expressions. 
	 Expanding this argument by linearity we obtain that, at finite temperatures, the Tan's contact reads 
	 \begin{equation}\label{ctan}
	     C(g,\beta)  = \frac{\int d\Lambda e^{-\beta \mathcal{F}_0(\Lambda)}\left[S_0(\Lambda)S_2(\Lambda)-S_1(\Lambda)^2\right]}{\pi \int d\Lambda e^{-\beta \mathcal{F}_0(\Lambda)} S_0(\Lambda)},
	 \end{equation}
	 where 
	 \begin{equation}
	     S_n(\Lambda) = \int\frac{dk}{\pi} \frac{k^n \sigma(k)}{(\alpha k-\Lambda)^2+1}.
	 \end{equation}
	 One can check that the contact \eqref{ctan}, following the general principles \cite{Tan_2008,Barth2011}, can be written as a derivative of a thermodynamic quantity over the coupling constant. Indeed, 
	 taking into account 
	 \begin{equation}
	     \partial_\Lambda \mathcal{F}_0 =2 S_1,\,\, \partial_\alpha \mathcal{F}_0  = -2S_2,\,\, \partial_\alpha S_0 + \partial_\Lambda S_1 = 0,
	 \end{equation}
	 and after an integration by parts, one can arrive at 
	 \begin{equation}
	     C(g,\beta)  = \frac{1}{2\pi \beta} \partial_\alpha \ln \int d\Lambda e^{-\beta \mathcal{F}_0(\Lambda)} S_0(\Lambda).
	 \end{equation}
	 We plot Eq. \eqref{ctan} in Fig. \eqref{figTan} for three different interaction regimes. We observe the increase of the contact with both the temperature and the interaction strength at least  to very low temperatures.
	 Recently, the finite temperature polaron's behavior in three-dimensional gases were addressed with the help of varional methods \cite{PhysRevLett.125.065301,PhysRevA.102.023304,hu2022fermi}. In particular, 
	 the temperature dependence of the Tan's contact was studied. The qualitative behavior found in these papers is different from ours, and at least partially explainable by different dimensionalities of the problems. 
At the very specific point of the infinite coupling constant, i.e. $\alpha=0$, the Tan's contact is proportional to the average kinetic energy of in the Fermi gas 
\begin{equation}
    C(g=\infty,\beta) = \int \frac{dk}{2\pi^2}k^2 \sigma(k) = -\frac{\rm{Li}_{\frac{3}{2}}\left(-e^{\beta  \mu }\right)}{4 (\pi\beta)^{3/2}},
\end{equation}
  which makes its growth with the temperature manifest. Similar behavior for the Tan's contact in a one-dimensional gas was also observed in~\cite{PhysRevLett.111.025302}.

%

    \section{Conclusions}\label{sec:f}
    
    In this work we tackled the problem of calculating the one-body correlation function of an interacting mobile impurity submerged in a free fermionic gas at arbitrary finite temperatures. We formulated the problem exactly in terms of Fredholm determinants and inspected its large distance asymptotics through the effective form factors approach. We found that the effective form-factors provide an efficient way in extracting the asymptotics of the Fredholm determinant circumventing the need of studying the technically very involved matrix Riemann-Hilbert problem. 
    
    We observed that depending on the value of the spin rapidity $\Lambda$ the asymptotic behavior looks structurally very different. The reason for this is topological as for $|\Lambda|<\Lambda_c$, the effective phase shift has the winding number $1$, while for $|\Lambda|>\Lambda_c$ the winding is absent. Similar situation happens of the Ising model \cite{Sachdev1996} or the XY spin chain \cite{EffectiveXY} where the role of $\Lambda$ is played by the magnetic field. Our asymptotics are still continuous when crossing over $\Lambda_c$ as there are no quantum phase transitions in one-dimensional systems at zero temperature \cite{sachdev2011quantum}.  
    
    
    With the one-body correlation function computed we investigated the momentum distribution of the impurity. This is the quantity directly accessible in cold-atoms experiments \cite{PhysRevLett.105.070402,PhysRevLett.122.203402,PhysRevLett.110.055305, PhysRevLett.104.235301,PhysRevLett.109.220402}. We observed the characteristic narrowing of the curve with the decrease of temperature. 
    Finally, by performing the short distance expansion of the Fredholm determinants, we evaluated the Tan's contact of the impurity and showed its growth with both the temperature and the interaction strength.

    The next step would be to generalize this approach to the time-dependent case and access the spectral functions. 
    It would be interesting to analyze not only the Green's function of the mobile impurity \cite{Gamayun2015,Gamayun2016} but also  
    the case of a static impurity in a three-dimensional gas. Indeed, it is known that various spectral observables can be expressed via the Fredholm determinants with kernels very similar to ours ~\cite{Schmidt2018}, that allows to their analysis with the effective form factor approach. The time-dependent case is important from the experimental point of view \cite{Cetina2016} but also 
    for the capturing universal contributions of the highly excited states in the generic correlation functions in quantum one-dimensional systems, within the non-linear Luttinger liquid paradigm 
\cite{Imambekov2009,RevModPhys.84.1253,10.21468/SciPostPhys.7.4.047}. For fixed $\Lambda$ we expect an additional power law prefactor for large times \cite{chernowitz2022dynamics,zhuravlev2021large}.
In the infinite coupling constant case we expect to recover the predicted logarithmic diffusion \cite{PhysRevLett.99.240404,PhysRevLett.103.110401}. 

    In this work we focused on thermal equilibrium as it is the most experimentally relevant case. However, the presented techniques do not rely on the gas distribution to be thermal and the presented results can be generalized to stationary non-equilibrium ensembles, such as the ones arising in the quench action~\cite{QA_Caux} or generalized hydrodynamics~\cite{Bastianello_2022}. 

    \vspace{0.2cm}
    \noindent
    {\bf Acknowledgments:} 
    We are grateful to Oleg Lychkovskiy for useful discussions.
    OG acknowledges the support from the Polish National
Agency for Academic Exchange (NAWA) through the Grant No. PPN/ULM/2020/1/00247 and the support from the National Research Foundation of Ukraine grant 2020.02/0296. MP and FS acknowledge the support from the National Science Centre, Poland, under the SONATA grant 2018/31/D/ST3/03588.

    \appendix

    \section{Thermodynamics of the impurity} \label{app:thermodynamics}
    
    In this Appendix we evaluate the impurity dependent contribution to the free energy of the model. The main result is that the impurity affects only the intensive part of the free energy and the contribution comes purely from the energy. There is no impurity dependent contribution to the entropy. To derive this result we start by recalling the thermodynamic Bethe ansatz approach to the thermodynamics of integrable models. 
      
	\subsection{Thermodynamic functions}
	
	In the thermodynamic limit, $N, L \rightarrow \infty$ with $\rho_0 = N/L$ fixed. Following standard procedure~\cite{Korepin1993} we introduce function $k(x)$ defined by the following relation
	\begin{equation}
		L k(x) + 2 \delta(k(x)) = 2\pi L x.
	\end{equation}
	By the definition $k(n_j/L) = k_j$, see Bethe equations~\eqref{bethe}. It is customary to introduce two density functions. The total density $\rho_{\rm tot} (k)$ and the particle density $\rho_{\rm p}(k)$. The former is defined through
	\begin{equation}
		\rho_{\rm tot} (k) = \frac{dx(k)}{dk},
	\end{equation}
	and, including the subleading corrections in the system size, reads
	\begin{equation}
		\rho_{\rm tot, L} (k) = \frac{1}{2\pi}\rho_{\rm tot} (k) \left(1 + \frac{2}{ L} \partial_k \delta(k) \right). \label{rho_tot}
	\end{equation}
	The density of particles is defined as
	\begin{equation}
		\rho_{\rm p}(k_j) = \lim_{\rm th} \frac{1}{L(k_{j+1} - k_j)},
	\end{equation}
	such that sum of over $k_j$ becomes an integral over the density
	\begin{equation}
		\frac{1}{L} \sum_{j=1}^{N+1} f(k_j) = \int {\rm d}k \rho_{\rm p}(k) f(k).
	\end{equation}
	We derive now an expression for $\rho_{\rm p}(k)$, including $1/L$ correction depending on $\Lambda$. To this end we consider again the sum and include corrections to $k_j$ coming from the impurity. We have
	\begin{align}
		\frac{1}{L} \sum_{j=1}^{N+1} f(k_j) = \frac{1}{L} \sum_{j=1}^{N+1} f\left(\frac{2\pi}{L} n_j - \frac{2}{L}\delta_j \right) \nonumber \\
	    = \frac{1}{L} \sum_{j=1}^{N+1} \left( f\left(\frac{2\pi}{L} n_j \right) -  \frac{2\delta_j}{L}   f'\left(\frac{2\pi}{L} n_j\right) \right).
	\end{align}
	Using that $\delta_j = \delta(k_j) \approx \delta (2\pi n_j /L)$, we have
	\begin{equation}
		\frac{1}{L} \sum_{j=1}^{N+1} f(k_j) =  \int {\rm d}k\, \rho_{\rm p}(k) \left( f(k) - \frac{2}{L} \delta(k) f'(k) \right)
	\end{equation}
	We can now incorporate the $1/L$ term as a correction to the particle density by integrating by parts. Neglecting here the boundary terms and obtain
	\begin{equation}
		\frac{1}{L} \sum_{j=1}^{N+1} f(k_j) = \int {\rm d}k\, \rho_{\rm p, L}(k) f(k), \label{density_finite}
	\end{equation}
	with
	\begin{equation}
		\rho_{\rm p, L}(k) = \rho_{\rm p}(k) + \frac{2}{L} \partial_k \left[ \delta (k) \rho_{\rm p}(k)\right].
	\end{equation}
	We will also need the filling function, which including the $1/L$ correction, is 
	\begin{equation}
		\sigma_L(k) = \frac{\rho_{\rm p, L}(k)}{\rho_{\rm tot, L}(k)} = \sigma(k) + \frac{2}{L} \delta (k)\left(\partial_k \sigma(k)\right) .
	\end{equation}
	
	Concluding, the states of the system in the thermodynamic limit are characterised by density functions which do not depend on the impurity. The dependence comes only in the subleading in the system size terms.

	\subsection{Thermal equilibrium}
	
	We consider now thermal equilibrium by minimising the free energy $F = E - TS$. Both energy and entropy have extensive parts, independent of the impurity, and intensive part which depends on it. There are also intensive parts that do not depend on the impurity. As we are interested in the physics of impurity, those can be neglected. As we shall see they don't influence the saddle point distribution and lead only to a multiplicative constant for a partition function. 
	
	The energy~\eqref{energy_finite}, in the large system, becomes
	\begin{equation}
		E = L E_{\rm th}[\sigma] + E_{0}[\sigma, \Lambda] + \mathcal{O}(L^0)
	\end{equation}
	where
	\begin{align} 
		E_{\rm th}[\sigma] &= \int\frac{{\rm d}k}{2\pi}\, \sigma(k) \left(\frac{k^2}{2} - h\right), \\
		E_{0}[\sigma, \Lambda] &= -2 \int\frac{{\rm d}k}{2\pi}\, k \sigma(k) \delta(k).
	\end{align}
	We used here eq.~\eqref{energy_finite} for the energy in a finite system and eq.~\eqref{density_finite} for $\rho_{\rm p, L}(k)$. We have also integrated by parts in $E_{0}[\rho_{\rm p}, \Lambda]$.
	The entropy has the same structure
	\begin{equation}
		S = L S_{\rm th}[\sigma] + S_{0}[\sigma, \Lambda] + \mathcal{O}(L^0),  \label{entropy_th}
	\end{equation}
	where
	\begin{align}
		S_{\rm th}[\rho_{\rm p}, \rho_{\rm tot}] &= -\int \frac{{\rm d}k}{2\pi}\, G(\sigma(k)), \\
		S_{0}[\sigma, \Lambda] &= 0,
	\end{align}
	with
	\begin{equation}
		G(\sigma) = \sigma \ln \sigma + (1-\sigma) \ln (1 - \sigma).
	\end{equation}
	In the following we show that $S_{0}[\sigma, \Lambda] $, the subleading contribution depending on $\Lambda$, is zero and in the process we derive also the leading expression $S_{\rm th}[\rho_{\rm p}, \rho_{\rm tot}]$.
	
	\emph{Derivation of eq.~\eqref{entropy_th}:}
	The entropy density $dS(k)$ in the interval $[k, k +{\rm d}k]$ is~\cite{Korepin1993}
	\begin{align}
		dS(k) = \ln \left(\frac{\left[L \rho_{\rm tot, L} (k) dk\right]!}{\left[L \rho_{\rm p, L} (k) dk\right]! \left[L \rho_{\rm h, L} (k) dk\right]!}\right),
	\end{align}
	with the total entropy $S = \int {\rm d}S(k)$ and with the holes density $\rho_{\rm h, L}(k)$ defined as
	\begin{equation}
		\rho_{\rm h, L}(k) = \rho_{\rm tot, L}(k) - \rho_{\rm p,L}(k) 
	\end{equation}
	Using Stirling approximation for the factorial, $\ln n! \approx n \ln n - n + \dots$, we find
	\begin{align}
		{\rm d}S(k) = - L \rho_{\rm tot, L}(k) G(\sigma_L(k)) {\rm d}k,
	\end{align}
	from which $s_{\rm th}[\rho_{\rm p}, \rho_{\rm tot}]$ follows as the leading term in $L$. 
	We expand now $G(\sigma_L(k))$ in $L$, the first two orders are
	\begin{equation}
	    G(\sigma_L(k)) = G(\sigma(k)) + \frac{2}{L} \delta(k)\left(\partial_k G(\sigma(k))\right).
	\end{equation}
	Using now the expression for $\rho_{\rm tot, L}(k)$, the subleading in system size contribution to ${\rm d}S(k)$ is a total derivative with respect to $k$ and therefore does not contribute to the total density. Therefore $S_0[\Lambda] = 0$.
	
	We consider now the free energy
	\begin{align}
		\mathcal{F}= L \mathcal{F}_{\rm th} + \mathcal{F}_{0} + \mathcal{O}(L^0),
	\end{align}
	where $\mathcal{F}_0$ captures all system size independent contribution to the free energy that depends on $\Lambda$. According to the computation presented above
	\begin{equation}
	    \mathcal{F}_{\rm th} =  E_{\rm th} - T S_{\rm th}, \qquad \mathcal{F}_0 = E_0(\Lambda).
	\end{equation}
	Therefore, in the thermodynamic limit, the saddle point configuration comes from minimising $(E_{\rm th} - T S_{\rm th})$ which leads to the Fermi-Dirac distribution
	\begin{equation}
		\sigma(k) = \frac{1}{1 + e^{\epsilon(k)}}, \qquad \epsilon(k) = \frac{\frac{k^2}{2} - h}{T}.
	\end{equation}
	The subleading contribution to the free energy, the correlation energy, is then
	\begin{align} \label{corr_energy_app}
		\mathcal{F}_0 = - 2 \int\frac{{\rm d}k}{2\pi}\, k \sigma(k) \delta(k).
	\end{align}
	
	For further convenience we will redefine the temperature and the chemical potential such that $\epsilon(k)$ takes the ``standard'' form
	\begin{equation}
		\epsilon(k) = \beta(k^2 - \mu).
	\end{equation}
	This rescaling of temperature affects the contribution from the correlation energy~\eqref{corr_energy_app} which leads to formula~\eqref{corr_energy_intro} of the main text.
	
	

    \section{Thermodynamic limit of the form factors and spectral series}	\label{app:factorD}

\subsection{Non-winding case \texorpdfstring{$w=0$}{w=0}} \label{app:factorD1}

\begin{figure}[H]
 	    \centering
 	    \includegraphics[width=\columnwidth]{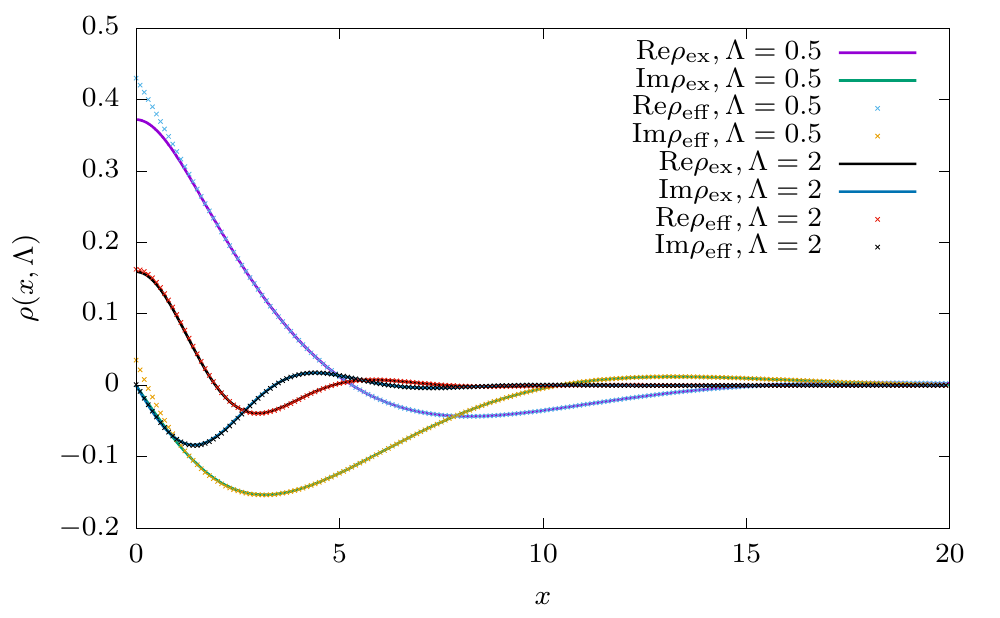}
 	    \caption{One-body correlation function $\rho(x,\Lambda)$ for $\alpha = 1$ and $\beta = 1$. We evaluate it for two different values of $\Lambda$ values, 
 	    $\Lambda=1/2$ - the winding region, and $\Lambda=2$ - the non-winding region. Notice that the critical value of the spin rapidity for such parameters is $\Lambda_c = 1$.
 	    $\rho_{\rm ex}$ refers to the exact formulation in terms of Fredholm determinants and is shown in solid lines, while $\rho_{\rm eff}$ refers to effective form-factors calculation and is shown with dots.}
 	    \label{fig:rho_x_fixed_Lambda}
 	\end{figure}

In this appendix we study contribution of the single hole excitations to the sum  \eqref{def_tau_N}. 
Namely, let us denote the position of the hole by $h$ and the corresponding set of $q$'s by ${\bf q}_h$, see Fig.~(\ref{fig:excitations}a). By ${\bf \bar{q}}$ we denote the set ${\bf q}_h$ with the hole filled in. 
Notice that the determinant in \eqref{D_def} can be presented as 
\begin{equation}
    \det D = \frac{\prod_{i > j}^{N+1} (k_i - k_j) \prod_{i > j}^N (q_i - q_j)}{\prod_{i=1}^{N+1} \prod_{j=1}^N (k_i - q_j)},
\end{equation}
so the effective form factor~\eqref{effective_ff} reads
\begin{equation}\label{over1}
	\frac{|\langle {\bf k}|{\bf q}_h\rangle|^2}{\widetilde{|\langle {\bf k}|{\bf \bar{q}}\rangle|}^2 } =  \frac{L }{2}  e^{g(q_h)}
	\left(\frac{2\pi \nu_h}{L}\right)^2 \prod_{j \neq h} \left(\frac{k_j - q_h}{q_j - q_h}\right)^2,
\end{equation}
where we have introduced
\begin{equation}\label{over2}
	\widetilde{|\langle {\bf k}|{\bf \bar{q}}\rangle|}^2 = 
	\prod\limits_{i=1}^{N+1} \frac{4e^{g(k_i)-g(q_i)}\sin^2 \pi \nu(k_i)}{L^2(1+\frac{2\pi}{L}\nu'(k_i))}\left(\det_{N+1}\frac{1}{k_i-q_j}\right)^2,
    \end{equation}
which is the bulk contribution independent of the hole position. 
First let us perform the summation over $h$. To this end we approximate the discrete product from $|\langle {\bf k}|{\bf q}_h\rangle|^2$ in the following way
\begin{align}\nonumber
&	\prod_{j \neq h}  \left(\frac{k_j - q_h}{q_j - q_h}\right)^2 = \prod_{\substack{j=-M \\ j \neq h}}^M \left(1 + \frac{\nu_j}{h-j}\right)^2 \\
	=& \prod_{\substack{j=-M \\ j \neq h}}^M \left(1 + \frac{\nu_j-\nu_h}{h-j+\nu_h}\right)^2  \prod_{\substack{j=-M \\ j \neq h}}^M \left(1 + \frac{\nu_h}{h-j}\right)^2 \nonumber \\
	\approx& \prod_{\substack{j=-M \\ j \neq h}}^M \left(1 + \frac{\nu_j-\nu_h}{h-j}\right)^2  \prod_{\substack{j=-M \\ j \neq h}}^M \left(1 + \frac{\nu_h}{h-j}\right)^2.
\end{align}
In going to the third line we neglected $\nu_h$ in the denominator. For $h-j$ large it gives a subleading correction, whereas for $h-j$ small the whole fraction vanishes. The first product turns then in the thermodynamic limit into an integral. The second product instead can be rewritten in terms of the $\Gamma$ functions. The result~is
\begin{widetext}
\begin{equation}
	\prod_{j \neq h}  \left(\frac{k_j - q_h}{q_j - q_h}\right)^2  \approx \left(\frac{\sin \pi \nu_h}{\pi \nu_h}\right)^2 
	\exp\left( -2 \int dp \frac{\nu(p)- \nu(q_h)}{p-q_h}
	\right)\left( \frac{\Gamma(M-h - \nu_h+1)}{\Gamma(M-h +1)}\frac{\Gamma(M+h + \nu_h+1)}{\Gamma(M+h +1)} \right)^2.
\end{equation}
\end{widetext}
The integration is over the range $[-q_F, q_F]$ but we have already taken $q_F \rightarrow \infty$ limit in this part. Moreover, understanding the integral in the principal value sense, we obtain
\begin{multline}
	\frac{	|\langle {\bf k}|{\bf q}_h\rangle|^2 }{ \widetilde{|\langle {\bf k}|{\bf \bar{q}}\rangle|}^2   }  \approx \frac{2}{L} e^{g(q_h)} \sin^2 (\pi \nu_h)\exp\left(
 \fint dp \frac{ 2\nu(p)}{q_h-p} 
	\right)\\ \times \left( \frac{\Gamma(M-h - \nu_h+1)}{\Gamma(M-h +1)}\frac{\Gamma(M+h + \nu_h+1)}{\Gamma(M+h +1)}  \right)^2.
\end{multline}
It is important that this ratio is $O(1/L)$ for all $h$, therefore in the summation we can consider only bulk contributions where the hole is far from the edges. That is $M\gg h \gg 1$, which implies that in the leading order the ratio of the $\Gamma$-functions is $1$. We then have
\begin{equation}
|\langle {\bf k}|{\bf q}_h\rangle|^2 	  \approx \frac{2}{L} \widetilde{|\langle {\bf k}|{\bf {\bf q}}\rangle|}^2 \sigma(q_h )\sin^2(\delta(q_h)) \Phi(q_h)
\end{equation}
where we used~\eqref{matching_kernels} to rewrite the formula in terms of $\delta(q)$ and $\sigma(q)$ instead of $g(q)$ and $\nu(q)$, and denoted 
\begin{equation}\label{PHI}
    \Phi(q_h) = \exp\left(	 \fint dp \frac{ 2\nu(p)}{q_h-p}	\right),
\end{equation}
This way, the tau-function on the one-hole states reads
\begin{multline}
	\tau(x) = \widetilde{|\langle {\bf k}|{\bf \bar{q}}\rangle|}^2   \frac{2}{L} \sum_{q} \sigma(q)\sin^2(\delta(q)) \times\\ \times\Phi(q)\exp\left(ix \int \nu(p) {\rm d}p -i x q\right).
\end{multline}
The sum over $q$ in  can be rewritten as an integral
\begin{equation}
	J(x) =  \int \frac{dq}{\pi} \sigma(q) \sin^2(\delta(q))\Phi(q)e^{-i x q},
\end{equation}
and the overall factor $\widetilde{|\langle {\bf k}|{\bf \bar{q}}\rangle|}^2 $, in the thermodynamic and ${q_F \rightarrow \infty}$ limits, can be evaluated, for instance, as in Ref.~\cite{chernowitz2022dynamics}.
This leads to the final answer
\begin{multline}\label{eq.77}
	\tau(x) = J(x) \exp\left(
	ix \int \nu(p) dp -\int \nu(q) g'(q) dq 
	\right) \times\\ 
	\exp\left(- \frac{1}{2} \int dk dq \left(\frac{\nu(k)-\nu(q)}{k-q}\right)^2\right).
\end{multline}
Integrating by parts we arrive at the expression \eqref{asymptNON}. 
We compare the exact the Fredholm determinants and asymptotics in Fig. \eqref{fig:rho_x_fixed_Lambda}. 
We see that for chosen typical parameters it is almost impossible to distinguish the exact and approximate expressions.

    \subsection{Winding case \texorpdfstring{$w=1$}{w=1} }\label{app:factorD2}
    
	
    If winding $w=1$ then the state $|{\bf k}\rangle$ is effectively compressed, see Fig.~(\ref{fig:excitations}b).
    and contrary to the non-winding case, there is no extra space for a hole in the state $|{\bf q}\rangle$. The possible excitations are then only particle-hole excitations which we neglect in view of the $q_F \rightarrow \infty$ limit, as discussed in the main text. Therefore in the spectral sum there is only one state contributing
    \begin{equation}
    	\rho(x, \Lambda) = |\langle {\bf k}|  {\bf q}\rangle|^2 \exp\left(-i x \int k \nu'(k) {\rm d}k\right),
    \end{equation}
    where states $|{\bf k}\rangle$ and $|{\bf q} \rangle$ are both defined with the ground state quantum numbers and
	\begin{align}
		q_j &= \frac{2\pi}{L}\left(-\frac{N-1}{2} + j-1\right) ,\qquad j=1,\dots N, \\
		k_j &= \frac{2\pi}{L}\left(-\frac{N-1}{2} + j-1- \nu_j \right) ,\qquad j=1,\dots N+1,
	\end{align}
	with $\nu_1 \approx 0$ and  $\nu_{N+1} \approx 1$.	We present the overlap $|\langle {\bf k}|{\bf q}\rangle|^2$ as 
	\begin{equation}\label{over}
		|\langle {\bf k}|{\bf q}\rangle|^2 =  \mathcal{D}  \exp\left(\int (g(k)-1)\nu'(k)dk  \right),
	\end{equation}
	where we have defined 
	\begin{equation}
		\mathcal{D} = \prod\limits_{i=1}^{N+1}
		\frac{2  \sin^2 \pi \nu_i}{L}
		\prod\limits_{i=1}^N \frac{2}{L} 
		(\det D)^2 .
	\end{equation}
Further we can present it as $\mathcal{D} =\mathcal{D}_0  \mathcal{F}$, where 
\begin{equation}
	\mathcal{D}_0  =
	\frac{2}{L}\sin^2(\pi \nu_m) \frac{\prod\limits_{j\neq m}^{N+1} (k_m-k_j)^2}{\prod\limits_{j=1}^{N} (k_m-q_j)^2},
\end{equation}
and 
\begin{equation}
	\mathcal{F} = \left(\det\limits_{1\le i,j \le N}\frac{\sin(\pi \eta_i)}{\pi (i-j-\eta_i)}\right)^2.
\end{equation}
In the expression for $\mathcal{F}$ we use $\eta_i$
\begin{equation}
	\eta_i = \begin{cases}
		\nu_i,\qquad &i\le m \\
		\nu_i-1,\qquad &i>m,
	\end{cases}
\end{equation}
which has a jump at $i = m$. In doing so we trade the continuous non-zero winding function $\nu(q)$ into discontinuous but zero winding function $\eta(q)$. It turns out the in computing the thermodynamic limit it is easier to deal with a discontinuous rather that non-zero winding functions. The point $m$ of the discontinuity is chosen arbitrarily (but far from the edges) and the final answer does not depend on it.
	
First we evaluate $\mathcal{D}_0$ assuming that  $N \gg m \gg 1$ and $N\gg L \gg 1$. We find
\begin{equation}
	\mathcal{D}_0  \approx 
	\frac{2}{L}(N-m)^2\frac{\sin^2(\pi \nu_m)}{\nu_m^2} \prod\limits_{j\neq m}^{N} \left(\frac{1- \frac{\nu_m-\nu_j}{m-j}}{1- \frac{\nu_m}{m-j}}\right)^2.
\end{equation}
The numerator in the product can be evaluated as 
\begin{equation}
	\prod\limits_{j\neq m}^{N}\left(1- \frac{\nu_m-\nu_j}{m-j}\right) = \exp\left(
	-\int\limits_{-q_F}^{q_F} \frac{\nu(p_m)-\nu(k)}{p_m-k}{\rm d}k
	\right),
\end{equation}
here $q_F = \frac{\pi N}{L}$, $p_m =2\pi/L(-N/2+m)$ and the phase shifts are defined in the usual way $\nu_i=\nu(k_i) = \nu(2\pi/L(-N/2+i))$.
Integrating this expression by parts we obtain 
\begin{multline}
	\prod\limits_{j\neq m}^{N}\left(1- \frac{\nu_m-\nu_j}{m-j}\right) = \frac{1}{q_F-p_m}\left(\frac{q_F-p_m}{q_F+p_m}\right)^{\nu_m}\times\\\exp\left(\int\limits_{-q_F}^{q_F} \nu'(k) \ln|p_m-k|{\rm d} k\right).
\end{multline}
Note that in the integral we can already send $q_F\to \infty$. The product in the denominator of $\mathcal{D}_0$ can be evaluated explicitly and the result reads 
\begin{multline}
	\prod\limits_{j\neq m}^{N}\left(1- \frac{\nu_m}{m-j}\right)=\frac{\Gamma(N-m+1+\nu_m)}{\Gamma(1+\nu_m)\Gamma(N-m+1)}\times \\ \frac{\Gamma(m-\nu_m)}{\Gamma(1-\nu_m)\Gamma(m)}\approx 
	\frac{\sin(\pi \nu_m)}{\pi \nu_m} \left(\frac{N-m}{m}\right)^{\nu_m}.
\end{multline}
This way we obtain
\begin{equation}
	\mathcal{D}_0 =\frac{L}{2}\exp\left(
	2\int \nu'(k) \ln|p_m-k| {\rm d}k
	\right).
\end{equation}
Further for simplicity we can put $p_m=0$, which means that $m\approx N/2$, so 
\begin{equation}
	\mathcal{D}_0 =\frac{L}{2} \exp\left(
	2\int \nu'(k) \ln|k|{\rm d}k
	\right).
\end{equation}
To estimate $\mathcal{F}$ we use the result listed in Appendix B in \cite{chernowitz2022dynamics}. It states that if 
function $\eta(q)$ is discontinuous at $q=0$ and the discontinuity is $\delta$, that is 
	\begin{equation}
		\lim_{q \rightarrow 0^+} \eta(q) - \lim_{q\rightarrow 0^-} \eta(q) = \delta,
	\end{equation}
	smooth everywhere else and vanishing fast enough at $q\rightarrow \pm \infty$ (in our case this vanishing is exponential due to $\sigma(q)$), then
	\begin{widetext}
	\begin{equation}
		\mathcal{F} = \left(\frac{2 \pi}{L}\right)^{\delta^2} G(1-\delta)^2 \left(\frac{2\pi}{e}\right)^{\delta}
		\exp\left(
		\int {\rm d}q \int {\rm d}p[\eta'](q)[\eta'](p)\ln|q-p|+2\delta \int [\eta'](q)\ln|q|{\rm d}q  \right),
	\end{equation}	
	\end{widetext}
	where $G(x)$ is Barnes Gamma function and $[\eta]'(q)$ is the piece-wise derivative of $\eta(q)$,
	\begin{equation}
		[\eta]'(q) = \Theta(-q) \eta'(q) + \Theta(q) \eta'(q).
	\end{equation}
	In particular, if $\eta(q)$ can be described with the help of smooth function $\nu(q)$ such that
	\begin{equation}
		\eta(q) = \nu(q) + \delta\, \Theta(q),
	\end{equation}
	then $[\eta'](q) = \nu'(q)$. In our case $\delta=-1$ and $\eta(q) = \nu(q) - \Theta(q)$. This leads to 
\begin{multline}
	\mathcal{F} = \frac{e}{L} 
	\exp\left(- 2\int \nu'(q)\ln|q|{\rm d}q  \right)\times \\
	\exp\left(\int {\rm d}q \int {\rm d}p\,\nu'(p)\nu'(q)\ln|q-p|\right).
\end{multline}
The final answer for $\mathcal{D}$ is then
\begin{equation}
	\mathcal{D}  = \frac{e}{2}\exp\left(
	\int {\rm d}q \int {\rm d}p\,\nu'(p)\nu'(q)\ln|q-p| \right).
\end{equation}	
Combining this with the prefactor \eqref{over} we arrive at the expression \eqref{asymptW}. 

We compare the exact the Fredholm determinants and asymptotics in Fig. \eqref{fig:rho_x_fixed_Lambda}. 
We see that the asymptotic expression works decently even for small distancnes.

\section{Further analysis of the asymptotics} \label{appC}

In this appendix we analyse further the asymptotics of $\rho_T(x)$ discussed in Section~\ref{effective}. Specifically, we compute an asymptotic expansion of $J(x)$ in the non-winding region. We then use this result to analyse the behavior of $\rho(x, \Lambda)$ for $\Lambda \approx \Lambda_c$ and show that $\rho(x, \Lambda)$ is a continuous function of $\Lambda$.
	
\subsection{Asymptotic expansion of \texorpdfstring{$J(x)$}{J(x)}}
	
\begin{figure}
    	\includegraphics[scale=.9]{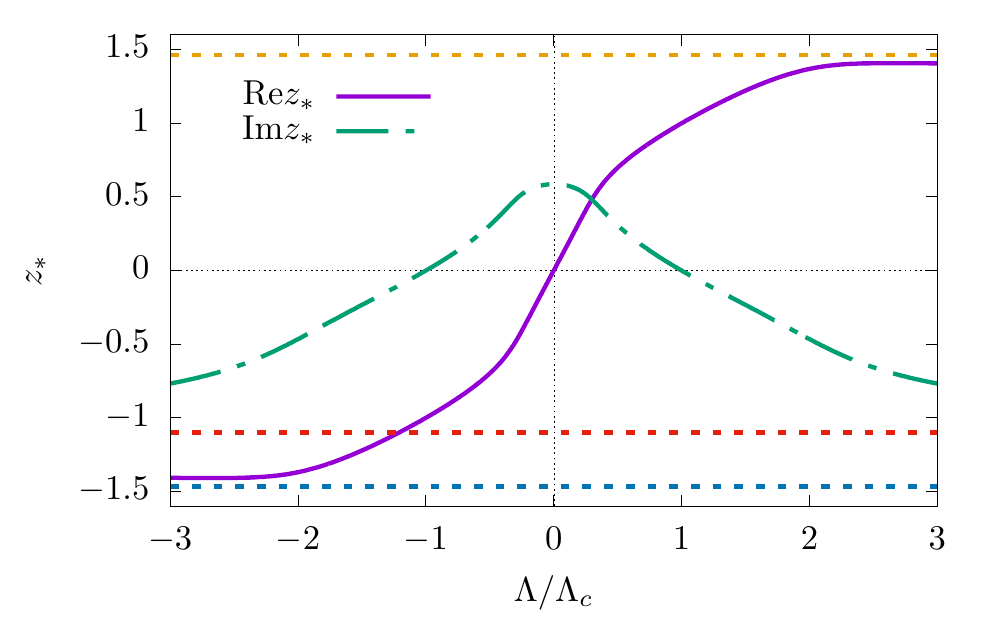}
	    \caption{Solution to $w(z)=0$ as a function of $\Lambda$ for $\alpha=1$, $\beta=1$ and $\mu = 1$. The critical value is $\Lambda_c = 1$. For $\Lambda = \pm \Lambda_c$ the solution is real. For large values of $\Lambda$ the solution approaches finite asymptotes $\pm \sqrt{\beta \mu \pm i \pi}/\sqrt{\beta}$. For $\Lambda > \Lambda_c$, the imaginary part of $z_*$ is negative.}
	    \label{fig:zero_w}
    \end{figure}

	In this section we perform asymptotic expansion of $J(x)$ defined in eq.~\eqref{J_def} which for convenience we repeat here
	\begin{equation}
	    J(x) =  \int \frac{dq}{\pi} \frac{\sigma(q)e^{-iqx}}{(\alpha q - \Lambda)^2+1} 
		\exp\left(\fint dp \frac{ 2\nu(p)}{q-p}\right).
	\end{equation}
	To extract the asymptotics  we rewrite the integral as an integral over a closed semicircle contour $\mathcal{C}_-$ in the lower part of the complex plane. The large $x$ behavior of $J(x)$ is then determined by the pole (with negative imaginary part) closest to the real axis. To this end we first rewrite the principal value integral as
    \begin{equation}
    	\fint dp \frac{ \nu(p)}{q-p} = \int dp \frac{\nu(p)}{q-p-i \epsilon} - i \pi \nu(q),
    \end{equation}
    where $\epsilon >0$ is a small number taken to zero at the end of the computations. This integral, as a function of $q$, has now a simple pole above the real axis, outside of the contour $\mathcal{C}_-$. We also define function $w(q)$ through the following relation
    \begin{equation} \label{def_w}
        \frac{1}{w(q)} = \frac{\sigma(q) e^{-2\pi i \nu(q)}}{\alpha q - \Lambda + i}.
    \end{equation}
    The formula for $w(q)$ can be simplified to
    \begin{equation}
        w(q) = (\alpha q - \Lambda + i)e^{\beta (q^2 - \mu)} + \alpha q - \Lambda - i.
    \end{equation}
    For $J(x)$ we then have
 	\begin{equation}
 		J(x) =  \int_{\mathcal{C}_-} \frac{dz}{\pi} \frac{e^{-izx }}{w(z)} 
		\frac{\exp\left(\int\frac{2\nu(p)dp}{z-p-i \epsilon}\right)}{\alpha z - \Lambda - i}.
    \end{equation} 
where the potential poles with the negative imaginary part come solely from $w(z) = 0$. Let us denote by $z_*$ the solution to this equation with the smallest negative imaginary part. In the vicinity of this solution, $w(z) \approx w'(z_*) (z - z_*) + \mathcal{O}((z-z_*)^2)$,
where
\begin{equation}
	w'(z_*) = 2i \frac{\alpha + i z_* \beta (1 + (z_*\alpha - \Lambda)^2)}{\alpha z_* - \Lambda + i}.
\end{equation}
    Therefore, the leading asymptotic contribution to $J(x)$~is 
    \begin{equation}
    	J(x) =  - \frac{\exp\left(- i z_* x - 2 i \delta(z_*)+
	 2\int \frac{\nu(p)dp}{z_*-p}\right)}{\alpha + i z_* \beta (1 + (z_*\alpha - \Lambda)^2)}, \label{J_asymptote}
\end{equation}
    with the exponential decay rate set by the imaginary part of $z_*$, which therefore has to be negative. Note that in writing the integral we took $\epsilon \rightarrow 0$ limit because $z_*$ has a negative imaginary part which makes the integral well defined, see Fig.~\ref{fig:zero_w}.
    

\subsection{Vicinity of the threshold \texorpdfstring{$\Lambda_c$}{Lambda c}}

The formula for $J(x)$ derived above is valid for ${|\Lambda| > \Lambda_c}$. In the following we will analyse it for $\Lambda$ close to $\Lambda_c$. We start by solving for the pole $z_*$. 
 
 We put $\Lambda = \Lambda_c+\delta$ with $\delta > 0$. To find the pole $z_*$ we put $z= \sqrt{\mu} + \delta v $. This way, in the first order in $\delta$, we obtain
\begin{equation}
w(z) =  2\delta( v (\alpha+i\beta\sqrt{\mu})-1) + O(\delta^2),
\end{equation}
and the pole is indeed located in the lower half plane for $\Lambda>\Lambda_c$,
\begin{equation}
z_* = \sqrt{\mu }+ \frac{\Lambda- \Lambda_c}{\alpha + i\beta \sqrt{\mu}} +O\left((\Lambda - \Lambda_c)^2\right).
\end{equation}
We can now substitute $z_*$ to formula~\eqref{J_asymptote}. The integral appearing there requires a separate treatment because $\nu(p)$ has a logarithmic singularity for $p = \sqrt{\mu}$ when ${\Lambda = \Lambda_c}$. We denote the integral 
\begin{equation}
    \mathcal{I}(\Lambda) = 2 \int \frac{\nu(p) {\rm d}p}{z^* - p},
\end{equation}
and in the following we will show that for $\Lambda \rightarrow \Lambda_c^+$, 
\begin{equation} \label{app_to_proof}
    \mathcal{I}(\Lambda_c^+) = \ln \left(\alpha + i \sqrt{\mu}\beta \right) + \mathcal{J}(\Lambda_c^+),
\end{equation}
where
\begin{align} \label{app_cal_J}
    \mathcal{J}(\Lambda) = \int_0^{\infty} {\rm d}p\, \left(\nu'(\sqrt{\mu} - p) + \nu'(\sqrt{\mu} + p) \right) \ln p.
\end{align}
The logarithmic term in $\mathcal{I}(\Lambda)$ exactly cancels the prefactor in $J(x)$ and the limiting expression for $J(x)$ when $\Lambda \rightarrow \Lambda_c^+$ is
\begin{align}
&J(x) = e^{- i \sqrt{\mu} x + \mathcal{J}(\Lambda)}.
\end{align}

We now derive eq.~\eqref{app_to_proof}. For convenience we introduce $v_* = z_* - \sqrt{\mu}$ and define $\bar{\nu}_{\Lambda}(p) = \nu_{\Lambda}(\sqrt{\mu} + p)$. Then
\begin{align}
\mathcal{I}(\Lambda) = 2 \int{\rm d}p \frac{\bar{\nu}(p) }{v_* - p}.
\end{align}
We rewrite $\mathcal{I}$ by first integrating by parts and then write it as a sum of two integrals
\begin{equation}
\mathcal{I}(\Lambda) = 2 \pi i \bar{\nu}(0) + \mathcal{I}_+(\Lambda) + \mathcal{I}_-(\Lambda),
\end{equation}
where
\begin{equation}
\mathcal{I}_{\pm}(\Lambda) = 2\int_0^{\infty} {\rm d}p\, \bar{\nu}'(\pm p) \ln (p \mp v_*).
\end{equation}
In writing the second contribution we used $\ln(-z) = \ln z - i \pi$.  
We rewrite now $\mathcal{I}_{\pm}(\Lambda)$ as 
\begin{align}
\mathcal{I}_{\pm}(\Lambda) = \int_0^{\infty} {\rm d}p\, (p \pm v_*)\bar{\nu}'(\mp p) \partial_p \left(\ln (p \pm v_*)\right)^2,
\end{align}
which, upon integrating by parts, leads to
\begin{equation}
\mathcal{I}_{\pm}(\Lambda) = \pm v_* \bar{\nu}'(0) \left( \ln (\pm v_*)\right)^2  + \mathcal{J}_{\pm}(\Lambda),
\end{equation}
where
\begin{equation} \label{app:J_vicinity}
    \mathcal{J}_{\pm}(\Lambda) = - \int_0^{\infty} {\rm d}p\, \partial_p\left[(p \pm v_*)\bar{\nu}'(\mp p)\right] \left(\ln (p \pm v_*)\right)^2.
\end{equation}
Collecting the integrals under $\mathcal{J}(\Lambda)$ and other terms under $\mathcal{C}(\Lambda)$ we find
\begin{equation}
    \mathcal{I}(\Lambda) = \mathcal{C} + \mathcal{J}(\Lambda),
\end{equation}
where $\mathcal{J}(\Lambda) = \mathcal{J}_+(\Lambda) + \mathcal{J}_-(\Lambda)$ and 
\begin{equation}
    \mathcal{C}(\Lambda) = 2\pi i \bar{\nu}(0) + v_* \bar{\nu}'(0) (\ln v_*)^2 - v_* \bar{\nu}'(0) (\ln (-v_*))^2.
\end{equation}
We consider now $\Lambda \rightarrow \Lambda_c^+$ for both expressions separately. For $\mathcal{C}(\Lambda)$ we use that
\begin{equation}
\lim_{\Lambda \rightarrow \Lambda_c} v_* \bar{\nu}(0) =  - \frac{1}{2\pi i},
\end{equation}
to find $\mathcal{C}(\Lambda_c) = \ln \left(\alpha + i \sqrt{\mu}\beta \right)$ in agreement with the first part of~\eqref{app_to_proof}. In the integral in $\mathcal{J}_{\pm}$ there is only a square logarithmic singularity, the first part of the integrand is regular for $p\rightarrow 0$ with $\Lambda \rightarrow  \Lambda_c$. Such singularity is integrable and therefore limit $\Lambda \rightarrow \Lambda_c$ of $\mathcal{J}_{\pm}(\Lambda)$ can be safely taken. The result is 
\begin{equation}
    \mathcal{J}_{\pm}(\Lambda) = - \int_0^{\infty} {\rm d}p\, \partial_p\left[p \bar{\nu}'(\mp p)\right] \left(\ln p \right)^2.
\end{equation}
Considering now $\mathcal{J}(\Lambda) = \mathcal{J}_+(\Lambda) + \mathcal{J}_-(\Lambda)$, we integrate back by parts and obtain~\eqref{app_cal_J}. This finishes the derivation of eq.~\eqref{app_to_proof}.

    \subsection{Continuity of the asymptotics} 
    
    In this section we show that $\rho(x, \Lambda)$ is continuous across $\Lambda_c$, namely
    \begin{equation}
    	\lim_{\Lambda \rightarrow \Lambda_c^+} \rho(x, \Lambda) = \lim_{\Lambda \rightarrow \Lambda_c^-} \rho(x, \Lambda).
    \end{equation}
    We start by recalling the relevant expressions for $\rho(x, \Lambda)$ in the winding and non-winding regions from Section~\ref{effective} of the main text. In the former, for $|\Lambda| < \Lambda_c$
    \begin{equation}
        \rho(x,\Lambda) \sim A(\Lambda)\exp\left(-i x \int k \nu_{\Lambda}'(k) {\rm d}k \right),
    \end{equation}
    with
    \begin{align}
    	\ln A(\Lambda) &= \int g(k)\nu_{\Lambda}'(k)dk - \ln 2 \nonumber \\
    	&+\int dq \int dp\,\nu_{\Lambda}'(p)\nu_{\Lambda}'(q)\ln|q-p|.
    \end{align}
    Instead, in the non-winding region with $|\Lambda| > \Lambda_c$,
    \begin{equation}
	\rho(x,\Lambda) \sim 2 A(\Lambda) J(x)\exp\left(- ix \int k \nu_{\Lambda}'(k) {\rm d}k  \right).
    \end{equation}
    In these expression we added $\Lambda$ to $\nu_{\Lambda}(q)$ to highlight its dependence on this parameter and $\nu_{\Lambda_c^{\pm}}(q)$ means the limiting expression when $\Lambda$ approaches $\Lambda_c$ from either above or below. 
    As shown in the previous section, for $\Lambda \sim \Lambda_c^+$, $J(x)$ can be approximated by formula~\eqref{app:J_vicinity}. In this case we can write
    \begin{equation}
        \rho(x, \Lambda_c^+) \sim B(\Lambda_c^+) \exp\left(\! - ix \left(\sqrt{\mu} +  \int k \nu_{\Lambda_c^+}'(k) {\rm d}k  \right)\!\right),
    \end{equation}
    where
    \begin{align}
        &B(\Lambda) = 2 A(\Lambda) e^{\mathcal{J}(\Lambda)}.
    \end{align}

    To confirm the continuity of $\rho(x, \Lambda)$ we start with the $x$-dependent part. We should verify the following relation
    \begin{equation}
    	 \int k \nu_{\Lambda_c^+}'(k) {\rm d}k + \sqrt{\mu} =  \int k \nu_{\Lambda_c^-}'(k) {\rm d}k.   \label{x-continuity}
    \end{equation}
    To this end consider $\nu'(q)$. From the definition~\eqref{nug}, 
    \begin{equation}
    	\nu'(q) = \frac{1}{2\pi i}\frac{\varphi'(q)}{\varphi(q)},
    \end{equation}
    Function $\varphi(q)$ defined in \eqref{varphi} is a bounded function for real $q$. Therefore $\varphi'(q)$ is also bounded and any non-analyticities of $\nu'(q)$ must come from points where $\varphi(q) = 0$. We can rewrite $\varphi(q)$ with the help of $w(q)$, defined in~\eqref{def_w}, as follows
    \begin{equation}
    	w(q) = \frac{\varphi(q) \sigma(q)}{i + \alpha q - \Lambda},
    \end{equation}
    which shows that for real $q$ the set of zeroes of $\varphi(q)$ is the same as the set of zeroes of $w(q)$. Specifically, if we extend to the complex plane, there is a special zero $z_*$ that, as shown in the computation of the asymptotic of $J(x)$ approaches real line from below as $\Lambda \rightarrow \Lambda_c^+$ and for $\Lambda = \Lambda_c$, $z_* = \sqrt{\mu}$.
    
    \begin{figure}
        \centering
        \includegraphics[scale=0.8, trim = 0 40 0 0, clip]{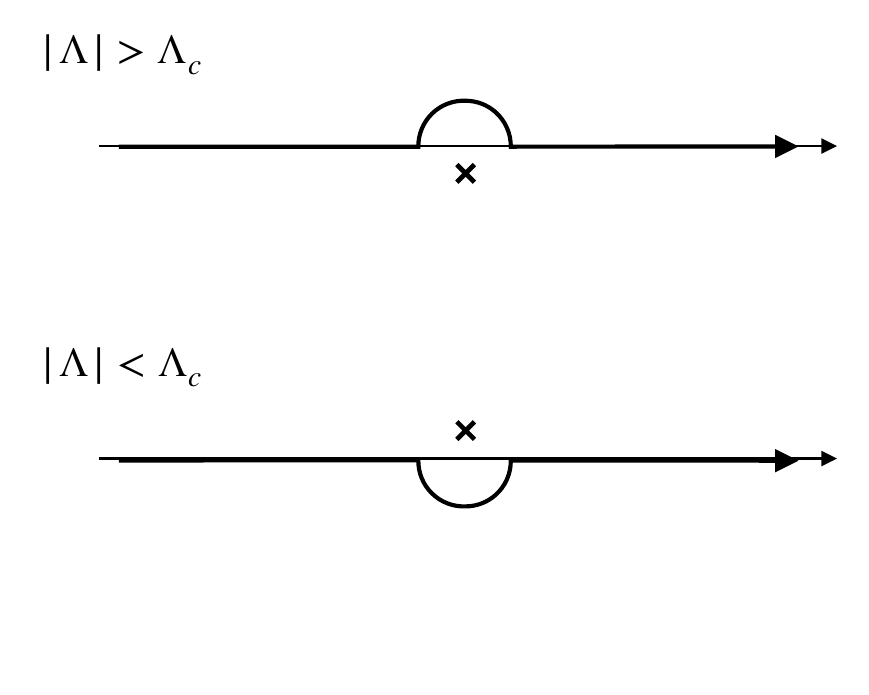}
        \caption{Contours of integrations used in showing the continuity of $\rho(x, \Lambda)$.}
        \label{fig:contours}
    \end{figure}
    The presence of this singularity can be taken into account by deforming the integration contours in~\eqref{x-continuity}. For the first integral, we avoid the pole from above whereas for the second integral we avoid the pole by deflecting the contour below the real axis, see fig.~\ref{fig:contours}. Then the difference of the integrals can be written as a single integral over a closed contour around the pole at $z = \sqrt{\mu}$,
    \begin{equation}
	\int k \nu_{\Lambda_c^+}'(k) {\rm d}k - \int k \nu_{\Lambda_c^-}'(k) {\rm d}k = - \oint {\rm d}z z \nu_{\Lambda_c}'(z).
    \end{equation} 
    The minus sign appears to compensate for the clockwise orientation of the initial contour. The orientation of the contour in the final integral is then counterclockwise.
    Performing now this integral with the help of the residue theorem we find 
    \begin{equation}
    	\oint {\rm d}z z \nu_{\Lambda_c}'(z) = \sqrt{\mu},
    \end{equation}
    and therefore, the $x$-dependent part is continuous. 
    
    We consider now the prefactors. For the ratio of them we have
	\begin{equation}
		\frac{B(\Lambda_c^+)}{A(\Lambda_c^-)} = \exp\left( \mathcal{I}_1 + \mathcal{I}_2 + \ln 2 + \mathcal{J} \right),
	\end{equation}
	where we defined
	    \begin{equation}
    	\mathcal{I}_1 = \int g(k) \nu_{\Lambda_c^+}'(k) {\rm d}k - \int g(k) \nu_{\Lambda_c^-}'(k) {\rm d}k,
	 \end{equation}
	 and
	\begin{align}
	\mathcal{I}_2 = \int dq \int dp\,\nu_{\Lambda_c^+}'(p)\nu_{\Lambda_c^+}'(q)\ln |q-p| \nonumber \\
	- \int dq \int dp\,\nu_{\Lambda_c^-}'(p)\nu_{\Lambda_c^-}'(q)\ln|q-p|.
	\end{align}	 
	For the single integral we use that
	\begin{equation}
		g(k) = 2\pi i\, \nu(k) - \ln a(k), \qquad a(k) = e^{2i \delta(k)}\sigma(k),
	\end{equation}
	to find
	\begin{equation}
	\mathcal{I}_1 = \pi i (w^+ - w^-) + \oint \ln a(k) \nu_{\Lambda_c}'(k) {\rm d}k.
	\end{equation}
	Here $w^+ = 0$ is the winding number of $\nu_{\Lambda}(k)$ for $|\Lambda| > \Lambda_c$ and $w^- = 1$ is the winding number for $|\Lambda| < \Lambda_c$. 
	The remaining integral can be again evaluated by the residue theorem with the result
	\begin{equation}
	\mathcal{I}_1 = \pi i (w^+ - w^-) + \ln a(\sqrt{\mu}) = - \ln 2, 
	\end{equation}
	where we used that $\sigma(\sqrt{\mu}) = 1/2$ and $\delta(\sqrt{\mu}) = \pi/2$ for $\Lambda = \Lambda_c$.  
	
	We consider now a difference of the double integrals. We rewrite them as 
	\begin{align}
		\int dq \int dp\,\nu'(p)\nu'(q)\ln |q-p| \nonumber \\
		= 2 \int\limits_{0}^\infty dp\, \ln p \int dq\, \nu'(p+q)\nu'(q).
	\end{align}	
	The outer integral is the same in both regimes. Therefore 
	\begin{equation}
		\mathcal{I}_2 = 2 \int\limits_{0}^\infty dp\, \ln p\; \mathcal{I}_3(p),
	\end{equation}
	where
	\begin{align}
		\mathcal{I}_3(p) = \int dq\, \nu_{\Lambda_c^+}'(q + p)\nu_{\Lambda_c^+}'(q) \nonumber \\
		- \int dq\, \nu_{\Lambda_c^-}'(q+p)\nu_{\Lambda_c^-}'(q).
	\end{align}
	The difference of the two integrals can be now analysed in a similar manner as before. The integrands have two poles, at $q = \sqrt{\mu}$ and $q = \sqrt{\mu} - p$, and
	\begin{equation}
		\mathcal{I}_3(p) = -\nu_{\Lambda_c}'(\sqrt{\mu} +p) - \nu_{\Lambda_c}'(\sqrt{\mu} - p).
	\end{equation}
	Therefore $\mathcal{I}_2 = -\mathcal{J}$ and the ratio $B(\Lambda_c^+)/A(\Lambda_c^-) = 1$ showing that the $\rho(x, \Lambda)$ is a continuous function of $\Lambda$. 
	
	\subsection{Alternative representation}
	
	The presented about proof of continuity of $\rho(x, \Lambda)$ suggest a possibility to write $\rho(x, \Lambda)$ in a uniform manner for winding and non-winding regimes. In this section we demonstrate this for the $x$-dependent part. In the winding region the $x$-dependent part is given by
	\begin{equation}
		\xi^{-1}(\Lambda) = - \int {\rm d}k\, k \nu'(k) = - \oint_{\mathcal{C}} {\rm d}z z \nu'(z), \label{winding_x_part}
	\end{equation}
	where the closed contour extends in the lower part of the complex plane. As we have seen in the previous section function $\nu'(z)$ has a simple pole which, for $\Lambda =  \pm \Lambda_c$, is located at $z = \pm \sqrt{\mu}$ and otherwise continuously depends on $\Lambda$, see Fig.~\eqref{fig:zero_w}. For $|\Lambda| < \Lambda_c$ the pole is in the upper half of the complex plane and does not affect the integral in~\ref{winding_x_part}. 
	
	In the non-winding region, the $x$-dependent part is 
	\begin{equation}
		\xi^{-1}(\Lambda) = - z_* - \int {\rm d}k\, k \nu'(k) = - z_* - \oint_{\mathcal{C}} {\rm d}z z \nu'(z), \label{non-winding_x_part}
	\end{equation}
	The extra contribution $-z_*$, can be now taken into the account by deforming the contour $\mathcal{C}$ such that it excludes the point $z_*$. With the new contour $\mathcal{C}(\Lambda)$ we have
	\begin{equation}
		\oint_{\mathcal{C}} {\rm d}z z \nu'(z) = \oint_{\mathcal{C}(\Lambda)} {\rm d}z z \nu'(z) - \oint_{z_*} {\rm d}z z \nu'(z),
	\end{equation}
	where the second integral is around the counterclockwise contour including point $z_*$ and no other poles of $\nu'(z)$. The latter integral evaluates to $z^*$ such that in both regions, for the $x$-dependent part we have
	\begin{equation}
		\xi^{-1}(\Lambda) = - \oint_{\mathcal{C}(\Lambda)} {\rm d}z z \nu'(z).
	\end{equation}
	This is the sought after expression for $\xi^{-1}(\Lambda)$ which is valid in the winding and non-winding regimes. As we change $\Lambda$ it possible to adjust contour $\mathcal{C}(\Lambda)$ in a continuous manner and as the result the integral varies in a smooth way.

\bibliography{litra}

\begin{thebibliography}{89}%
\makeatletter
\providecommand \@ifxundefined [1]{%
 \@ifx{#1\undefined}
}%
\providecommand \@ifnum [1]{%
 \ifnum #1\expandafter \@firstoftwo
 \else \expandafter \@secondoftwo
 \fi
}%
\providecommand \@ifx [1]{%
 \ifx #1\expandafter \@firstoftwo
 \else \expandafter \@secondoftwo
 \fi
}%
\providecommand \natexlab [1]{#1}%
\providecommand \enquote  [1]{``#1''}%
\providecommand \bibnamefont  [1]{#1}%
\providecommand \bibfnamefont [1]{#1}%
\providecommand \citenamefont [1]{#1}%
\providecommand \href@noop [0]{\@secondoftwo}%
\providecommand \href [0]{\begingroup \@sanitize@url \@href}%
\providecommand \@href[1]{\@@startlink{#1}\@@href}%
\providecommand \@@href[1]{\endgroup#1\@@endlink}%
\providecommand \@sanitize@url [0]{\catcode `\\12\catcode `\$12\catcode
  `\&12\catcode `\#12\catcode `\^12\catcode `\_12\catcode `\%12\relax}%
\providecommand \@@startlink[1]{}%
\providecommand \@@endlink[0]{}%
\providecommand \url  [0]{\begingroup\@sanitize@url \@url }%
\providecommand \@url [1]{\endgroup\@href {#1}{\urlprefix }}%
\providecommand \urlprefix  [0]{URL }%
\providecommand \Eprint [0]{\href }%
\providecommand \doibase [0]{https://doi.org/}%
\providecommand \selectlanguage [0]{\@gobble}%
\providecommand \bibinfo  [0]{\@secondoftwo}%
\providecommand \bibfield  [0]{\@secondoftwo}%
\providecommand \translation [1]{[#1]}%
\providecommand \BibitemOpen [0]{}%
\providecommand \bibitemStop [0]{}%
\providecommand \bibitemNoStop [0]{.\EOS\space}%
\providecommand \EOS [0]{\spacefactor3000\relax}%
\providecommand \BibitemShut  [1]{\csname bibitem#1\endcsname}%
\let\auto@bib@innerbib\@empty
\bibitem [{lan(1965)}]{landau1965}%
  \BibitemOpen
  \bibfield  {title} {\bibinfo {title} {The effective mass of the polaron},\
  }in\ \href {https://doi.org/10.1016/b978-0-08-010586-4.50072-9} {\emph
  {\bibinfo {booktitle} {Collected Papers of L.D. Landau}}}\ (\bibinfo
  {publisher} {Elsevier},\ \bibinfo {year} {1965})\ pp.\ \bibinfo {pages}
  {478--483}\BibitemShut {NoStop}%
\bibitem [{\citenamefont {Devreese}\ and\ \citenamefont
  {Alexandrov}(2009)}]{Devreese_2009}%
  \BibitemOpen
  \bibfield  {author} {\bibinfo {author} {\bibfnamefont {J.~T.}\ \bibnamefont
  {Devreese}}\ and\ \bibinfo {author} {\bibfnamefont {A.~S.}\ \bibnamefont
  {Alexandrov}},\ }\bibfield  {title} {\bibinfo {title} {Fr{\"o}hlich polaron
  and bipolaron: recent developments},\ }\href
  {https://doi.org/10.1088/0034-4885/72/6/066501} {\bibfield  {journal}
  {\bibinfo  {journal} {Reports on Progress in Physics}\ }\textbf {\bibinfo
  {volume} {72}},\ \bibinfo {pages} {066501} (\bibinfo {year}
  {2009})}\BibitemShut {NoStop}%
\bibitem [{\citenamefont {{C. G. Kuper and G. D. Whitfield
  (eds)}}(1963)}]{kuper_book}%
  \BibitemOpen
  \bibfield  {author} {\bibinfo {author} {\bibnamefont {{C. G. Kuper and G. D.
  Whitfield (eds)}}},\ }\href@noop {} {\emph {\bibinfo {title} {Polarons and
  excitons: Scottish Universities' Summer School 1962}}}\ (\bibinfo
  {publisher} {Oliver and Boyd},\ \bibinfo {address} {Edinburgh and London},\
  \bibinfo {year} {1963})\BibitemShut {NoStop}%
\bibitem [{\citenamefont {Bloch}\ \emph {et~al.}(2008)\citenamefont {Bloch},
  \citenamefont {Dalibard},\ and\ \citenamefont {Zwerger}}]{RevModPhys.80.885}%
  \BibitemOpen
  \bibfield  {author} {\bibinfo {author} {\bibfnamefont {I.}~\bibnamefont
  {Bloch}}, \bibinfo {author} {\bibfnamefont {J.}~\bibnamefont {Dalibard}},\
  and\ \bibinfo {author} {\bibfnamefont {W.}~\bibnamefont {Zwerger}},\
  }\bibfield  {title} {\bibinfo {title} {Many-body physics with ultracold
  gases},\ }\href {https://doi.org/10.1103/RevModPhys.80.885} {\bibfield
  {journal} {\bibinfo  {journal} {Rev. Mod. Phys.}\ }\textbf {\bibinfo {volume}
  {80}},\ \bibinfo {pages} {885} (\bibinfo {year} {2008})}\BibitemShut
  {NoStop}%
\bibitem [{\citenamefont {T\"{o}rm\"{a}}\ and\ \citenamefont
  {Sengstock}(2014)}]{Trm2014}%
  \BibitemOpen
  \bibfield  {author} {\bibinfo {author} {\bibfnamefont {P.}~\bibnamefont
  {T\"{o}rm\"{a}}}\ and\ \bibinfo {author} {\bibfnamefont {K.}~\bibnamefont
  {Sengstock}},\ }\href {https://doi.org/10.1142/p945} {\emph {\bibinfo {title}
  {Quantum Gas Experiments}}}\ (\bibinfo  {publisher} {{IMPERIAL} {COLLEGE}
  {PRESS}},\ \bibinfo {year} {2014})\BibitemShut {NoStop}%
\bibitem [{\citenamefont {Modugno}\ \emph {et~al.}(2002)\citenamefont
  {Modugno}, \citenamefont {Modugno}, \citenamefont {Riboli}, \citenamefont
  {Roati},\ and\ \citenamefont {Inguscio}}]{PhysRevLett.89.190404}%
  \BibitemOpen
  \bibfield  {author} {\bibinfo {author} {\bibfnamefont {G.}~\bibnamefont
  {Modugno}}, \bibinfo {author} {\bibfnamefont {M.}~\bibnamefont {Modugno}},
  \bibinfo {author} {\bibfnamefont {F.}~\bibnamefont {Riboli}}, \bibinfo
  {author} {\bibfnamefont {G.}~\bibnamefont {Roati}},\ and\ \bibinfo {author}
  {\bibfnamefont {M.}~\bibnamefont {Inguscio}},\ }\bibfield  {title} {\bibinfo
  {title} {Two atomic species superfluid},\ }\href
  {https://doi.org/10.1103/PhysRevLett.89.190404} {\bibfield  {journal}
  {\bibinfo  {journal} {Phys. Rev. Lett.}\ }\textbf {\bibinfo {volume} {89}},\
  \bibinfo {pages} {190404} (\bibinfo {year} {2002})}\BibitemShut {NoStop}%
\bibitem [{\citenamefont {Myatt}\ \emph {et~al.}(1997)\citenamefont {Myatt},
  \citenamefont {Burt}, \citenamefont {Ghrist}, \citenamefont {Cornell},\ and\
  \citenamefont {Wieman}}]{PhysRevLett.78.586}%
  \BibitemOpen
  \bibfield  {author} {\bibinfo {author} {\bibfnamefont {C.~J.}\ \bibnamefont
  {Myatt}}, \bibinfo {author} {\bibfnamefont {E.~A.}\ \bibnamefont {Burt}},
  \bibinfo {author} {\bibfnamefont {R.~W.}\ \bibnamefont {Ghrist}}, \bibinfo
  {author} {\bibfnamefont {E.~A.}\ \bibnamefont {Cornell}},\ and\ \bibinfo
  {author} {\bibfnamefont {C.~E.}\ \bibnamefont {Wieman}},\ }\bibfield  {title}
  {\bibinfo {title} {Production of two overlapping bose-einstein condensates by
  sympathetic cooling},\ }\href {https://doi.org/10.1103/PhysRevLett.78.586}
  {\bibfield  {journal} {\bibinfo  {journal} {Phys. Rev. Lett.}\ }\textbf
  {\bibinfo {volume} {78}},\ \bibinfo {pages} {586} (\bibinfo {year}
  {1997})}\BibitemShut {NoStop}%
\bibitem [{\citenamefont {Stenger}\ \emph {et~al.}(1998)\citenamefont
  {Stenger}, \citenamefont {Inouye}, \citenamefont {Stamper-Kurn},
  \citenamefont {Miesner}, \citenamefont {Chikkatur},\ and\ \citenamefont
  {Ketterle}}]{Stenger1998}%
  \BibitemOpen
  \bibfield  {author} {\bibinfo {author} {\bibfnamefont {J.}~\bibnamefont
  {Stenger}}, \bibinfo {author} {\bibfnamefont {S.}~\bibnamefont {Inouye}},
  \bibinfo {author} {\bibfnamefont {D.~M.}\ \bibnamefont {Stamper-Kurn}},
  \bibinfo {author} {\bibfnamefont {H.-J.}\ \bibnamefont {Miesner}}, \bibinfo
  {author} {\bibfnamefont {A.~P.}\ \bibnamefont {Chikkatur}},\ and\ \bibinfo
  {author} {\bibfnamefont {W.}~\bibnamefont {Ketterle}},\ }\bibfield  {title}
  {\bibinfo {title} {Spin domains in ground-state bose{\textendash}einstein
  condensates},\ }\href {https://doi.org/10.1038/24567} {\bibfield  {journal}
  {\bibinfo  {journal} {Nature}\ }\textbf {\bibinfo {volume} {396}},\ \bibinfo
  {pages} {345} (\bibinfo {year} {1998})}\BibitemShut {NoStop}%
\bibitem [{\citenamefont {Schirotzek}\ \emph {et~al.}(2009)\citenamefont
  {Schirotzek}, \citenamefont {Wu}, \citenamefont {Sommer},\ and\ \citenamefont
  {Zwierlein}}]{PhysRevLett.102.230402}%
  \BibitemOpen
  \bibfield  {author} {\bibinfo {author} {\bibfnamefont {A.}~\bibnamefont
  {Schirotzek}}, \bibinfo {author} {\bibfnamefont {C.-H.}\ \bibnamefont {Wu}},
  \bibinfo {author} {\bibfnamefont {A.}~\bibnamefont {Sommer}},\ and\ \bibinfo
  {author} {\bibfnamefont {M.~W.}\ \bibnamefont {Zwierlein}},\ }\bibfield
  {title} {\bibinfo {title} {Observation of fermi polarons in a tunable fermi
  liquid of ultracold atoms},\ }\href
  {https://doi.org/10.1103/PhysRevLett.102.230402} {\bibfield  {journal}
  {\bibinfo  {journal} {Phys. Rev. Lett.}\ }\textbf {\bibinfo {volume} {102}},\
  \bibinfo {pages} {230402} (\bibinfo {year} {2009})}\BibitemShut {NoStop}%
\bibitem [{\citenamefont {Kohstall}\ \emph {et~al.}(2012)\citenamefont
  {Kohstall}, \citenamefont {Zaccanti}, \citenamefont {Jag}, \citenamefont
  {Trenkwalder}, \citenamefont {Massignan}, \citenamefont {Bruun},
  \citenamefont {Schreck},\ and\ \citenamefont {Grimm}}]{Kohstall2012}%
  \BibitemOpen
  \bibfield  {author} {\bibinfo {author} {\bibfnamefont {C.}~\bibnamefont
  {Kohstall}}, \bibinfo {author} {\bibfnamefont {M.}~\bibnamefont {Zaccanti}},
  \bibinfo {author} {\bibfnamefont {M.}~\bibnamefont {Jag}}, \bibinfo {author}
  {\bibfnamefont {A.}~\bibnamefont {Trenkwalder}}, \bibinfo {author}
  {\bibfnamefont {P.}~\bibnamefont {Massignan}}, \bibinfo {author}
  {\bibfnamefont {G.~M.}\ \bibnamefont {Bruun}}, \bibinfo {author}
  {\bibfnamefont {F.}~\bibnamefont {Schreck}},\ and\ \bibinfo {author}
  {\bibfnamefont {R.}~\bibnamefont {Grimm}},\ }\bibfield  {title} {\bibinfo
  {title} {Metastability and coherence of repulsive polarons in a strongly
  interacting fermi mixture},\ }\href {https://doi.org/10.1038/nature11065}
  {\bibfield  {journal} {\bibinfo  {journal} {Nature}\ }\textbf {\bibinfo
  {volume} {485}},\ \bibinfo {pages} {615} (\bibinfo {year}
  {2012})}\BibitemShut {NoStop}%
\bibitem [{\citenamefont {Hu}\ \emph {et~al.}(2016)\citenamefont {Hu},
  \citenamefont {Van~de Graaff}, \citenamefont {Kedar}, \citenamefont {Corson},
  \citenamefont {Cornell},\ and\ \citenamefont {Jin}}]{PhysRevLett.117.055301}%
  \BibitemOpen
  \bibfield  {author} {\bibinfo {author} {\bibfnamefont {M.-G.}\ \bibnamefont
  {Hu}}, \bibinfo {author} {\bibfnamefont {M.~J.}\ \bibnamefont {Van~de
  Graaff}}, \bibinfo {author} {\bibfnamefont {D.}~\bibnamefont {Kedar}},
  \bibinfo {author} {\bibfnamefont {J.~P.}\ \bibnamefont {Corson}}, \bibinfo
  {author} {\bibfnamefont {E.~A.}\ \bibnamefont {Cornell}},\ and\ \bibinfo
  {author} {\bibfnamefont {D.~S.}\ \bibnamefont {Jin}},\ }\bibfield  {title}
  {\bibinfo {title} {Bose polarons in the strongly interacting regime},\ }\href
  {https://doi.org/10.1103/PhysRevLett.117.055301} {\bibfield  {journal}
  {\bibinfo  {journal} {Phys. Rev. Lett.}\ }\textbf {\bibinfo {volume} {117}},\
  \bibinfo {pages} {055301} (\bibinfo {year} {2016})}\BibitemShut {NoStop}%
\bibitem [{\citenamefont {J\o{}rgensen}\ \emph {et~al.}(2016)\citenamefont
  {J\o{}rgensen}, \citenamefont {Wacker}, \citenamefont {Skalmstang},
  \citenamefont {Parish}, \citenamefont {Levinsen}, \citenamefont
  {Christensen}, \citenamefont {Bruun},\ and\ \citenamefont
  {Arlt}}]{PhysRevLett.117.055302}%
  \BibitemOpen
  \bibfield  {author} {\bibinfo {author} {\bibfnamefont {N.~B.}\ \bibnamefont
  {J\o{}rgensen}}, \bibinfo {author} {\bibfnamefont {L.}~\bibnamefont
  {Wacker}}, \bibinfo {author} {\bibfnamefont {K.~T.}\ \bibnamefont
  {Skalmstang}}, \bibinfo {author} {\bibfnamefont {M.~M.}\ \bibnamefont
  {Parish}}, \bibinfo {author} {\bibfnamefont {J.}~\bibnamefont {Levinsen}},
  \bibinfo {author} {\bibfnamefont {R.~S.}\ \bibnamefont {Christensen}},
  \bibinfo {author} {\bibfnamefont {G.~M.}\ \bibnamefont {Bruun}},\ and\
  \bibinfo {author} {\bibfnamefont {J.~J.}\ \bibnamefont {Arlt}},\ }\bibfield
  {title} {\bibinfo {title} {Observation of attractive and repulsive polarons
  in a bose-einstein condensate},\ }\href
  {https://doi.org/10.1103/PhysRevLett.117.055302} {\bibfield  {journal}
  {\bibinfo  {journal} {Phys. Rev. Lett.}\ }\textbf {\bibinfo {volume} {117}},\
  \bibinfo {pages} {055302} (\bibinfo {year} {2016})}\BibitemShut {NoStop}%
\bibitem [{\citenamefont {Yan}\ \emph {et~al.}(2019)\citenamefont {Yan},
  \citenamefont {Patel}, \citenamefont {Mukherjee}, \citenamefont {Fletcher},
  \citenamefont {Struck},\ and\ \citenamefont
  {Zwierlein}}]{PhysRevLett.122.093401}%
  \BibitemOpen
  \bibfield  {author} {\bibinfo {author} {\bibfnamefont {Z.}~\bibnamefont
  {Yan}}, \bibinfo {author} {\bibfnamefont {P.~B.}\ \bibnamefont {Patel}},
  \bibinfo {author} {\bibfnamefont {B.}~\bibnamefont {Mukherjee}}, \bibinfo
  {author} {\bibfnamefont {R.~J.}\ \bibnamefont {Fletcher}}, \bibinfo {author}
  {\bibfnamefont {J.}~\bibnamefont {Struck}},\ and\ \bibinfo {author}
  {\bibfnamefont {M.~W.}\ \bibnamefont {Zwierlein}},\ }\bibfield  {title}
  {\bibinfo {title} {Boiling a unitary fermi liquid},\ }\href
  {https://doi.org/10.1103/PhysRevLett.122.093401} {\bibfield  {journal}
  {\bibinfo  {journal} {Phys. Rev. Lett.}\ }\textbf {\bibinfo {volume} {122}},\
  \bibinfo {pages} {093401} (\bibinfo {year} {2019})}\BibitemShut {NoStop}%
\bibitem [{\citenamefont {Moritz}\ \emph {et~al.}(2003)\citenamefont {Moritz},
  \citenamefont {St\"oferle}, \citenamefont {K\"ohl},\ and\ \citenamefont
  {Esslinger}}]{PhysRevLett.91.250402}%
  \BibitemOpen
  \bibfield  {author} {\bibinfo {author} {\bibfnamefont {H.}~\bibnamefont
  {Moritz}}, \bibinfo {author} {\bibfnamefont {T.}~\bibnamefont {St\"oferle}},
  \bibinfo {author} {\bibfnamefont {M.}~\bibnamefont {K\"ohl}},\ and\ \bibinfo
  {author} {\bibfnamefont {T.}~\bibnamefont {Esslinger}},\ }\bibfield  {title}
  {\bibinfo {title} {Exciting collective oscillations in a trapped 1d gas},\
  }\href {https://doi.org/10.1103/PhysRevLett.91.250402} {\bibfield  {journal}
  {\bibinfo  {journal} {Phys. Rev. Lett.}\ }\textbf {\bibinfo {volume} {91}},\
  \bibinfo {pages} {250402} (\bibinfo {year} {2003})}\BibitemShut {NoStop}%
\bibitem [{\citenamefont {Kinoshita}\ \emph {et~al.}(2004)\citenamefont
  {Kinoshita}, \citenamefont {Wenger},\ and\ \citenamefont
  {Weiss}}]{Kinoshita2004}%
  \BibitemOpen
  \bibfield  {author} {\bibinfo {author} {\bibfnamefont {T.}~\bibnamefont
  {Kinoshita}}, \bibinfo {author} {\bibfnamefont {T.}~\bibnamefont {Wenger}},\
  and\ \bibinfo {author} {\bibfnamefont {D.~S.}\ \bibnamefont {Weiss}},\
  }\bibfield  {title} {\bibinfo {title} {Observation of a one-dimensional
  tonks-girardeau gas},\ }\href {https://doi.org/10.1126/science.1100700}
  {\bibfield  {journal} {\bibinfo  {journal} {Science}\ }\textbf {\bibinfo
  {volume} {305}},\ \bibinfo {pages} {1125} (\bibinfo {year}
  {2004})}\BibitemShut {NoStop}%
\bibitem [{\citenamefont {Paredes}\ \emph {et~al.}(2004)\citenamefont
  {Paredes}, \citenamefont {Widera}, \citenamefont {Murg}, \citenamefont
  {Mandel}, \citenamefont {F\"{o}lling}, \citenamefont {Cirac}, \citenamefont
  {Shlyapnikov}, \citenamefont {H\"{a}nsch},\ and\ \citenamefont
  {Bloch}}]{Paredes2004}%
  \BibitemOpen
  \bibfield  {author} {\bibinfo {author} {\bibfnamefont {B.}~\bibnamefont
  {Paredes}}, \bibinfo {author} {\bibfnamefont {A.}~\bibnamefont {Widera}},
  \bibinfo {author} {\bibfnamefont {V.}~\bibnamefont {Murg}}, \bibinfo {author}
  {\bibfnamefont {O.}~\bibnamefont {Mandel}}, \bibinfo {author} {\bibfnamefont
  {S.}~\bibnamefont {F\"{o}lling}}, \bibinfo {author} {\bibfnamefont
  {I.}~\bibnamefont {Cirac}}, \bibinfo {author} {\bibfnamefont {G.~V.}\
  \bibnamefont {Shlyapnikov}}, \bibinfo {author} {\bibfnamefont {T.~W.}\
  \bibnamefont {H\"{a}nsch}},\ and\ \bibinfo {author} {\bibfnamefont
  {I.}~\bibnamefont {Bloch}},\ }\bibfield  {title} {\bibinfo {title}
  {Tonks{\textendash}girardeau gas of ultracold atoms in an optical lattice},\
  }\href {https://doi.org/10.1038/nature02530} {\bibfield  {journal} {\bibinfo
  {journal} {Nature}\ }\textbf {\bibinfo {volume} {429}},\ \bibinfo {pages}
  {277} (\bibinfo {year} {2004})}\BibitemShut {NoStop}%
\bibitem [{\citenamefont {Catani}\ \emph {et~al.}(2012)\citenamefont {Catani},
  \citenamefont {Lamporesi}, \citenamefont {Naik}, \citenamefont {Gring},
  \citenamefont {Inguscio}, \citenamefont {Minardi}, \citenamefont {Kantian},\
  and\ \citenamefont {Giamarchi}}]{PhysRevA.85.023623}%
  \BibitemOpen
  \bibfield  {author} {\bibinfo {author} {\bibfnamefont {J.}~\bibnamefont
  {Catani}}, \bibinfo {author} {\bibfnamefont {G.}~\bibnamefont {Lamporesi}},
  \bibinfo {author} {\bibfnamefont {D.}~\bibnamefont {Naik}}, \bibinfo {author}
  {\bibfnamefont {M.}~\bibnamefont {Gring}}, \bibinfo {author} {\bibfnamefont
  {M.}~\bibnamefont {Inguscio}}, \bibinfo {author} {\bibfnamefont
  {F.}~\bibnamefont {Minardi}}, \bibinfo {author} {\bibfnamefont
  {A.}~\bibnamefont {Kantian}},\ and\ \bibinfo {author} {\bibfnamefont
  {T.}~\bibnamefont {Giamarchi}},\ }\bibfield  {title} {\bibinfo {title}
  {Quantum dynamics of impurities in a one-dimensional bose gas},\ }\href
  {https://doi.org/10.1103/PhysRevA.85.023623} {\bibfield  {journal} {\bibinfo
  {journal} {Phys. Rev. A}\ }\textbf {\bibinfo {volume} {85}},\ \bibinfo
  {pages} {023623} (\bibinfo {year} {2012})}\BibitemShut {NoStop}%
\bibitem [{\citenamefont {Kinoshita}\ \emph {et~al.}(2006)\citenamefont
  {Kinoshita}, \citenamefont {Wenger},\ and\ \citenamefont
  {Weiss}}]{Kinoshita2006}%
  \BibitemOpen
  \bibfield  {author} {\bibinfo {author} {\bibfnamefont {T.}~\bibnamefont
  {Kinoshita}}, \bibinfo {author} {\bibfnamefont {T.}~\bibnamefont {Wenger}},\
  and\ \bibinfo {author} {\bibfnamefont {D.~S.}\ \bibnamefont {Weiss}},\
  }\bibfield  {title} {\bibinfo {title} {A quantum newton{\textquotesingle}s
  cradle},\ }\href {https://doi.org/10.1038/nature04693} {\bibfield  {journal}
  {\bibinfo  {journal} {Nature}\ }\textbf {\bibinfo {volume} {440}},\ \bibinfo
  {pages} {900} (\bibinfo {year} {2006})}\BibitemShut {NoStop}%
\bibitem [{\citenamefont {Meinert}\ \emph {et~al.}(2017)\citenamefont
  {Meinert}, \citenamefont {Knap}, \citenamefont {Kirilov}, \citenamefont
  {Jag-Lauber}, \citenamefont {Zvonarev}, \citenamefont {Demler},\ and\
  \citenamefont {N\"{a}gerl}}]{Meinert2017}%
  \BibitemOpen
  \bibfield  {author} {\bibinfo {author} {\bibfnamefont {F.}~\bibnamefont
  {Meinert}}, \bibinfo {author} {\bibfnamefont {M.}~\bibnamefont {Knap}},
  \bibinfo {author} {\bibfnamefont {E.}~\bibnamefont {Kirilov}}, \bibinfo
  {author} {\bibfnamefont {K.}~\bibnamefont {Jag-Lauber}}, \bibinfo {author}
  {\bibfnamefont {M.~B.}\ \bibnamefont {Zvonarev}}, \bibinfo {author}
  {\bibfnamefont {E.}~\bibnamefont {Demler}},\ and\ \bibinfo {author}
  {\bibfnamefont {H.-C.}\ \bibnamefont {N\"{a}gerl}},\ }\bibfield  {title}
  {\bibinfo {title} {Bloch oscillations in the absence of a lattice},\ }\href
  {https://doi.org/10.1126/science.aah6616} {\bibfield  {journal} {\bibinfo
  {journal} {Science}\ }\textbf {\bibinfo {volume} {356}},\ \bibinfo {pages}
  {945} (\bibinfo {year} {2017})}\BibitemShut {NoStop}%
\bibitem [{\citenamefont {Panochko}\ and\ \citenamefont
  {Pastukhov}(2019)}]{Panochko2019}%
  \BibitemOpen
  \bibfield  {author} {\bibinfo {author} {\bibfnamefont {G.}~\bibnamefont
  {Panochko}}\ and\ \bibinfo {author} {\bibfnamefont {V.}~\bibnamefont
  {Pastukhov}},\ }\bibfield  {title} {\bibinfo {title} {Mean-field construction
  for spectrum of one-dimensional bose polaron},\ }\href
  {https://doi.org/10.1016/j.aop.2019.167933} {\bibfield  {journal} {\bibinfo
  {journal} {Annals of Physics}\ }\textbf {\bibinfo {volume} {409}},\ \bibinfo
  {pages} {167933} (\bibinfo {year} {2019})}\BibitemShut {NoStop}%
\bibitem [{\citenamefont {Panochko}\ and\ \citenamefont
  {Pastukhov}(2021)}]{panochko2021static}%
  \BibitemOpen
  \bibfield  {author} {\bibinfo {author} {\bibfnamefont {G.}~\bibnamefont
  {Panochko}}\ and\ \bibinfo {author} {\bibfnamefont {V.}~\bibnamefont
  {Pastukhov}},\ }\href@noop {} {\bibinfo {title} {Static impurities in a
  weakly-interacting bose gas}} (\bibinfo {year} {2021}),\ \Eprint
  {https://arxiv.org/abs/2109.14931} {arXiv:2109.14931 [cond-mat.quant-gas]}
  \BibitemShut {NoStop}%
\bibitem [{\citenamefont {Hryhorchak}\ and\ \citenamefont
  {Pastukhov}(2021)}]{hryhorchak2021polaron}%
  \BibitemOpen
  \bibfield  {author} {\bibinfo {author} {\bibfnamefont {O.}~\bibnamefont
  {Hryhorchak}}\ and\ \bibinfo {author} {\bibfnamefont {V.}~\bibnamefont
  {Pastukhov}},\ }\href@noop {} {\bibinfo {title} {Polaron in almost ideal
  molecular bose-einstein condensate}} (\bibinfo {year} {2021}),\ \Eprint
  {https://arxiv.org/abs/2111.07095} {arXiv:2111.07095 [cond-mat.quant-gas]}
  \BibitemShut {NoStop}%
\bibitem [{\citenamefont {Koutentakis}\ \emph {et~al.}(2021)\citenamefont
  {Koutentakis}, \citenamefont {Mistakidis},\ and\ \citenamefont
  {Schmelcher}}]{Koutentakis2021}%
  \BibitemOpen
  \bibfield  {author} {\bibinfo {author} {\bibfnamefont {G.~M.}\ \bibnamefont
  {Koutentakis}}, \bibinfo {author} {\bibfnamefont {S.~I.}\ \bibnamefont
  {Mistakidis}},\ and\ \bibinfo {author} {\bibfnamefont {P.}~\bibnamefont
  {Schmelcher}},\ }\bibfield  {title} {\bibinfo {title} {Pattern formation in
  one-dimensional polaron systems and temporal orthogonality catastrophe},\
  }\href {https://doi.org/10.3390/atoms10010003} {\bibfield  {journal}
  {\bibinfo  {journal} {Atoms}\ }\textbf {\bibinfo {volume} {10}},\ \bibinfo
  {pages} {3} (\bibinfo {year} {2021})}\BibitemShut {NoStop}%
\bibitem [{\citenamefont {Mistakidis}\ \emph {et~al.}(2019)\citenamefont
  {Mistakidis}, \citenamefont {Katsimiga}, \citenamefont {Koutentakis},
  \citenamefont {Busch},\ and\ \citenamefont
  {Schmelcher}}]{PhysRevLett.122.183001}%
  \BibitemOpen
  \bibfield  {author} {\bibinfo {author} {\bibfnamefont {S.~I.}\ \bibnamefont
  {Mistakidis}}, \bibinfo {author} {\bibfnamefont {G.~C.}\ \bibnamefont
  {Katsimiga}}, \bibinfo {author} {\bibfnamefont {G.~M.}\ \bibnamefont
  {Koutentakis}}, \bibinfo {author} {\bibfnamefont {T.}~\bibnamefont {Busch}},\
  and\ \bibinfo {author} {\bibfnamefont {P.}~\bibnamefont {Schmelcher}},\
  }\bibfield  {title} {\bibinfo {title} {Quench dynamics and orthogonality
  catastrophe of bose polarons},\ }\href
  {https://doi.org/10.1103/PhysRevLett.122.183001} {\bibfield  {journal}
  {\bibinfo  {journal} {Phys. Rev. Lett.}\ }\textbf {\bibinfo {volume} {122}},\
  \bibinfo {pages} {183001} (\bibinfo {year} {2019})}\BibitemShut {NoStop}%
\bibitem [{\citenamefont {Massignan}\ \emph {et~al.}(2014)\citenamefont
  {Massignan}, \citenamefont {Zaccanti},\ and\ \citenamefont
  {Bruun}}]{Massignan2014}%
  \BibitemOpen
  \bibfield  {author} {\bibinfo {author} {\bibfnamefont {P.}~\bibnamefont
  {Massignan}}, \bibinfo {author} {\bibfnamefont {M.}~\bibnamefont
  {Zaccanti}},\ and\ \bibinfo {author} {\bibfnamefont {G.~M.}\ \bibnamefont
  {Bruun}},\ }\bibfield  {title} {\bibinfo {title} {Polarons, dressed molecules
  and itinerant ferromagnetism in ultracold fermi gases},\ }\href
  {https://doi.org/10.1088/0034-4885/77/3/034401} {\bibfield  {journal}
  {\bibinfo  {journal} {Reports on Progress in Physics}\ }\textbf {\bibinfo
  {volume} {77}},\ \bibinfo {pages} {034401} (\bibinfo {year}
  {2014})}\BibitemShut {NoStop}%
\bibitem [{\citenamefont {Burovski}\ \emph {et~al.}(2014)\citenamefont
  {Burovski}, \citenamefont {Cheianov}, \citenamefont {Gamayun},\ and\
  \citenamefont {Lychkovskiy}}]{Burovski2014}%
  \BibitemOpen
  \bibfield  {author} {\bibinfo {author} {\bibfnamefont {E.}~\bibnamefont
  {Burovski}}, \bibinfo {author} {\bibfnamefont {V.}~\bibnamefont {Cheianov}},
  \bibinfo {author} {\bibfnamefont {O.}~\bibnamefont {Gamayun}},\ and\ \bibinfo
  {author} {\bibfnamefont {O.}~\bibnamefont {Lychkovskiy}},\ }\bibfield
  {title} {\bibinfo {title} {Momentum relaxation of a mobile impurity in a
  one-dimensional quantum gas},\ }\bibfield  {journal} {\bibinfo  {journal}
  {Physical Review A}\ }\textbf {\bibinfo {volume} {89}},\ \href
  {https://doi.org/10.1103/physreva.89.041601} {10.1103/physreva.89.041601}
  (\bibinfo {year} {2014})\BibitemShut {NoStop}%
\bibitem [{\citenamefont {Gamayun}(2014)}]{PhysRevA.89.063627}%
  \BibitemOpen
  \bibfield  {author} {\bibinfo {author} {\bibfnamefont {O.}~\bibnamefont
  {Gamayun}},\ }\bibfield  {title} {\bibinfo {title} {Quantum boltzmann
  equation for a mobile impurity in a degenerate tonks-girardeau gas},\ }\href
  {https://doi.org/10.1103/PhysRevA.89.063627} {\bibfield  {journal} {\bibinfo
  {journal} {Phys. Rev. A}\ }\textbf {\bibinfo {volume} {89}},\ \bibinfo
  {pages} {063627} (\bibinfo {year} {2014})}\BibitemShut {NoStop}%
\bibitem [{\citenamefont {Gamayun}\ \emph {et~al.}(2014)\citenamefont
  {Gamayun}, \citenamefont {Lychkovskiy},\ and\ \citenamefont
  {Cheianov}}]{PhysRevE.90.032132}%
  \BibitemOpen
  \bibfield  {author} {\bibinfo {author} {\bibfnamefont {O.}~\bibnamefont
  {Gamayun}}, \bibinfo {author} {\bibfnamefont {O.}~\bibnamefont
  {Lychkovskiy}},\ and\ \bibinfo {author} {\bibfnamefont {V.}~\bibnamefont
  {Cheianov}},\ }\bibfield  {title} {\bibinfo {title} {Kinetic theory for a
  mobile impurity in a degenerate tonks-girardeau gas},\ }\href
  {https://doi.org/10.1103/PhysRevE.90.032132} {\bibfield  {journal} {\bibinfo
  {journal} {Phys. Rev. E}\ }\textbf {\bibinfo {volume} {90}},\ \bibinfo
  {pages} {032132} (\bibinfo {year} {2014})}\BibitemShut {NoStop}%
\bibitem [{\citenamefont {Petkovi\ifmmode~\acute{c}\else
  \'{c}\fi{}}(2020)}]{PhysRevB.101.104503}%
  \BibitemOpen
  \bibfield  {author} {\bibinfo {author} {\bibfnamefont {A.}~\bibnamefont
  {Petkovi\ifmmode~\acute{c}\else \'{c}\fi{}}},\ }\bibfield  {title} {\bibinfo
  {title} {Microscopic theory of the friction force exerted on a quantum
  impurity in one-dimensional quantum liquids},\ }\href
  {https://doi.org/10.1103/PhysRevB.101.104503} {\bibfield  {journal} {\bibinfo
   {journal} {Phys. Rev. B}\ }\textbf {\bibinfo {volume} {101}},\ \bibinfo
  {pages} {104503} (\bibinfo {year} {2020})}\BibitemShut {NoStop}%
\bibitem [{\citenamefont {Peotta}\ \emph {et~al.}(2013)\citenamefont {Peotta},
  \citenamefont {Rossini}, \citenamefont {Polini}, \citenamefont {Minardi},\
  and\ \citenamefont {Fazio}}]{PhysRevLett.110.015302}%
  \BibitemOpen
  \bibfield  {author} {\bibinfo {author} {\bibfnamefont {S.}~\bibnamefont
  {Peotta}}, \bibinfo {author} {\bibfnamefont {D.}~\bibnamefont {Rossini}},
  \bibinfo {author} {\bibfnamefont {M.}~\bibnamefont {Polini}}, \bibinfo
  {author} {\bibfnamefont {F.}~\bibnamefont {Minardi}},\ and\ \bibinfo {author}
  {\bibfnamefont {R.}~\bibnamefont {Fazio}},\ }\bibfield  {title} {\bibinfo
  {title} {Quantum breathing of an impurity in a one-dimensional bath of
  interacting bosons},\ }\href {https://doi.org/10.1103/PhysRevLett.110.015302}
  {\bibfield  {journal} {\bibinfo  {journal} {Phys. Rev. Lett.}\ }\textbf
  {\bibinfo {volume} {110}},\ \bibinfo {pages} {015302} (\bibinfo {year}
  {2013})}\BibitemShut {NoStop}%
\bibitem [{\citenamefont {Massel}\ \emph {et~al.}(2013)\citenamefont {Massel},
  \citenamefont {Kantian}, \citenamefont {Daley}, \citenamefont {Giamarchi},\
  and\ \citenamefont {T\"{o}rm\"{a}}}]{Massel2013}%
  \BibitemOpen
  \bibfield  {author} {\bibinfo {author} {\bibfnamefont {F.}~\bibnamefont
  {Massel}}, \bibinfo {author} {\bibfnamefont {A.}~\bibnamefont {Kantian}},
  \bibinfo {author} {\bibfnamefont {A.~J.}\ \bibnamefont {Daley}}, \bibinfo
  {author} {\bibfnamefont {T.}~\bibnamefont {Giamarchi}},\ and\ \bibinfo
  {author} {\bibfnamefont {P.}~\bibnamefont {T\"{o}rm\"{a}}},\ }\bibfield
  {title} {\bibinfo {title} {Dynamics of an impurity in a one-dimensional
  lattice},\ }\href {https://doi.org/10.1088/1367-2630/15/4/045018} {\bibfield
  {journal} {\bibinfo  {journal} {New Journal of Physics}\ }\textbf {\bibinfo
  {volume} {15}},\ \bibinfo {pages} {045018} (\bibinfo {year}
  {2013})}\BibitemShut {NoStop}%
\bibitem [{\citenamefont {Knap}\ \emph {et~al.}(2014)\citenamefont {Knap},
  \citenamefont {Mathy}, \citenamefont {Ganahl}, \citenamefont {Zvonarev},\
  and\ \citenamefont {Demler}}]{PhysRevLett.112.015302}%
  \BibitemOpen
  \bibfield  {author} {\bibinfo {author} {\bibfnamefont {M.}~\bibnamefont
  {Knap}}, \bibinfo {author} {\bibfnamefont {C.~J.~M.}\ \bibnamefont {Mathy}},
  \bibinfo {author} {\bibfnamefont {M.}~\bibnamefont {Ganahl}}, \bibinfo
  {author} {\bibfnamefont {M.~B.}\ \bibnamefont {Zvonarev}},\ and\ \bibinfo
  {author} {\bibfnamefont {E.}~\bibnamefont {Demler}},\ }\bibfield  {title}
  {\bibinfo {title} {Quantum flutter: Signatures and robustness},\ }\href
  {https://doi.org/10.1103/PhysRevLett.112.015302} {\bibfield  {journal}
  {\bibinfo  {journal} {Phys. Rev. Lett.}\ }\textbf {\bibinfo {volume} {112}},\
  \bibinfo {pages} {015302} (\bibinfo {year} {2014})}\BibitemShut {NoStop}%
\bibitem [{\citenamefont {Chevy}(2006)}]{Chevy2006}%
  \BibitemOpen
  \bibfield  {author} {\bibinfo {author} {\bibfnamefont {F.}~\bibnamefont
  {Chevy}},\ }\bibfield  {title} {\bibinfo {title} {Universal phase diagram of
  a strongly interacting fermi gas with unbalanced spin populations},\
  }\bibfield  {journal} {\bibinfo  {journal} {Physical Review A}\ }\textbf
  {\bibinfo {volume} {74}},\ \href {https://doi.org/10.1103/physreva.74.063628}
  {10.1103/physreva.74.063628} (\bibinfo {year} {2006})\BibitemShut {NoStop}%
\bibitem [{\citenamefont {Combescot}\ \emph {et~al.}(2007)\citenamefont
  {Combescot}, \citenamefont {Recati}, \citenamefont {Lobo},\ and\
  \citenamefont {Chevy}}]{PhysRevLett.98.180402}%
  \BibitemOpen
  \bibfield  {author} {\bibinfo {author} {\bibfnamefont {R.}~\bibnamefont
  {Combescot}}, \bibinfo {author} {\bibfnamefont {A.}~\bibnamefont {Recati}},
  \bibinfo {author} {\bibfnamefont {C.}~\bibnamefont {Lobo}},\ and\ \bibinfo
  {author} {\bibfnamefont {F.}~\bibnamefont {Chevy}},\ }\bibfield  {title}
  {\bibinfo {title} {Normal state of highly polarized fermi gases: Simple
  many-body approaches},\ }\href
  {https://doi.org/10.1103/PhysRevLett.98.180402} {\bibfield  {journal}
  {\bibinfo  {journal} {Phys. Rev. Lett.}\ }\textbf {\bibinfo {volume} {98}},\
  \bibinfo {pages} {180402} (\bibinfo {year} {2007})}\BibitemShut {NoStop}%
\bibitem [{\citenamefont {Giraud}\ and\ \citenamefont
  {Combescot}(2009)}]{PhysRevA.79.043615}%
  \BibitemOpen
  \bibfield  {author} {\bibinfo {author} {\bibfnamefont {S.}~\bibnamefont
  {Giraud}}\ and\ \bibinfo {author} {\bibfnamefont {R.}~\bibnamefont
  {Combescot}},\ }\bibfield  {title} {\bibinfo {title} {Highly polarized fermi
  gases: One-dimensional case},\ }\href
  {https://doi.org/10.1103/PhysRevA.79.043615} {\bibfield  {journal} {\bibinfo
  {journal} {Phys. Rev. A}\ }\textbf {\bibinfo {volume} {79}},\ \bibinfo
  {pages} {043615} (\bibinfo {year} {2009})}\BibitemShut {NoStop}%
\bibitem [{\citenamefont {Shchadilova}\ \emph {et~al.}(2016)\citenamefont
  {Shchadilova}, \citenamefont {Grusdt}, \citenamefont {Rubtsov},\ and\
  \citenamefont {Demler}}]{PhysRevA.93.043606}%
  \BibitemOpen
  \bibfield  {author} {\bibinfo {author} {\bibfnamefont {Y.~E.}\ \bibnamefont
  {Shchadilova}}, \bibinfo {author} {\bibfnamefont {F.}~\bibnamefont {Grusdt}},
  \bibinfo {author} {\bibfnamefont {A.~N.}\ \bibnamefont {Rubtsov}},\ and\
  \bibinfo {author} {\bibfnamefont {E.}~\bibnamefont {Demler}},\ }\bibfield
  {title} {\bibinfo {title} {Polaronic mass renormalization of impurities in
  bose-einstein condensates: Correlated gaussian-wave-function approach},\
  }\href {https://doi.org/10.1103/PhysRevA.93.043606} {\bibfield  {journal}
  {\bibinfo  {journal} {Phys. Rev. A}\ }\textbf {\bibinfo {volume} {93}},\
  \bibinfo {pages} {043606} (\bibinfo {year} {2016})}\BibitemShut {NoStop}%
\bibitem [{\citenamefont {Grusdt}\ \emph {et~al.}(2015)\citenamefont {Grusdt},
  \citenamefont {Shchadilova}, \citenamefont {Rubtsov},\ and\ \citenamefont
  {Demler}}]{Grusdt2015}%
  \BibitemOpen
  \bibfield  {author} {\bibinfo {author} {\bibfnamefont {F.}~\bibnamefont
  {Grusdt}}, \bibinfo {author} {\bibfnamefont {Y.~E.}\ \bibnamefont
  {Shchadilova}}, \bibinfo {author} {\bibfnamefont {A.~N.}\ \bibnamefont
  {Rubtsov}},\ and\ \bibinfo {author} {\bibfnamefont {E.}~\bibnamefont
  {Demler}},\ }\bibfield  {title} {\bibinfo {title} {Renormalization group
  approach to the fr\"{o}hlich polaron model: application to impurity-{BEC}
  problem},\ }\bibfield  {journal} {\bibinfo  {journal} {Scientific Reports}\
  }\textbf {\bibinfo {volume} {5}},\ \href {https://doi.org/10.1038/srep12124}
  {10.1038/srep12124} (\bibinfo {year} {2015})\BibitemShut {NoStop}%
\bibitem [{\citenamefont {Parish}\ and\ \citenamefont
  {Levinsen}(2013)}]{PhysRevA.87.033616}%
  \BibitemOpen
  \bibfield  {author} {\bibinfo {author} {\bibfnamefont {M.~M.}\ \bibnamefont
  {Parish}}\ and\ \bibinfo {author} {\bibfnamefont {J.}~\bibnamefont
  {Levinsen}},\ }\bibfield  {title} {\bibinfo {title} {Highly polarized fermi
  gases in two dimensions},\ }\href
  {https://doi.org/10.1103/PhysRevA.87.033616} {\bibfield  {journal} {\bibinfo
  {journal} {Phys. Rev. A}\ }\textbf {\bibinfo {volume} {87}},\ \bibinfo
  {pages} {033616} (\bibinfo {year} {2013})}\BibitemShut {NoStop}%
\bibitem [{\citenamefont {Dolgirev}\ \emph {et~al.}(2021)\citenamefont
  {Dolgirev}, \citenamefont {Qu}, \citenamefont {Zvonarev}, \citenamefont
  {Shi},\ and\ \citenamefont {Demler}}]{PhysRevX.11.041015}%
  \BibitemOpen
  \bibfield  {author} {\bibinfo {author} {\bibfnamefont {P.~E.}\ \bibnamefont
  {Dolgirev}}, \bibinfo {author} {\bibfnamefont {Y.-F.}\ \bibnamefont {Qu}},
  \bibinfo {author} {\bibfnamefont {M.~B.}\ \bibnamefont {Zvonarev}}, \bibinfo
  {author} {\bibfnamefont {T.}~\bibnamefont {Shi}},\ and\ \bibinfo {author}
  {\bibfnamefont {E.}~\bibnamefont {Demler}},\ }\bibfield  {title} {\bibinfo
  {title} {Emergence of a sharp quantum collective mode in a one-dimensional
  fermi polaron},\ }\href {https://doi.org/10.1103/PhysRevX.11.041015}
  {\bibfield  {journal} {\bibinfo  {journal} {Phys. Rev. X}\ }\textbf {\bibinfo
  {volume} {11}},\ \bibinfo {pages} {041015} (\bibinfo {year}
  {2021})}\BibitemShut {NoStop}%
\bibitem [{\citenamefont {Mathy}\ \emph {et~al.}(2012)\citenamefont {Mathy},
  \citenamefont {Zvonarev},\ and\ \citenamefont {Demler}}]{Mathy2012}%
  \BibitemOpen
  \bibfield  {author} {\bibinfo {author} {\bibfnamefont {C.~J.~M.}\
  \bibnamefont {Mathy}}, \bibinfo {author} {\bibfnamefont {M.~B.}\ \bibnamefont
  {Zvonarev}},\ and\ \bibinfo {author} {\bibfnamefont {E.}~\bibnamefont
  {Demler}},\ }\bibfield  {title} {\bibinfo {title} {Quantum flutter of
  supersonic particles in one-dimensional quantum liquids},\ }\href
  {https://doi.org/10.1038/nphys2455} {\bibfield  {journal} {\bibinfo
  {journal} {Nature Physics}\ }\textbf {\bibinfo {volume} {8}},\ \bibinfo
  {pages} {881} (\bibinfo {year} {2012})}\BibitemShut {NoStop}%
\bibitem [{\citenamefont {Prokofev}(1993)}]{PROKOF_EV_1993}%
  \BibitemOpen
  \bibfield  {author} {\bibinfo {author} {\bibfnamefont {N.}~\bibnamefont
  {Prokofev}},\ }\bibfield  {title} {\bibinfo {title} {Diffusion of a heavy
  particle in a fermi-liquid theory},\ }\href
  {https://doi.org/10.1142/s0217979293003255} {\bibfield  {journal} {\bibinfo
  {journal} {International Journal of Modern Physics B}\ }\textbf {\bibinfo
  {volume} {07}},\ \bibinfo {pages} {3327} (\bibinfo {year}
  {1993})}\BibitemShut {NoStop}%
\bibitem [{\citenamefont {Prokof'ev}\ and\ \citenamefont
  {Svistunov}(2008)}]{PhysRevB.77.020408}%
  \BibitemOpen
  \bibfield  {author} {\bibinfo {author} {\bibfnamefont {N.}~\bibnamefont
  {Prokof'ev}}\ and\ \bibinfo {author} {\bibfnamefont {B.}~\bibnamefont
  {Svistunov}},\ }\bibfield  {title} {\bibinfo {title} {Fermi-polaron problem:
  Diagrammatic monte carlo method for divergent sign-alternating series},\
  }\href {https://doi.org/10.1103/PhysRevB.77.020408} {\bibfield  {journal}
  {\bibinfo  {journal} {Phys. Rev. B}\ }\textbf {\bibinfo {volume} {77}},\
  \bibinfo {pages} {020408} (\bibinfo {year} {2008})}\BibitemShut {NoStop}%
\bibitem [{\citenamefont {Parisi}\ and\ \citenamefont
  {Giorgini}(2017)}]{PhysRevA.95.023619}%
  \BibitemOpen
  \bibfield  {author} {\bibinfo {author} {\bibfnamefont {L.}~\bibnamefont
  {Parisi}}\ and\ \bibinfo {author} {\bibfnamefont {S.}~\bibnamefont
  {Giorgini}},\ }\bibfield  {title} {\bibinfo {title} {Quantum monte carlo
  study of the bose-polaron problem in a one-dimensional gas with contact
  interactions},\ }\href {https://doi.org/10.1103/PhysRevA.95.023619}
  {\bibfield  {journal} {\bibinfo  {journal} {Phys. Rev. A}\ }\textbf {\bibinfo
  {volume} {95}},\ \bibinfo {pages} {023619} (\bibinfo {year}
  {2017})}\BibitemShut {NoStop}%
\bibitem [{\citenamefont {Grusdt}\ \emph {et~al.}(2017)\citenamefont {Grusdt},
  \citenamefont {Astrakharchik},\ and\ \citenamefont {Demler}}]{Grusdt2017}%
  \BibitemOpen
  \bibfield  {author} {\bibinfo {author} {\bibfnamefont {F.}~\bibnamefont
  {Grusdt}}, \bibinfo {author} {\bibfnamefont {G.~E.}\ \bibnamefont
  {Astrakharchik}},\ and\ \bibinfo {author} {\bibfnamefont {E.}~\bibnamefont
  {Demler}},\ }\bibfield  {title} {\bibinfo {title} {Bose polarons in ultracold
  atoms in one dimension: beyond the fr\"{o}hlich paradigm},\ }\href
  {https://doi.org/10.1088/1367-2630/aa8a2e} {\bibfield  {journal} {\bibinfo
  {journal} {New Journal of Physics}\ }\textbf {\bibinfo {volume} {19}},\
  \bibinfo {pages} {103035} (\bibinfo {year} {2017})}\BibitemShut {NoStop}%
\bibitem [{\citenamefont {Grusdt}\ and\ \citenamefont
  {Demler}(2015)}]{grusdt2015new}%
  \BibitemOpen
  \bibfield  {author} {\bibinfo {author} {\bibfnamefont {F.}~\bibnamefont
  {Grusdt}}\ and\ \bibinfo {author} {\bibfnamefont {E.}~\bibnamefont
  {Demler}},\ }\href@noop {} {\bibinfo {title} {New theoretical approaches to
  bose polarons}} (\bibinfo {year} {2015}),\ \Eprint
  {https://arxiv.org/abs/1510.04934} {arXiv:1510.04934 [cond-mat.quant-gas]}
  \BibitemShut {NoStop}%
\bibitem [{\citenamefont {McGuire}(1965)}]{McGuire1965}%
  \BibitemOpen
  \bibfield  {author} {\bibinfo {author} {\bibfnamefont {J.~B.}\ \bibnamefont
  {McGuire}},\ }\bibfield  {title} {\bibinfo {title} {Interacting fermions in
  one dimension. i. repulsive potential},\ }\href
  {https://doi.org/10.1063/1.1704291} {\bibfield  {journal} {\bibinfo
  {journal} {Journal of Mathematical Physics}\ }\textbf {\bibinfo {volume}
  {6}},\ \bibinfo {pages} {432} (\bibinfo {year} {1965})}\BibitemShut {NoStop}%
\bibitem [{\citenamefont {McGuire}(1966)}]{McGuire1966}%
  \BibitemOpen
  \bibfield  {author} {\bibinfo {author} {\bibfnamefont {J.~B.}\ \bibnamefont
  {McGuire}},\ }\bibfield  {title} {\bibinfo {title} {Interacting fermions in
  one dimension. {II}. attractive potential},\ }\href
  {https://doi.org/10.1063/1.1704798} {\bibfield  {journal} {\bibinfo
  {journal} {Journal of Mathematical Physics}\ }\textbf {\bibinfo {volume}
  {7}},\ \bibinfo {pages} {123} (\bibinfo {year} {1966})}\BibitemShut {NoStop}%
\bibitem [{\citenamefont {Gaudin}(1967)}]{Gaudin1967}%
  \BibitemOpen
  \bibfield  {author} {\bibinfo {author} {\bibfnamefont {M.}~\bibnamefont
  {Gaudin}},\ }\bibfield  {title} {\bibinfo {title} {Un systeme a une dimension
  de fermions en interaction},\ }\href
  {https://doi.org/10.1016/0375-9601(67)90193-4} {\bibfield  {journal}
  {\bibinfo  {journal} {Physics Letters A}\ }\textbf {\bibinfo {volume} {24}},\
  \bibinfo {pages} {55} (\bibinfo {year} {1967})}\BibitemShut {NoStop}%
\bibitem [{\citenamefont {Yang}(1967)}]{PhysRevLett.19.1312}%
  \BibitemOpen
  \bibfield  {author} {\bibinfo {author} {\bibfnamefont {C.~N.}\ \bibnamefont
  {Yang}},\ }\bibfield  {title} {\bibinfo {title} {Some exact results for the
  many-body problem in one dimension with repulsive delta-function
  interaction},\ }\href {https://doi.org/10.1103/PhysRevLett.19.1312}
  {\bibfield  {journal} {\bibinfo  {journal} {Phys. Rev. Lett.}\ }\textbf
  {\bibinfo {volume} {19}},\ \bibinfo {pages} {1312} (\bibinfo {year}
  {1967})}\BibitemShut {NoStop}%
\bibitem [{\citenamefont {Edwards}(1990)}]{Edwards1990}%
  \BibitemOpen
  \bibfield  {author} {\bibinfo {author} {\bibfnamefont {D.~M.}\ \bibnamefont
  {Edwards}},\ }\bibfield  {title} {\bibinfo {title} {Magnetism in single-band
  models},\ }\href {https://doi.org/10.1143/ptps.101.453} {\bibfield  {journal}
  {\bibinfo  {journal} {Progress of Theoretical Physics Supplement}\ }\textbf
  {\bibinfo {volume} {101}},\ \bibinfo {pages} {453} (\bibinfo {year}
  {1990})}\BibitemShut {NoStop}%
\bibitem [{\citenamefont {Castella}\ and\ \citenamefont
  {Zotos}(1993)}]{PhysRevB.47.16186}%
  \BibitemOpen
  \bibfield  {author} {\bibinfo {author} {\bibfnamefont {H.}~\bibnamefont
  {Castella}}\ and\ \bibinfo {author} {\bibfnamefont {X.}~\bibnamefont
  {Zotos}},\ }\bibfield  {title} {\bibinfo {title} {Exact calculation of
  spectral properties of a particle interacting with a one-dimensional
  fermionic system},\ }\href {https://doi.org/10.1103/PhysRevB.47.16186}
  {\bibfield  {journal} {\bibinfo  {journal} {Phys. Rev. B}\ }\textbf {\bibinfo
  {volume} {47}},\ \bibinfo {pages} {16186} (\bibinfo {year}
  {1993})}\BibitemShut {NoStop}%
\bibitem [{\citenamefont {Gamayun}\ \emph {et~al.}(2018)\citenamefont
  {Gamayun}, \citenamefont {Lychkovskiy}, \citenamefont {Burovski},
  \citenamefont {Malcomson}, \citenamefont {Cheianov},\ and\ \citenamefont
  {Zvonarev}}]{PhysRevLett.120.220605}%
  \BibitemOpen
  \bibfield  {author} {\bibinfo {author} {\bibfnamefont {O.}~\bibnamefont
  {Gamayun}}, \bibinfo {author} {\bibfnamefont {O.}~\bibnamefont
  {Lychkovskiy}}, \bibinfo {author} {\bibfnamefont {E.}~\bibnamefont
  {Burovski}}, \bibinfo {author} {\bibfnamefont {M.}~\bibnamefont {Malcomson}},
  \bibinfo {author} {\bibfnamefont {V.~V.}\ \bibnamefont {Cheianov}},\ and\
  \bibinfo {author} {\bibfnamefont {M.~B.}\ \bibnamefont {Zvonarev}},\
  }\bibfield  {title} {\bibinfo {title} {Impact of the injection protocol on an
  impurity's stationary state},\ }\href
  {https://doi.org/10.1103/PhysRevLett.120.220605} {\bibfield  {journal}
  {\bibinfo  {journal} {Phys. Rev. Lett.}\ }\textbf {\bibinfo {volume} {120}},\
  \bibinfo {pages} {220605} (\bibinfo {year} {2018})}\BibitemShut {NoStop}%
\bibitem [{\citenamefont {Gamayun}\ \emph {et~al.}(2015)\citenamefont
  {Gamayun}, \citenamefont {Pronko},\ and\ \citenamefont
  {Zvonarev}}]{Gamayun2015}%
  \BibitemOpen
  \bibfield  {author} {\bibinfo {author} {\bibfnamefont {O.}~\bibnamefont
  {Gamayun}}, \bibinfo {author} {\bibfnamefont {A.~G.}\ \bibnamefont
  {Pronko}},\ and\ \bibinfo {author} {\bibfnamefont {M.~B.}\ \bibnamefont
  {Zvonarev}},\ }\bibfield  {title} {\bibinfo {title} {Impurity
  green{\textquotesingle}s function of a one-dimensional fermi gas},\ }\href
  {https://doi.org/10.1016/j.nuclphysb.2015.01.004} {\bibfield  {journal}
  {\bibinfo  {journal} {Nuclear Physics B}\ }\textbf {\bibinfo {volume}
  {892}},\ \bibinfo {pages} {83} (\bibinfo {year} {2015})}\BibitemShut
  {NoStop}%
\bibitem [{\citenamefont {Gamayun}\ \emph {et~al.}(2016)\citenamefont
  {Gamayun}, \citenamefont {Pronko},\ and\ \citenamefont
  {Zvonarev}}]{Gamayun2016}%
  \BibitemOpen
  \bibfield  {author} {\bibinfo {author} {\bibfnamefont {O.}~\bibnamefont
  {Gamayun}}, \bibinfo {author} {\bibfnamefont {A.~G.}\ \bibnamefont
  {Pronko}},\ and\ \bibinfo {author} {\bibfnamefont {M.~B.}\ \bibnamefont
  {Zvonarev}},\ }\bibfield  {title} {\bibinfo {title} {Time and
  temperature-dependent correlation function of an impurity in one-dimensional
  fermi and tonks{\textendash}girardeau gases as a fredholm determinant},\
  }\href {https://doi.org/10.1088/1367-2630/18/4/045005} {\bibfield  {journal}
  {\bibinfo  {journal} {New Journal of Physics}\ }\textbf {\bibinfo {volume}
  {18}},\ \bibinfo {pages} {045005} (\bibinfo {year} {2016})}\BibitemShut
  {NoStop}%
\bibitem [{\citenamefont {Recher}\ and\ \citenamefont
  {Kohler}(2012)}]{Recher2012}%
  \BibitemOpen
  \bibfield  {author} {\bibinfo {author} {\bibfnamefont {C.}~\bibnamefont
  {Recher}}\ and\ \bibinfo {author} {\bibfnamefont {H.}~\bibnamefont
  {Kohler}},\ }\bibfield  {title} {\bibinfo {title} {From hardcore bosons to
  free fermions with painlev{\'{e}} v},\ }\href
  {https://doi.org/10.1007/s10955-012-0482-1} {\bibfield  {journal} {\bibinfo
  {journal} {Journal of Statistical Physics}\ }\textbf {\bibinfo {volume}
  {147}},\ \bibinfo {pages} {542} (\bibinfo {year} {2012})}\BibitemShut
  {NoStop}%
\bibitem [{\citenamefont {Gamayun}\ \emph {et~al.}(2020)\citenamefont
  {Gamayun}, \citenamefont {Lychkovskiy},\ and\ \citenamefont
  {Zvonarev}}]{10.21468/SciPostPhys.8.4.053}%
  \BibitemOpen
  \bibfield  {author} {\bibinfo {author} {\bibfnamefont {O.}~\bibnamefont
  {Gamayun}}, \bibinfo {author} {\bibfnamefont {O.}~\bibnamefont
  {Lychkovskiy}},\ and\ \bibinfo {author} {\bibfnamefont {M.~B.}\ \bibnamefont
  {Zvonarev}},\ }\bibfield  {title} {\bibinfo {title} {{Zero temperature
  momentum distribution of an impurity in a polaron state of one-dimensional
  Fermi and Tonks-Girardeau gases}},\ }\href
  {https://doi.org/10.21468/SciPostPhys.8.4.053} {\bibfield  {journal}
  {\bibinfo  {journal} {SciPost Phys.}\ }\textbf {\bibinfo {volume} {8}},\
  \bibinfo {pages} {53} (\bibinfo {year} {2020})}\BibitemShut {NoStop}%
\bibitem [{\citenamefont {Gamayun}\ \emph {et~al.}(2021)\citenamefont
  {Gamayun}, \citenamefont {Iorgov},\ and\ \citenamefont
  {Zhuravlev}}]{EffectiveXY}%
  \BibitemOpen
  \bibfield  {author} {\bibinfo {author} {\bibfnamefont {O.}~\bibnamefont
  {Gamayun}}, \bibinfo {author} {\bibfnamefont {N.}~\bibnamefont {Iorgov}},\
  and\ \bibinfo {author} {\bibfnamefont {Y.}~\bibnamefont {Zhuravlev}},\
  }\bibfield  {title} {\bibinfo {title} {{Effective free-fermionic form factors
  and the XY spin chain}},\ }\href
  {https://doi.org/10.21468/SciPostPhys.10.3.070} {\bibfield  {journal}
  {\bibinfo  {journal} {SciPost Phys.}\ }\textbf {\bibinfo {volume} {10}},\
  \bibinfo {pages} {70} (\bibinfo {year} {2021})}\BibitemShut {NoStop}%
\bibitem [{\citenamefont {Zhuravlev}\ \emph {et~al.}(2021)\citenamefont
  {Zhuravlev}, \citenamefont {Naichuk}, \citenamefont {Iorgov},\ and\
  \citenamefont {Gamayun}}]{zhuravlev2021large}%
  \BibitemOpen
  \bibfield  {author} {\bibinfo {author} {\bibfnamefont {Y.}~\bibnamefont
  {Zhuravlev}}, \bibinfo {author} {\bibfnamefont {E.}~\bibnamefont {Naichuk}},
  \bibinfo {author} {\bibfnamefont {N.}~\bibnamefont {Iorgov}},\ and\ \bibinfo
  {author} {\bibfnamefont {O.}~\bibnamefont {Gamayun}},\ }\href@noop {}
  {\bibinfo {title} {Large time and long distance asymptotics of the thermal
  correlators of the impenetrable anyonic lattice gas}} (\bibinfo {year}
  {2021}),\ \Eprint {https://arxiv.org/abs/2110.06860} {arXiv:2110.06860
  [cond-mat.quant-gas]} \BibitemShut {NoStop}%
\bibitem [{\citenamefont {Chernowitz}\ and\ \citenamefont
  {Gamayun}(2022)}]{chernowitz2022dynamics}%
  \BibitemOpen
  \bibfield  {author} {\bibinfo {author} {\bibfnamefont {D.}~\bibnamefont
  {Chernowitz}}\ and\ \bibinfo {author} {\bibfnamefont {O.}~\bibnamefont
  {Gamayun}},\ }\href@noop {} {\bibinfo {title} {On the dynamics of
  free-fermionic tau-functions at finite temperature}} (\bibinfo {year}
  {2022}),\ \Eprint {https://arxiv.org/abs/2110.08194} {arXiv:2110.08194
  [cond-mat.stat-mech]} \BibitemShut {NoStop}%
\bibitem [{\citenamefont {De~Nardis}\ \emph {et~al.}(2014)\citenamefont
  {De~Nardis}, \citenamefont {Wouters}, \citenamefont {Brockmann},\ and\
  \citenamefont {Caux}}]{PhysRevA.89.033601}%
  \BibitemOpen
  \bibfield  {author} {\bibinfo {author} {\bibfnamefont {J.}~\bibnamefont
  {De~Nardis}}, \bibinfo {author} {\bibfnamefont {B.}~\bibnamefont {Wouters}},
  \bibinfo {author} {\bibfnamefont {M.}~\bibnamefont {Brockmann}},\ and\
  \bibinfo {author} {\bibfnamefont {J.-S.}\ \bibnamefont {Caux}},\ }\bibfield
  {title} {\bibinfo {title} {Solution for an interaction quench in the
  lieb-liniger bose gas},\ }\href {https://doi.org/10.1103/PhysRevA.89.033601}
  {\bibfield  {journal} {\bibinfo  {journal} {Phys. Rev. A}\ }\textbf {\bibinfo
  {volume} {89}},\ \bibinfo {pages} {033601} (\bibinfo {year}
  {2014})}\BibitemShut {NoStop}%
\bibitem [{\citenamefont {Castella}\ \emph {et~al.}(1995)\citenamefont
  {Castella}, \citenamefont {Zotos},\ and\ \citenamefont
  {Prelov\ifmmode~\check{s}\else \v{s}\fi{}ek}}]{PhysRevLett.74.972}%
  \BibitemOpen
  \bibfield  {author} {\bibinfo {author} {\bibfnamefont {H.}~\bibnamefont
  {Castella}}, \bibinfo {author} {\bibfnamefont {X.}~\bibnamefont {Zotos}},\
  and\ \bibinfo {author} {\bibfnamefont {P.}~\bibnamefont
  {Prelov\ifmmode~\check{s}\else \v{s}\fi{}ek}},\ }\bibfield  {title} {\bibinfo
  {title} {Integrability and ideal conductance at finite temperatures},\ }\href
  {https://doi.org/10.1103/PhysRevLett.74.972} {\bibfield  {journal} {\bibinfo
  {journal} {Phys. Rev. Lett.}\ }\textbf {\bibinfo {volume} {74}},\ \bibinfo
  {pages} {972} (\bibinfo {year} {1995})}\BibitemShut {NoStop}%
\bibitem [{\citenamefont {Korepin}\ \emph {et~al.}(1993)\citenamefont
  {Korepin}, \citenamefont {Bogoliubov},\ and\ \citenamefont
  {Izergin}}]{Korepin1993}%
  \BibitemOpen
  \bibfield  {author} {\bibinfo {author} {\bibfnamefont {V.~E.}\ \bibnamefont
  {Korepin}}, \bibinfo {author} {\bibfnamefont {N.~M.}\ \bibnamefont
  {Bogoliubov}},\ and\ \bibinfo {author} {\bibfnamefont {A.~G.}\ \bibnamefont
  {Izergin}},\ }\href {https://doi.org/10.1017/cbo9780511628832} {\emph
  {\bibinfo {title} {Quantum Inverse Scattering Method and Correlation
  Functions}}}\ (\bibinfo  {publisher} {Cambridge University Press},\ \bibinfo
  {year} {1993})\BibitemShut {NoStop}%
\bibitem [{\citenamefont {{Panfil}}\ and\ \citenamefont
  {{Caux}}(2014)}]{2014_PRA_Panfil}%
  \BibitemOpen
  \bibfield  {author} {\bibinfo {author} {\bibfnamefont {M.}~\bibnamefont
  {{Panfil}}}\ and\ \bibinfo {author} {\bibfnamefont {J.-S.}\ \bibnamefont
  {{Caux}}},\ }\bibfield  {title} {\bibinfo {title} {{Finite-temperature
  correlations in the Lieb-Liniger one-dimensional Bose gas}},\ }\href
  {https://doi.org/10.1103/PhysRevA.89.033605} {\bibfield  {journal} {\bibinfo
  {journal} {Phys. Rev. A}\ }\textbf {\bibinfo {volume} {89}},\ \bibinfo {eid}
  {033605} (\bibinfo {year} {2014})},\ \Eprint
  {https://arxiv.org/abs/1308.2887} {arXiv:1308.2887 [cond-mat.quant-gas]}
  \BibitemShut {NoStop}%
\bibitem [{\citenamefont {Bornemann}(2009)}]{Bornemann2009}%
  \BibitemOpen
  \bibfield  {author} {\bibinfo {author} {\bibfnamefont {F.}~\bibnamefont
  {Bornemann}},\ }\bibfield  {title} {\bibinfo {title} {On the numerical
  evaluation of fredholm determinants},\ }\href
  {https://doi.org/10.1090/s0025-5718-09-02280-7} {\bibfield  {journal}
  {\bibinfo  {journal} {Mathematics of Computation}\ }\textbf {\bibinfo
  {volume} {79}},\ \bibinfo {pages} {871} (\bibinfo {year} {2009})}\BibitemShut
  {NoStop}%
\bibitem [{\citenamefont {Kitanine}\ \emph {et~al.}(2009)\citenamefont
  {Kitanine}, \citenamefont {Kozlowski}, \citenamefont {Maillet}, \citenamefont
  {Slavnov},\ and\ \citenamefont {Terras}}]{Kitanine_2009}%
  \BibitemOpen
  \bibfield  {author} {\bibinfo {author} {\bibfnamefont {N.}~\bibnamefont
  {Kitanine}}, \bibinfo {author} {\bibfnamefont {K.~K.}\ \bibnamefont
  {Kozlowski}}, \bibinfo {author} {\bibfnamefont {J.~M.}\ \bibnamefont
  {Maillet}}, \bibinfo {author} {\bibfnamefont {N.~A.}\ \bibnamefont
  {Slavnov}},\ and\ \bibinfo {author} {\bibfnamefont {V.}~\bibnamefont
  {Terras}},\ }\bibfield  {title} {\bibinfo {title}
  {{Riemann{\textendash}Hilbert Approach to a Generalised Sine Kernel and
  Applications}},\ }\href {https://doi.org/10.1007/s00220-009-0878-1}
  {\bibfield  {journal} {\bibinfo  {journal} {Communications in Mathematical
  Physics}\ }\textbf {\bibinfo {volume} {291}},\ \bibinfo {pages} {691}
  (\bibinfo {year} {2009})}\BibitemShut {NoStop}%
\bibitem [{\citenamefont {Slavnov}(2010)}]{Slavnov_2010}%
  \BibitemOpen
  \bibfield  {author} {\bibinfo {author} {\bibfnamefont {N.~A.}\ \bibnamefont
  {Slavnov}},\ }\bibfield  {title} {\bibinfo {title} {Integral operators with
  the generalized sine kernel on the real axis},\ }\href
  {https://doi.org/10.1007/s11232-010-0108-1} {\bibfield  {journal} {\bibinfo
  {journal} {Theoretical and Mathematical Physics}\ }\textbf {\bibinfo {volume}
  {165}},\ \bibinfo {pages} {1262} (\bibinfo {year} {2010})}\BibitemShut
  {NoStop}%
\bibitem [{\citenamefont {Shafarevich}\ and\ \citenamefont
  {Remizov}(2013)}]{Shafarevich2013}%
  \BibitemOpen
  \bibfield  {author} {\bibinfo {author} {\bibfnamefont {I.~R.}\ \bibnamefont
  {Shafarevich}}\ and\ \bibinfo {author} {\bibfnamefont {A.~O.}\ \bibnamefont
  {Remizov}},\ }\href {https://doi.org/10.1007/978-3-642-30994-6} {\emph
  {\bibinfo {title} {Linear Algebra and Geometry}}}\ (\bibinfo  {publisher}
  {Springer Berlin Heidelberg},\ \bibinfo {year} {2013})\BibitemShut {NoStop}%
\bibitem [{\citenamefont {Sachdev}(1996)}]{Sachdev1996}%
  \BibitemOpen
  \bibfield  {author} {\bibinfo {author} {\bibfnamefont {S.}~\bibnamefont
  {Sachdev}},\ }\bibfield  {title} {\bibinfo {title} {Universal,
  finite-temperature, crossover functions of the quantum transition in the
  ising chain in a transverse field},\ }\href
  {https://doi.org/10.1016/0550-3213(95)00657-5} {\bibfield  {journal}
  {\bibinfo  {journal} {Nuclear Physics B}\ }\textbf {\bibinfo {volume}
  {464}},\ \bibinfo {pages} {576} (\bibinfo {year} {1996})}\BibitemShut
  {NoStop}%
\bibitem [{\citenamefont {Tan}(2008)}]{Tan_2008}%
  \BibitemOpen
  \bibfield  {author} {\bibinfo {author} {\bibfnamefont {S.}~\bibnamefont
  {Tan}},\ }\bibfield  {title} {\bibinfo {title} {Large momentum part of a
  strongly correlated fermi gas},\ }\href
  {https://doi.org/10.1016/j.aop.2008.03.005} {\bibfield  {journal} {\bibinfo
  {journal} {Annals of Physics}\ }\textbf {\bibinfo {volume} {323}},\ \bibinfo
  {pages} {2971} (\bibinfo {year} {2008})}\BibitemShut {NoStop}%
\bibitem [{\citenamefont {Barth}\ and\ \citenamefont
  {Zwerger}(2011)}]{Barth2011}%
  \BibitemOpen
  \bibfield  {author} {\bibinfo {author} {\bibfnamefont {M.}~\bibnamefont
  {Barth}}\ and\ \bibinfo {author} {\bibfnamefont {W.}~\bibnamefont
  {Zwerger}},\ }\bibfield  {title} {\bibinfo {title} {Tan relations in one
  dimension},\ }\href {https://doi.org/10.1016/j.aop.2011.05.010} {\bibfield
  {journal} {\bibinfo  {journal} {Annals of Physics}\ }\textbf {\bibinfo
  {volume} {326}},\ \bibinfo {pages} {2544} (\bibinfo {year}
  {2011})}\BibitemShut {NoStop}%
\bibitem [{\citenamefont {Liu}\ \emph {et~al.}(2020{\natexlab{a}})\citenamefont
  {Liu}, \citenamefont {Shi}, \citenamefont {Levinsen},\ and\ \citenamefont
  {Parish}}]{PhysRevLett.125.065301}%
  \BibitemOpen
  \bibfield  {author} {\bibinfo {author} {\bibfnamefont {W.~E.}\ \bibnamefont
  {Liu}}, \bibinfo {author} {\bibfnamefont {Z.-Y.}\ \bibnamefont {Shi}},
  \bibinfo {author} {\bibfnamefont {J.}~\bibnamefont {Levinsen}},\ and\
  \bibinfo {author} {\bibfnamefont {M.~M.}\ \bibnamefont {Parish}},\ }\bibfield
   {title} {\bibinfo {title} {Radio-frequency response and contact of
  impurities in a quantum gas},\ }\href
  {https://doi.org/10.1103/PhysRevLett.125.065301} {\bibfield  {journal}
  {\bibinfo  {journal} {Phys. Rev. Lett.}\ }\textbf {\bibinfo {volume} {125}},\
  \bibinfo {pages} {065301} (\bibinfo {year} {2020}{\natexlab{a}})}\BibitemShut
  {NoStop}%
\bibitem [{\citenamefont {Liu}\ \emph {et~al.}(2020{\natexlab{b}})\citenamefont
  {Liu}, \citenamefont {Shi}, \citenamefont {Parish},\ and\ \citenamefont
  {Levinsen}}]{PhysRevA.102.023304}%
  \BibitemOpen
  \bibfield  {author} {\bibinfo {author} {\bibfnamefont {W.~E.}\ \bibnamefont
  {Liu}}, \bibinfo {author} {\bibfnamefont {Z.-Y.}\ \bibnamefont {Shi}},
  \bibinfo {author} {\bibfnamefont {M.~M.}\ \bibnamefont {Parish}},\ and\
  \bibinfo {author} {\bibfnamefont {J.}~\bibnamefont {Levinsen}},\ }\bibfield
  {title} {\bibinfo {title} {Theory of radio-frequency spectroscopy of
  impurities in quantum gases},\ }\href
  {https://doi.org/10.1103/PhysRevA.102.023304} {\bibfield  {journal} {\bibinfo
   {journal} {Phys. Rev. A}\ }\textbf {\bibinfo {volume} {102}},\ \bibinfo
  {pages} {023304} (\bibinfo {year} {2020}{\natexlab{b}})}\BibitemShut
  {NoStop}%
\bibitem [{\citenamefont {Hu}\ and\ \citenamefont {Liu}(2022)}]{hu2022fermi}%
  \BibitemOpen
  \bibfield  {author} {\bibinfo {author} {\bibfnamefont {H.}~\bibnamefont
  {Hu}}\ and\ \bibinfo {author} {\bibfnamefont {X.-J.}\ \bibnamefont {Liu}},\
  }\href@noop {} {\bibinfo {title} {Fermi polarons at finite temperature:
  Spectral function and rf-spectroscopy}} (\bibinfo {year} {2022}),\ \Eprint
  {https://arxiv.org/abs/2201.07872} {arXiv:2201.07872 [cond-mat.quant-gas]}
  \BibitemShut {NoStop}%
\bibitem [{\citenamefont {Doggen}\ and\ \citenamefont
  {Kinnunen}(2013)}]{PhysRevLett.111.025302}%
  \BibitemOpen
  \bibfield  {author} {\bibinfo {author} {\bibfnamefont {E.~V.~H.}\
  \bibnamefont {Doggen}}\ and\ \bibinfo {author} {\bibfnamefont {J.~J.}\
  \bibnamefont {Kinnunen}},\ }\bibfield  {title} {\bibinfo {title} {Energy and
  contact of the one-dimensional fermi polaron at zero and finite
  temperature},\ }\href {https://doi.org/10.1103/PhysRevLett.111.025302}
  {\bibfield  {journal} {\bibinfo  {journal} {Phys. Rev. Lett.}\ }\textbf
  {\bibinfo {volume} {111}},\ \bibinfo {pages} {025302} (\bibinfo {year}
  {2013})}\BibitemShut {NoStop}%
\bibitem [{\citenamefont {Sachdev}(2011)}]{sachdev2011quantum}%
  \BibitemOpen
  \bibfield  {author} {\bibinfo {author} {\bibfnamefont {S.}~\bibnamefont
  {Sachdev}},\ }\href@noop {} {\emph {\bibinfo {title} {Quantum phase
  transitions}}}\ (\bibinfo  {publisher} {Cambridge university press},\
  \bibinfo {year} {2011})\BibitemShut {NoStop}%
\bibitem [{\citenamefont {Kuhnle}\ \emph {et~al.}(2010)\citenamefont {Kuhnle},
  \citenamefont {Hu}, \citenamefont {Liu}, \citenamefont {Dyke}, \citenamefont
  {Mark}, \citenamefont {Drummond}, \citenamefont {Hannaford},\ and\
  \citenamefont {Vale}}]{PhysRevLett.105.070402}%
  \BibitemOpen
  \bibfield  {author} {\bibinfo {author} {\bibfnamefont {E.~D.}\ \bibnamefont
  {Kuhnle}}, \bibinfo {author} {\bibfnamefont {H.}~\bibnamefont {Hu}}, \bibinfo
  {author} {\bibfnamefont {X.-J.}\ \bibnamefont {Liu}}, \bibinfo {author}
  {\bibfnamefont {P.}~\bibnamefont {Dyke}}, \bibinfo {author} {\bibfnamefont
  {M.}~\bibnamefont {Mark}}, \bibinfo {author} {\bibfnamefont {P.~D.}\
  \bibnamefont {Drummond}}, \bibinfo {author} {\bibfnamefont {P.}~\bibnamefont
  {Hannaford}},\ and\ \bibinfo {author} {\bibfnamefont {C.~J.}\ \bibnamefont
  {Vale}},\ }\bibfield  {title} {\bibinfo {title} {Universal behavior of pair
  correlations in a strongly interacting fermi gas},\ }\href
  {https://doi.org/10.1103/PhysRevLett.105.070402} {\bibfield  {journal}
  {\bibinfo  {journal} {Phys. Rev. Lett.}\ }\textbf {\bibinfo {volume} {105}},\
  \bibinfo {pages} {070402} (\bibinfo {year} {2010})}\BibitemShut {NoStop}%
\bibitem [{\citenamefont {Mukherjee}\ \emph {et~al.}(2019)\citenamefont
  {Mukherjee}, \citenamefont {Patel}, \citenamefont {Yan}, \citenamefont
  {Fletcher}, \citenamefont {Struck},\ and\ \citenamefont
  {Zwierlein}}]{PhysRevLett.122.203402}%
  \BibitemOpen
  \bibfield  {author} {\bibinfo {author} {\bibfnamefont {B.}~\bibnamefont
  {Mukherjee}}, \bibinfo {author} {\bibfnamefont {P.~B.}\ \bibnamefont
  {Patel}}, \bibinfo {author} {\bibfnamefont {Z.}~\bibnamefont {Yan}}, \bibinfo
  {author} {\bibfnamefont {R.~J.}\ \bibnamefont {Fletcher}}, \bibinfo {author}
  {\bibfnamefont {J.}~\bibnamefont {Struck}},\ and\ \bibinfo {author}
  {\bibfnamefont {M.~W.}\ \bibnamefont {Zwierlein}},\ }\bibfield  {title}
  {\bibinfo {title} {Spectral response and contact of the unitary fermi gas},\
  }\href {https://doi.org/10.1103/PhysRevLett.122.203402} {\bibfield  {journal}
  {\bibinfo  {journal} {Phys. Rev. Lett.}\ }\textbf {\bibinfo {volume} {122}},\
  \bibinfo {pages} {203402} (\bibinfo {year} {2019})}\BibitemShut {NoStop}%
\bibitem [{\citenamefont {Hoinka}\ \emph {et~al.}(2013)\citenamefont {Hoinka},
  \citenamefont {Lingham}, \citenamefont {Fenech}, \citenamefont {Hu},
  \citenamefont {Vale}, \citenamefont {Drut},\ and\ \citenamefont
  {Gandolfi}}]{PhysRevLett.110.055305}%
  \BibitemOpen
  \bibfield  {author} {\bibinfo {author} {\bibfnamefont {S.}~\bibnamefont
  {Hoinka}}, \bibinfo {author} {\bibfnamefont {M.}~\bibnamefont {Lingham}},
  \bibinfo {author} {\bibfnamefont {K.}~\bibnamefont {Fenech}}, \bibinfo
  {author} {\bibfnamefont {H.}~\bibnamefont {Hu}}, \bibinfo {author}
  {\bibfnamefont {C.~J.}\ \bibnamefont {Vale}}, \bibinfo {author}
  {\bibfnamefont {J.~E.}\ \bibnamefont {Drut}},\ and\ \bibinfo {author}
  {\bibfnamefont {S.}~\bibnamefont {Gandolfi}},\ }\bibfield  {title} {\bibinfo
  {title} {Precise determination of the structure factor and contact in a
  unitary fermi gas},\ }\href {https://doi.org/10.1103/PhysRevLett.110.055305}
  {\bibfield  {journal} {\bibinfo  {journal} {Phys. Rev. Lett.}\ }\textbf
  {\bibinfo {volume} {110}},\ \bibinfo {pages} {055305} (\bibinfo {year}
  {2013})}\BibitemShut {NoStop}%
\bibitem [{\citenamefont {Stewart}\ \emph {et~al.}(2010)\citenamefont
  {Stewart}, \citenamefont {Gaebler}, \citenamefont {Drake},\ and\
  \citenamefont {Jin}}]{PhysRevLett.104.235301}%
  \BibitemOpen
  \bibfield  {author} {\bibinfo {author} {\bibfnamefont {J.~T.}\ \bibnamefont
  {Stewart}}, \bibinfo {author} {\bibfnamefont {J.~P.}\ \bibnamefont
  {Gaebler}}, \bibinfo {author} {\bibfnamefont {T.~E.}\ \bibnamefont {Drake}},\
  and\ \bibinfo {author} {\bibfnamefont {D.~S.}\ \bibnamefont {Jin}},\
  }\bibfield  {title} {\bibinfo {title} {Verification of universal relations in
  a strongly interacting fermi gas},\ }\href
  {https://doi.org/10.1103/PhysRevLett.104.235301} {\bibfield  {journal}
  {\bibinfo  {journal} {Phys. Rev. Lett.}\ }\textbf {\bibinfo {volume} {104}},\
  \bibinfo {pages} {235301} (\bibinfo {year} {2010})}\BibitemShut {NoStop}%
\bibitem [{\citenamefont {Sagi}\ \emph {et~al.}(2012)\citenamefont {Sagi},
  \citenamefont {Drake}, \citenamefont {Paudel},\ and\ \citenamefont
  {Jin}}]{PhysRevLett.109.220402}%
  \BibitemOpen
  \bibfield  {author} {\bibinfo {author} {\bibfnamefont {Y.}~\bibnamefont
  {Sagi}}, \bibinfo {author} {\bibfnamefont {T.~E.}\ \bibnamefont {Drake}},
  \bibinfo {author} {\bibfnamefont {R.}~\bibnamefont {Paudel}},\ and\ \bibinfo
  {author} {\bibfnamefont {D.~S.}\ \bibnamefont {Jin}},\ }\bibfield  {title}
  {\bibinfo {title} {Measurement of the homogeneous contact of a unitary fermi
  gas},\ }\href {https://doi.org/10.1103/PhysRevLett.109.220402} {\bibfield
  {journal} {\bibinfo  {journal} {Phys. Rev. Lett.}\ }\textbf {\bibinfo
  {volume} {109}},\ \bibinfo {pages} {220402} (\bibinfo {year}
  {2012})}\BibitemShut {NoStop}%
\bibitem [{\citenamefont {Schmidt}\ \emph {et~al.}(2018)\citenamefont
  {Schmidt}, \citenamefont {Knap}, \citenamefont {Ivanov}, \citenamefont {You},
  \citenamefont {Cetina},\ and\ \citenamefont {Demler}}]{Schmidt2018}%
  \BibitemOpen
  \bibfield  {author} {\bibinfo {author} {\bibfnamefont {R.}~\bibnamefont
  {Schmidt}}, \bibinfo {author} {\bibfnamefont {M.}~\bibnamefont {Knap}},
  \bibinfo {author} {\bibfnamefont {D.~A.}\ \bibnamefont {Ivanov}}, \bibinfo
  {author} {\bibfnamefont {J.-S.}\ \bibnamefont {You}}, \bibinfo {author}
  {\bibfnamefont {M.}~\bibnamefont {Cetina}},\ and\ \bibinfo {author}
  {\bibfnamefont {E.}~\bibnamefont {Demler}},\ }\bibfield  {title} {\bibinfo
  {title} {Universal many-body response of heavy impurities coupled to a fermi
  sea: a review of recent progress},\ }\href
  {https://doi.org/10.1088/1361-6633/aa9593} {\bibfield  {journal} {\bibinfo
  {journal} {Reports on Progress in Physics}\ }\textbf {\bibinfo {volume}
  {81}},\ \bibinfo {pages} {024401} (\bibinfo {year} {2018})}\BibitemShut
  {NoStop}%
\bibitem [{\citenamefont {Cetina}\ \emph {et~al.}(2016)\citenamefont {Cetina},
  \citenamefont {Jag}, \citenamefont {Lous}, \citenamefont {Fritsche},
  \citenamefont {Walraven}, \citenamefont {Grimm}, \citenamefont {Levinsen},
  \citenamefont {Parish}, \citenamefont {Schmidt}, \citenamefont {Knap},\ and\
  \citenamefont {Demler}}]{Cetina2016}%
  \BibitemOpen
  \bibfield  {author} {\bibinfo {author} {\bibfnamefont {M.}~\bibnamefont
  {Cetina}}, \bibinfo {author} {\bibfnamefont {M.}~\bibnamefont {Jag}},
  \bibinfo {author} {\bibfnamefont {R.~S.}\ \bibnamefont {Lous}}, \bibinfo
  {author} {\bibfnamefont {I.}~\bibnamefont {Fritsche}}, \bibinfo {author}
  {\bibfnamefont {J.~T.~M.}\ \bibnamefont {Walraven}}, \bibinfo {author}
  {\bibfnamefont {R.}~\bibnamefont {Grimm}}, \bibinfo {author} {\bibfnamefont
  {J.}~\bibnamefont {Levinsen}}, \bibinfo {author} {\bibfnamefont {M.~M.}\
  \bibnamefont {Parish}}, \bibinfo {author} {\bibfnamefont {R.}~\bibnamefont
  {Schmidt}}, \bibinfo {author} {\bibfnamefont {M.}~\bibnamefont {Knap}},\ and\
  \bibinfo {author} {\bibfnamefont {E.}~\bibnamefont {Demler}},\ }\bibfield
  {title} {\bibinfo {title} {Ultrafast many-body interferometry of impurities
  coupled to a fermi sea},\ }\href {https://doi.org/10.1126/science.aaf5134}
  {\bibfield  {journal} {\bibinfo  {journal} {Science}\ }\textbf {\bibinfo
  {volume} {354}},\ \bibinfo {pages} {96} (\bibinfo {year} {2016})}\BibitemShut
  {NoStop}%
\bibitem [{\citenamefont {Imambekov}\ and\ \citenamefont
  {Glazman}(2009)}]{Imambekov2009}%
  \BibitemOpen
  \bibfield  {author} {\bibinfo {author} {\bibfnamefont {A.}~\bibnamefont
  {Imambekov}}\ and\ \bibinfo {author} {\bibfnamefont {L.~I.}\ \bibnamefont
  {Glazman}},\ }\bibfield  {title} {\bibinfo {title} {{Universal Theory of
  Nonlinear Luttinger Liquids}},\ }\href
  {https://doi.org/10.1126/science.1165403} {\bibfield  {journal} {\bibinfo
  {journal} {Science}\ }\textbf {\bibinfo {volume} {323}},\ \bibinfo {pages}
  {228} (\bibinfo {year} {2009})}\BibitemShut {NoStop}%
\bibitem [{\citenamefont {Imambekov}\ \emph {et~al.}(2012)\citenamefont
  {Imambekov}, \citenamefont {Schmidt},\ and\ \citenamefont
  {Glazman}}]{RevModPhys.84.1253}%
  \BibitemOpen
  \bibfield  {author} {\bibinfo {author} {\bibfnamefont {A.}~\bibnamefont
  {Imambekov}}, \bibinfo {author} {\bibfnamefont {T.~L.}\ \bibnamefont
  {Schmidt}},\ and\ \bibinfo {author} {\bibfnamefont {L.~I.}\ \bibnamefont
  {Glazman}},\ }\bibfield  {title} {\bibinfo {title} {{One-dimensional quantum
  liquids: Beyond the Luttinger liquid paradigm}},\ }\href
  {https://doi.org/10.1103/RevModPhys.84.1253} {\bibfield  {journal} {\bibinfo
  {journal} {Rev. Mod. Phys.}\ }\textbf {\bibinfo {volume} {84}},\ \bibinfo
  {pages} {1253} (\bibinfo {year} {2012})}\BibitemShut {NoStop}%
\bibitem [{\citenamefont {Markhof}\ \emph {et~al.}(2019)\citenamefont
  {Markhof}, \citenamefont {Pletyukhov},\ and\ \citenamefont
  {Meden}}]{10.21468/SciPostPhys.7.4.047}%
  \BibitemOpen
  \bibfield  {author} {\bibinfo {author} {\bibfnamefont {L.}~\bibnamefont
  {Markhof}}, \bibinfo {author} {\bibfnamefont {M.}~\bibnamefont
  {Pletyukhov}},\ and\ \bibinfo {author} {\bibfnamefont {V.}~\bibnamefont
  {Meden}},\ }\bibfield  {title} {\bibinfo {title} {{Investigating the roots of
  the nonlinear Luttinger liquid phenomenology}},\ }\href
  {https://doi.org/10.21468/SciPostPhys.7.4.047} {\bibfield  {journal}
  {\bibinfo  {journal} {SciPost Phys.}\ }\textbf {\bibinfo {volume} {7}},\
  \bibinfo {pages} {47} (\bibinfo {year} {2019})}\BibitemShut {NoStop}%
\bibitem [{\citenamefont {Zvonarev}\ \emph {et~al.}(2007)\citenamefont
  {Zvonarev}, \citenamefont {Cheianov},\ and\ \citenamefont
  {Giamarchi}}]{PhysRevLett.99.240404}%
  \BibitemOpen
  \bibfield  {author} {\bibinfo {author} {\bibfnamefont {M.~B.}\ \bibnamefont
  {Zvonarev}}, \bibinfo {author} {\bibfnamefont {V.~V.}\ \bibnamefont
  {Cheianov}},\ and\ \bibinfo {author} {\bibfnamefont {T.}~\bibnamefont
  {Giamarchi}},\ }\bibfield  {title} {\bibinfo {title} {Spin dynamics in a
  one-dimensional ferromagnetic bose gas},\ }\href
  {https://doi.org/10.1103/PhysRevLett.99.240404} {\bibfield  {journal}
  {\bibinfo  {journal} {Phys. Rev. Lett.}\ }\textbf {\bibinfo {volume} {99}},\
  \bibinfo {pages} {240404} (\bibinfo {year} {2007})}\BibitemShut {NoStop}%
\bibitem [{\citenamefont {Zvonarev}\ \emph {et~al.}(2009)\citenamefont
  {Zvonarev}, \citenamefont {Cheianov},\ and\ \citenamefont
  {Giamarchi}}]{PhysRevLett.103.110401}%
  \BibitemOpen
  \bibfield  {author} {\bibinfo {author} {\bibfnamefont {M.~B.}\ \bibnamefont
  {Zvonarev}}, \bibinfo {author} {\bibfnamefont {V.~V.}\ \bibnamefont
  {Cheianov}},\ and\ \bibinfo {author} {\bibfnamefont {T.}~\bibnamefont
  {Giamarchi}},\ }\bibfield  {title} {\bibinfo {title} {Dynamical properties of
  the one-dimensional spin-$1/2$ bose-hubbard model near a mott-insulator to
  ferromagnetic-liquid transition},\ }\href
  {https://doi.org/10.1103/PhysRevLett.103.110401} {\bibfield  {journal}
  {\bibinfo  {journal} {Phys. Rev. Lett.}\ }\textbf {\bibinfo {volume} {103}},\
  \bibinfo {pages} {110401} (\bibinfo {year} {2009})}\BibitemShut {NoStop}%
\bibitem [{\citenamefont {Caux}(2016)}]{QA_Caux}%
  \BibitemOpen
  \bibfield  {author} {\bibinfo {author} {\bibfnamefont {J.-S.}\ \bibnamefont
  {Caux}},\ }\bibfield  {title} {\bibinfo {title} {The quench action},\ }\href
  {http://stacks.iop.org/1742-5468/2016/i=6/a=064006} {\bibfield  {journal}
  {\bibinfo  {journal} {J. Stat. Mech. Theor. Exp.}\ }\textbf {\bibinfo
  {volume} {2016}},\ \bibinfo {pages} {064006} (\bibinfo {year}
  {2016})}\BibitemShut {NoStop}%
\bibitem [{\citenamefont {Bastianello}\ \emph {et~al.}(2022)\citenamefont
  {Bastianello}, \citenamefont {Bertini}, \citenamefont {Doyon},\ and\
  \citenamefont {Vasseur}}]{Bastianello_2022}%
  \BibitemOpen
  \bibfield  {author} {\bibinfo {author} {\bibfnamefont {A.}~\bibnamefont
  {Bastianello}}, \bibinfo {author} {\bibfnamefont {B.}~\bibnamefont
  {Bertini}}, \bibinfo {author} {\bibfnamefont {B.}~\bibnamefont {Doyon}},\
  and\ \bibinfo {author} {\bibfnamefont {R.}~\bibnamefont {Vasseur}},\
  }\bibfield  {title} {\bibinfo {title} {Introduction to the special issue on
  emergent hydrodynamics in integrable many-body systems},\ }\href
  {https://doi.org/10.1088/1742-5468/ac3e6a} {\bibfield  {journal} {\bibinfo
  {journal} {Journal of Statistical Mechanics: Theory and Experiment}\ }\textbf
  {\bibinfo {volume} {2022}},\ \bibinfo {pages} {014001} (\bibinfo {year}
  {2022})}\BibitemShut {NoStop}%
\end{thebibliography}%
\end{document}